\documentclass[floatfix,%
 reprint,%
 amssymb, amsmath,%
 aip,jcp,
frontmatterverbose,
]{revtex4-1}

\usepackage{amsmath,amssymb,amsfonts,mathtools}
\usepackage{multirow}
\usepackage{float}
\usepackage{csquotes}
\usepackage{dsfont}
\usepackage[mathscr]{euscript}
\usepackage{graphicx}
\usepackage{amsfonts}
\usepackage{amsmath}
\usepackage{stmaryrd}
\usepackage{bm}%
\usepackage{longtable}
\usepackage{supertabular}
\usepackage{rotating}
\usepackage{comment}
\usepackage[normalem]{ulem}
\usepackage{ifthen}
\newboolean{includeMQDT}
\setboolean{includeMQDT}{false}  

\usepackage[dvipsnames]{xcolor}

\usepackage{dcolumn,ulem}  
\usepackage{xspace}

\begin{document}
%
%
%
\title{\large Non-adiabatic Ring Polymer Molecular Dynamics in the Phase Space of the ${SU}(N)$ Lie Group}
\author{Duncan Bossion}
\email{dbossion@ur.rochester.edu}
\affiliation{Department of Chemistry, University of Rochester, 120 Trustee Road, Rochester, New York 14627, USA}
\author{Sutirtha N. Chowdhury}
\affiliation{Department of Chemistry, University of Rochester, 120 Trustee Road, Rochester, New York 14627, USA}
\author{Pengfei Huo}
\email{pengfei.huo@rochester.edu}
\affiliation{Department of Chemistry, University of Rochester, 120 Trustee Road, Rochester, New York 14627, USA}
\affiliation{The Institute of Optics, Hajim School of Engineering, University of Rochester, Rochester, New York 14627, USA}

\date{\today}
\begin{abstract}
We derive the non-adiabatic ring polymer molecular dynamics (RPMD) approach in the phase space of the ${SU}(N)$ Lie Group. This method, which we refer to as the spin mapping non-adiabatic RPMD (SM-NRPMD), is based on the spin-mapping formalism for the electronic degrees of freedom (DOFs) and ring polymer path-integral description for the nuclear DOFs. Using the Stratonovich-Weyl transform for the electronic DOFs, and the Wigner transform for the nuclear DOFs, we derived an exact expression of the  Kubo-transformed time-correlation function (TCF). We further derive the spin mapping non-adiabatic  Matsubara dynamics using the Matsubara approximation that removes the high frequency nuclear normal modes in the TCF and derive the SM-NRPMD approach from the non-adiabatic  Matsubara dynamics by discarding the imaginary part of the Liouvillian. The SM-NRPMD method has numerical advantages compared to the original NRPMD method based on the MMST mapping formalism, due to a more natural mapping using the ${SU}(N)$ Lie Group that preserves the symmetry of the original system. We numerically compute the Kubo-transformed position auto-correlation function and electronic population correlation function for three-state model systems. The numerical results demonstrate the accuracy of the SM-NRPMD method, which outperforms the original MMST-based NRPMD. We envision that the SM-NRPMD  method will be a powerful approach to simulate electronic non-adiabatic dynamics and nuclear quantum effects accurately.
\end{abstract}
\maketitle
\section{Introduction}\label{intro}
Accurately simulating the quantum dynamics of molecular systems in condensed phase remains a challenge in theoretical chemistry, due to the difficulties of accurately describing electronically non-adiabatic dynamics and nuclear quantum effects.\cite{tully2012} Such effects are inherent to a lot of key reactions in biochemistry, catalysis, and energy applications that involve electron transfer or proton-coupled electron transfer processes.\cite{marcus1985,gray1996,reece2009} Despite the recent progress on new theoretical approaches to study those types of reactions, the challenge remains for large systems with many degrees of freedom (DOFs), and the exact quantum simulations remain computationally expensive due to the unfavorable numerical scaling. 

Recently emerged state-dependent RPMD approaches provide a unified description of both the electronically non-adiabatic dynamics and nuclear quantum effects, using a trajectory-based description. Ring-polymer molecular dynamics (RPMD)\cite{craig2004,craig2005,craig2005_2,suleimanov2013,suleimanov2013_2} based on Feynman's imaginary-time path-integral formalism was originally developed for electronically adiabatic systems to effectively capture the nuclear quantum effects through extended phase space quantization. State-dependent RPMD methods are developed based on the RPMD framework, to provide both accurate non-adiabatic dynamics with an explicit description of electronic states, and a reliable treatment of nuclear quantum dynamics through the ring polymer path-integral quantization. These methods include non-adiabatic RPMD (NRPMD),\cite{richardson2013,richardson2017,chowdhury2019} mapping variable RPMD (MV-RPMD),\cite{ananth2013,duke2015,pierre2017} ring-polymer Ehrenfest dynamics,\cite{yoshikawa2013} kinetically-constrained RPMD (KC-RPMD),\cite{menzeleev2014,kretchmer2016} coherent state RPMD (CS-RPMD),\cite{chowdhury2017} and ring-polymer surface hopping (RPSH) \cite{shushkov2012,shakib2017,tao2018,tao2019} to name a few.

Among these various state-dependent RPMD approaches, NRPMD\cite{richardson2013,richardson2017,chowdhury2019,chowdhury2021} provides both accurate nuclear quantum dynamics and accurate electronic Rabi oscillations, thus being an ideal approach to investigate quantum dynamics of non-adiabatic systems.\cite{chowdhury2019,Chowdhury2021JCP-pola} This approach is based on the Meyer-Miller-Stock-Thoss (MMST) mapping formalism\cite{meyer1979_2,stock1997,thoss1999} and has been formally derived using the non-adiabatic Matsubara dynamics formalism.\cite{chowdhury2021} This mapping formalism, despite its great success and broad applications,\cite{miller1997,miller2001,bonella01,kim2008,HuoPLDM,huo2012,Miller:2016b,richardson2017,richardson2013} has known flaws.\cite{ananth2010,kelly2012,saller2019} In particular, it maps a $N$-level system onto $N$ singly excited harmonic oscillators (SEO), resulting in mapping variables that are conjugate position and momentum of each oscillator. This leads to a symmetry  $\otimes_N\mathfrak{su}(2)$ for the mapped system as each harmonic oscillator in this subspace has a symmetry $\mathfrak{su}(2)$ (only the ground and first excited states of the oscillators belong to the SEO subspace), whereas the original symmetry of a $N$-level system should be $\mathfrak{su}(N)$. Due to this mapping procedure, the MMST mapping operators belong to a larger Hilbert space that contains states outside the SEO subspace, whereas the MMST mapping procedure tries to map the original electronic subspace onto the SEO subspace. It thus requires a projection back to the SEO subspace to obtain accurate results.\cite{ananth2010,kelly2012,saller2019} As a consequence, the identity operator is not preserved through the MMST mapping and there is an ambiguity on how to evaluate it.\cite{saller2019} Related to the problem of the non-conserving identity, the non-adiabatic dynamics is sensitive to the separation between the state-dependent and the state-independent Hamiltonian.\cite{kelly2012,cotton2013}

Recently, a new mapping formalism based on the phase space of the ${SU}(N)$ Lie Group has been proposed by Runeson and Richardson.\cite{runesonrichardson2019,runesonrichardson2020} This new mapping formalism, referred to as the generalized spin mapping, preserves the original $\mathfrak{su}(N)$ symmetry of the Hamiltonian. Runeson and Richardson used the spin operators (which are equivalent to the generators of the $\mathfrak{su}(N)$ Lie algebra up to a constant) and the $SU(N)$ Lie group to perform the non-adiabatic mapping dynamics of a $N$-state vibronic Hamiltonian and developed the spin-Linearized semi-classical (spin-LSC) approach.\cite{runesonrichardson2020} In particular, the Stratonovich-Weyl (SW) transform\cite{stratonovich1957,varilly1989,brif1999,Klimov2009} is used to map an operator in the Hilbert space described by the spin operators to a continuous function on the Lie group/manifold, resulting in a classical-like Hamiltonian. The SW transform evaluates the expectation values of the spin operators under the generalized spin coherent states.\cite{radcliffe1971,nemoto2000} We have also provided a detailed derivation of the quantum Liouvillian and the linearized Liouvillian based on the spin mapping framework.\cite{bossion2022} We have also proposed a spin mapping NRPMD dynamics approach\cite{bossion2021} for two-level systems ($N=2$).

In this work, we apply the spin mapping formalism to describe the electronic DOFs and Wigner representation for the nuclear DOFs and derive the expression of the exact Kubo-transformed time correlation function (TCF). Applying the Matsubara approximation\cite{althorpe2015} and ring polymer approximation\cite{althorpe2016} to this exact TCF leads to the spin-mapping NRPMD (SM-NRPMD) method. The spin mapping formalism respecting the symmetry of the original system, the SM-NRPMD approach explicitly addresses the limitations of the NRPMD method\cite{richardson2013} due to the deficiencies of the MMST mapping formalism. The outline of the paper is described as follows. In Sec.~\ref{sun},  a brief overview of the $SU(N)$ mapping formalism is provided, together with the procedure to map a general operator in the electronic and nuclear Hilbert space through a mixed Stratonovich-Weyl/Wigner transformation.
In Sec.~\ref{kubotcf}, we provide an exact expression of the Kubo-transformed TCF and the corresponding exact quantum Liouvillian. In Sec.~\ref{matnrpmd} we introduce the Matsubara approximation and the RPMD approximation, upon which we derive the SM-NRPMD correlation function. The accuracy of the SM-NRPMD method is tested using three-level systems in Secs.~\ref{comp}-\ref{result} and compared to the MMST-based NRPMD approach\cite{richardson2013} demonstrating a significant improvement due to the spin mapping formalism. The conclusions and future directions are provided in Sec.~\ref{conclusion}. 


\section{The $SU(N)$ mapping formalism}\label{sun}
In this section, we briefly outline the basic idea of the $SU(N)$ mapping formalism, where the details can be found in the previous works.\cite{runesonrichardson2020,bossion2022} We are interested in the quantum dynamics of a system of $N$ electronic states coupled to nuclear DOFs as follows
\begin{align}\label{eq:Ham}
\hat{H}=&\big[\hat{T}_{R}+U_{0}(\hat{R})\big]\otimes\hat{\mathcal{I}}+\hat{V}_\mathrm{e}(\hat{R})\\
=&\big[\hat{T}_{R}+U_{0}(\hat{R})\big]\otimes\hat{\mathcal{I}}+\sum_{n,m}V_{nm}(\hat{R})|n\rangle\langle m|,\nonumber
\end{align}
where $\hat{T}_{R}$ is the nuclear kinetic energy operator, $U_{0}(\hat{R})$ represents the state-independent part of the potential, and $\hat{V}_\mathrm{e}(\hat{R})$ is the state-dependent part of the potential. While we consider one nuclear DOF $\hat{R}$ for convenience, the theory in this work can be easily generalized to many nuclear DOFs. Furthermore, $\{|n\rangle\}$ represents a set of diabatic electronic states, and $V_{nm}(\hat{R})=\langle n|\hat{V}_\mathrm{e}(\hat{R})|m\rangle$ is the matrix element of $\hat{V}_\mathrm{e}(\hat{R})$ in this diabatic representation. The electronic identity operator $\hat{\mathcal{I}}=\sum_{n=1}^{N}|n\rangle\langle n|$ represents the identity in the electronic Hilbert space. Note that this is a rather general Hamiltonian, representing a quantum subsystem that has $N$ states coupled to a ``classical" subsystem.

We aim to evaluate the TCF governed by the Hamiltonian in Eq.~\ref{eq:Ham}. In the following sections, we use the $SU(N)$ mapping formalism to map discrete electronic states onto continuous variables in the $SU(N)$ phase space, and the Wigner representation to describe the nuclear DOFs. The detailed derivation of all expressions can be found in our previous work in Ref.~\citenum{bossion2022}. This mixed representation will be used to express the Kubo-transformed TCF in Sec.~\ref{matnrpmd}.

\subsection{The Spin Mapping Formalism in the ${SU}(N)$ Representation}
We briefly review the general expressions of the generators of the $\mathfrak{su}(N)$ Lie algebra, which will be used as a matrix basis to represent electronic operators. The commutation (and anti-commutation) relations among these generators are defined in the $\mathfrak{su}(N)$ Lie algebra, whereas the exponential functions of these generators construct the elements of the $SU(N)$ Lie group via the exponential map.\cite{georgi2000, GTM222} 

The generators, denoted as $\hat{\mathcal{S}}_{i}$, are expressed in Appendix~\ref{fijk}. These generators are traceless, $\mathrm{Tr_{e}}\big[\hat{\mathcal S}_{i}\big]=0$, and are orthonormal to each other as $\mathrm{Tr_e}\big[\hat{\mathcal{S}}_i\hat{\mathcal{S}}_j\big]=\frac{\hbar^2}{2}\delta_{ij}$. The commutation and anti-commutation relations among the generators of the $\mathfrak{su}(N)$ Lie algebra are presented as follows
\begin{subequations}\label{generator}
    \begin{align}
        \big[\hat{\mathcal{S}}_i,\hat{\mathcal{S}}_j\big]&=i\hbar\sum_{k=1}^{N^2-1}f_{ijk}\hat{\mathcal{S}}_k,\label{f-ijk}\\
        \big\{\hat{\mathcal{S}}_i,\hat{\mathcal{S}}_j\big\}_{+}&    =\frac{\hbar^2}{N}\delta_{ij}\hat{\mathcal{I}}+\hbar\sum_{k=1}^{N^2-1}d_{ijk}\hat{\mathcal{S}}_k,\label{d-ijk}
    \end{align}
\end{subequations}
where $\big\{\hat{\mathcal{S}}_i,\hat{\mathcal{S}}_j\big\}_{+}$ represents the anti-commutator between $\hat{\mathcal{S}}_i$ and $\hat{\mathcal{S}}_j$, and $f_{ijk}$ and $d_{ijk}$ are the totally anti-symmetric and totally symmetric {\it structure constants}, respectively. Using Eqs.~\ref{f-ijk}-\ref{d-ijk}, one can express the structure constants as
\begin{subequations}\label{fdtrace}
    \begin{align}
        f_{ijk}=-i\frac{2}{\hbar^3}\mathrm{Tr}\Big[\big[\hat{\mathcal{S}}_i,\hat{\mathcal{S}}_j\big]\hat{\mathcal{S}}_k\Big],\\
        d_{ijk}=\frac{2}{\hbar^3}\mathrm{Tr}\Big[\big\{\hat{\mathcal{S}}_i,\hat{\mathcal{S}}_j\big\}_{+}\hat{\mathcal{S}}_k\Big].
    \end{align}
\end{subequations}

The generators of an algebra $\{\hat{\mathcal S}_{k}\}$ can be obtained in different ways, but the most commonly used ones are based on a generalization of the Pauli matrices of $\mathfrak{su}(2)$ and of the Gell-Mann matrices\cite{gellmann1962} of $\mathfrak{su}(3)$, which is what is used in this work. This representation of the generators is referred to as the Generalized Gell–Mann matrix (GGM) basis.\cite{pfeifer2003,tilma2011} Their detailed expressions are provided in Appendix~\ref{fijk} (see Eq.~
\ref{ssym}-Eq.~\ref{sdiag}). One can also derive an analytic expression (closed formulas) of $f_{ijk}$ and $d_{ijk}$ for the GGM basis, which can also be found in Appendix~\ref{fijk}. The derivation of them can be found in our previous work in Ref.~\citenum{bossion2022} 

Using these generators, the Hamiltonian $\hat{H}$ (Eq.~\ref{eq:Ham}) is represented as follows\cite{hioeeberly1981,runesonrichardson2020}
\begin{equation}\label{eq:Hsun}
    \hat{H}={\mathcal{H}}_{0}(\hat{R},\hat{P})\cdot\hat{\mathcal{I}}+\frac{1}{\hbar}\sum_{k=1}^{N^2-1}{\mathcal{H}}_k(\hat{R})\cdot\hat{\mathcal{S}}_{k},
\end{equation}
where the elements ${\mathcal{H}}_{0}(\hat{R},\hat{P})$ and ${\mathcal{H}}_k(\hat{R})$ are expressed as
\begin{subequations}\label{SW_Ham}
    \begin{align}
        \mathcal{H}_0(\hat{R},\hat{P})&=\frac{1}{N}\mathrm{Tr}_\mathrm{e}\big[\hat{H}\cdot\hat{\mathcal I}\big]=\hat{T}_{R}+U_0(\hat{R})+\frac{1}{N}\sum_{n=1}^NV_{nn}(\hat{R}),\\
        \mathcal{H}_k(\hat{R})&=\frac{2}{\hbar}\mathrm{Tr_e}\big[\hat{H}\cdot\hat{\mathcal S}_k\big]=\frac{2}{\hbar}\mathrm{Tr_e}\big[\hat{V}_\mathrm{e}(\hat{R})\cdot\hat{\mathcal S}_k\big].\label{eq:hk}
    \end{align}
\end{subequations}
Note that Eq.~\ref{SW_Ham} has an explicit separation of the trace and traceless parts of the potential, due to the traceless definition of the generators. Furthermore, we can explicitly write $\mathcal{H}_k(\hat{R})$ in Eq.~\ref{eq:hk} as
\begin{subequations}\label{eq:Hk}
    \begin{align}
        &{\mathcal H}_{\alpha_{nm}}(\hat{R})={V}_{mn}(\hat{R})+{V}_{nm}(\hat{R}),\\
        &{\mathcal H}_{\beta_{nm}}(\hat{R})=i\big({V}_{mn}(\hat{R})-{V}_{nm}(\hat{R})\big),\\
        &{\mathcal H}_{\gamma_{n}}(\hat{R})=\sum_{l=1}^{n-1}\sqrt{\frac{2}{n(n-1)}}{V}_{ll}(\hat{R})-\sqrt{\frac{2(n-1)}{n}}{V}_{nn}(\hat{R}),
    \end{align}
\end{subequations}
where $\alpha_{nm}=n^2+2(m-n)-1$ is the index related to the symmetric non-diagonal generators, $\hat{\mathcal S}_{\alpha_{nm}}$ (Eq.~\ref{ssym}), $\beta_{nm}=n^2+2(m-n)$ is the index related to the asymmetric non-diagonal generators, $\hat{\mathcal S}_{\beta{nm}}$ (Eq.~\ref{sasym}), and $\gamma_{n}=n^2-1$ is the index related to the diagonal generators $\hat{\mathcal S}_{\gamma_{n}}$ (Eq.~\ref{sdiag}), with $1\leq m<n\leq N$ and $2\leq n\leq N$.

\subsection{The Stratonovich-Weyl Transform}
The Stratonovich-Weyl (SW) transform evaluates the expectation values of the spin operators under the generalized spin coherent states.\cite{radcliffe1971,nemoto2000} The generalized spin coherent states are expressed as
  \begin{equation}\label{eq:omega-expand}
    |\mathbf{\Omega}\rangle=\sum_{n=1}^{N}|n\rangle\langle n|\mathbf{\Omega}\rangle,
\end{equation}
where the detailed expression of the expansion coefficients, $\langle n|\mathbf{\Omega}\rangle$, are provided in Eq.~\ref{absscs}. The spin coherent states are normalized, $\langle\mathbf{\Omega}|\mathbf{\Omega}\rangle=1$, and they form a resolution of identity,\cite{nemoto2000,tilma2002,tilma2004,tilma2011}
\begin{equation}\label{idNlevel}
    \hat{\mathcal{I}}=\sum_{n}|n\rangle\langle n|=\int d \mathbf{\Omega}|\mathbf{\Omega}\rangle\langle\mathbf{\Omega}|,
\end{equation}
where the expression of the differential phase-space volume element $d\mathbf{\Omega}\equiv \mathbf{K}(\boldsymbol{\theta})d \boldsymbol{\theta} d \boldsymbol{\varphi}$ is provided in Eq.~\ref{dOmega} in terms of the $2N-2$ independent variables, $\{\boldsymbol{\theta},\boldsymbol{\varphi}\}$, which are the generalized Euler angles of the $N$-dimensional Bloch sphere. In the following, we define the expectation value of the generalized spin operators as
\begin{equation}\label{Omegak}
    \hbar\Omega_k\equiv\langle\mathbf{\Omega}|\hat{\mathcal{S}}_k|\mathbf{\Omega}\rangle,
\end{equation}
where $\hbar\mathbf{\Omega}$ plays the role of the Bloch vector,\cite{Kimura2003,Krammer2008} and is referred to as the generalized Bloch vector.\cite{Kimura2003,Krammer2008} Its detailed expression in terms of $\{\boldsymbol{\theta},\boldsymbol{\varphi}\}$ is provided in Eqs.~\ref{eq:Omega_a}-\ref{eq:Omega_g}. 

The SW transform of an operator $\hat{A}$ is defined as
\begin{equation}\label{A-sw-map}
    \big[\hat{A}\big]_\mathrm{s}(\mathbf{\Omega})=\mathrm{Tr_e}\big[\hat{A}\cdot\hat{w}_\mathrm{s}\big],
\end{equation} 
where $\hat{w}_\mathrm{s}$ is the kernel of the SW transform, with the following expression\cite{brif1998,brif1999,tilma2011}
\begin{align}\label{SWkernel}
\hat{w}_\mathrm{s}(\mathbf{\Omega})&=\frac{1}{N}\hat{\mathcal{I}}+r_\mathrm{s}\cdot \frac{2}{\hbar}\sum_{i=1}^{N^2-1}\Omega_{i}\cdot\hat{\mathcal{S}}_i.
\end{align}
The parameter $r_\mathrm{s}$ is related to the radius of the Bloch sphere\cite{bloch1946,bloch1953,schwinger1954} representing the electronic states of the system. The kernel also defines an identity as follows
\begin{equation}\label{SW_Id}
\int d \mathbf{\Omega}\hat{w}_\mathrm{s} =\hat{\mathcal{I}},
\end{equation}
where the detailed proof can be found in Eq.~28 of Ref.~\citenum{bossion2022}.

The SW transform in Eq.~\ref{A-sw-map} constructs a mapping between an operator in the Hilbert space and a continuous function whose variables are $\{{\boldsymbol\theta}, {\boldsymbol\varphi}\}$ or $\{{\boldsymbol\Omega}\}$ on the $SU(N)$ Lie group/manifold. More specifically, this mapping formalism establishes the following relation
\begin{equation}
    \hat{A} \longrightarrow \big[\hat{A}\big]_\mathrm{s}(\mathbf{\Omega}),
\end{equation}
which is the basic idea of the generalized spin-mapping approach.\cite{runesonrichardson2019,runesonrichardson2020,bossion2022}

To conveniently evaluate any operator $\hat{A}$ under the SW transform, one first writes it using the GGM basis, $\{\hat{\mathcal{S}}_k\}$, as follows
\begin{equation}\label{SW_op}
    \hat{A}={\mathcal A}_0\cdot\hat{\mathcal I}+\frac{1}{\hbar}\sum_{k=1}^{N^2-1}{\mathcal A}_{k}\cdot\hat{\mathcal{S}}_k,
\end{equation}
where $\mathcal{A}_0$ and $\mathcal{A}_k$ are expressed as
\begin{subequations}\label{A_SW}
    \begin{align}
        \mathcal{A}_0&=\frac{1}{N}\mathrm{Tr}_\mathrm{e}\big[\hat{A}\cdot\hat{\mathcal I}\big]=\frac{1}{N}\sum_{n=1}^N A_{nn},\label{A0}\\
        \mathcal{A}_k&=\frac{2}{\hbar}\mathrm{Tr_e}\big[\hat{A}\cdot\hat{\mathcal S}_k\big],
    \end{align}
\end{subequations}
with the following detailed expressions of $\mathcal{A}_k$ as follows
\begin{subequations}\label{eq:Ak}
    \begin{align}
        &{\mathcal A}_{\alpha_{nm}}={A}_{mn}+{A}_{nm},\\
        &{\mathcal A}_{\beta_{nm}}=i\big({A}_{mn}-{A}_{nm}\big),\\
        &{\mathcal A}_{\gamma_{n}}=\sum_{l=1}^{n-1}\sqrt{\frac{2}{n(n-1)}}{A}_{ll}-\sqrt{\frac{2(n-1)}{n}}{A}_{nn}.
    \end{align}
\end{subequations}

Using the expression of the SW kernel in Eq.~\ref{SWkernel}, the SW transform of $\hat{A}$ is expressed (using Eq.~\ref{A-sw-map}) as
\begin{align}\label{A-sw-map2}
    \big[\hat{A}\big]_\mathrm{s}(\mathbf{\Omega})={\mathcal{A}}_{0}+r_{\mathrm{s}}\sum_{k=1}^{N^2-1}{\mathcal{A}}_{k}\cdot \Omega_{k}.
\end{align}

One of the important properties of the SW transform is that it can be used to evaluate a quantum mechanical trace in the continuous phase space as follows
\begin{align}\label{SW_TrA}
    \int d \mathbf{\Omega}\big[\hat{A}\big]_\mathrm{s}(\mathbf{\Omega})=\mathrm{Tr_e}\big[\hat{A}\big].
\end{align}
For two operators $\hat{A}$ and $\hat{B}$, it can be shown that the SW transform has the following property
\begin{align}\label{SWssb}
&\mathrm{Tr_e}\big[\hat{A}\hat{B}\big]=\int d \mathbf{\Omega}\big[\hat{A}\hat{B}\big]_\mathrm{s}(\mathbf{\Omega})\\
&=\int d \mathbf{\Omega}\big[\hat{A}\big]_\mathrm{s}(\mathbf{\Omega})\cdot\big[\hat{B}\big]_{\bar{\mathrm{s}}}(\mathbf{\Omega})=\int d \mathbf{\Omega}\big[\hat{A}\big]_{\bar{\mathrm{s}}}(\mathbf{\Omega})\cdot\big[\hat{B}\big]_{\mathrm{s}}(\mathbf{\Omega}),\nonumber
\end{align}
where $[\cdots]_{\bar{\mathrm{s}}}(\mathbf{\Omega})$ is SW transformed through Eq.~\ref{A-sw-map2} using $r_\mathrm{\bar s}$ instead of $r_\mathrm{s}$. The sum of the squares of the generators (the so-called Casimir operator of $\mathfrak{su}(N)$) requires that\cite{runesonrichardson2020}
\begin{equation}\label{cscsbar}
    r_\mathrm{s}\cdot r_\mathrm{\bar s}=N+1.
\end{equation}
The commonly used values\cite{runesonrichardson2020,Klimov2009} of $r_\mathrm{s}$ and $r_\mathrm{\bar s}$ is the following symmetrical choice
\begin{equation}\label{wpq}
r_\mathrm{s}=r_\mathrm{\bar s}=\sqrt{N+1}
\end{equation}
These parameters are not restricted to the above special case, and they can take any value in the range $r_\mathrm{s}\in(0,\infty)$. The exact quantum dynamics is invariant under the choice of $r_\mathrm{s}$, but the approximate quantum dynamics methods are not. Previous studies from both the linearized method\cite{bossion2022,runesonrichardson2020} and the spin mapping NRPMD\cite{bossion2021} suggest that Eq.~\ref{wpq} gives the best numerical performances for computing regular TCF\cite{bossion2022,runesonrichardson2020} and Kubo-transformed TCF\cite{bossion2021} (for two-level systems).

Another useful relation of the SW transform is 
\begin{align}\label{SW_N-prod}
\big[\hat{A}\hat{B}\big]_\mathrm{s}(\mathbf{\Omega})=&\mathcal{A}_0\mathcal{B}_0+\frac{1}{2N}\sum_{i=1}^{N^2-1}\mathcal{A}_i\mathcal{B}_i\\
&+r_\mathrm{s}\sum_{i=1}^{N^2-1}(\Omega_i\mathcal{A}_i\mathcal{B}_0+\Omega_i\mathcal{A}_0\mathcal{B}_i)\nonumber\\
&+\frac{r_\mathrm{s}}{2}\sum_{i,j,k=1}^{N^2-1}\Omega_i\mathcal{A}_j\mathcal{B}_k(d_{ijk}+if_{ijk}).\nonumber
\end{align}
The detailed derivation of this expression can be found in Eqs.~41-42 of Ref.~\citenum{bossion2022}.

\subsection{The Mixed Stratonovich-Weyl/Wigner Representation}
For the nuclear DOF, one often use the following Wigner (W) transform\cite{wigner1932}
\begin{equation}
    \big[\hat{O}(\hat{R},\hat{P})\big]_\mathrm{w}=\int  d  \Delta  e^{\frac{i}{\hbar}P\Delta}\Big\langle R-\frac{\Delta}{2}\Big|\hat{O}(\hat{R},\hat{P})\Big|R+\frac{\Delta}{2}\Big\rangle,
\end{equation}
which converts an operator $\hat{O}$ into a phase space function $\big[\hat{O}(\hat{R},\hat{P})\big]_\mathrm{w}$.

For a general operator in the electronic and nuclear Hilbert space, we use the SW representation for the electronic subsystem that exhibits $SU(N)$ symmetry, and the Wigner transform for the nuclear DOF. This mixed Stratonovich-Weyl/Wigner (SW/W) formalism was introduced in our previous work to approximately evaluate the regular TCF.\cite{bossion2022} For an operator $\hat{A}(\hat{R},\hat{P})$, the mixed SW/W transform is
\begin{align}\label{A-w-sw}
\big[\hat{A}(\hat{R},\hat{P})\big]_\mathrm{ws}&=\big[{\mathcal{A}}_{0}(\hat{R},\hat{P})\big]_\mathrm{w}+r_{\mathrm{s}}\sum_{k=1}^{N^2-1}\big[{\mathcal{A}}_{k}(\hat{R},\hat{P})\big]_\mathrm{w}\cdot \Omega_{k},\nonumber\\
&= {\mathcal{A}}_{0}(R,P)+r_{\mathrm{s}}\sum_{k=1}^{N^2-1}{\mathcal{A}}_{k}(R,P)\cdot \Omega_{k}
\end{align}
where ${\mathcal{A}}_{0}(R,P)$ and ${\mathcal{A}}_{k}(R,P)$ are Wigner transforms of ${\mathcal{A}}_{0}(\hat{R},\hat{P})$ and ${\mathcal{A}}_{k}(\hat{R},\hat{P})$, respectively, with the detailed forms of $\mathcal{A}_0$ and $\mathcal{A}_k$ defined in Eq.~\ref{A_SW}.

Furthermore, for two operators, the mixed SW/W representation is expressed as
\begin{align}\label{W-SWprod}
    \big[\hat{A}\hat{B}\big]_\mathrm{ws} =& {\mathcal{A}}_0e^{-i\frac{\hat{\Lambda}\hbar}{2}}{\mathcal{B}}_0+\frac{1}{2N}\sum_{i=1}^{N^2-1}{\mathcal{A}}_ie^{-i\frac{\hat{\Lambda}\hbar}{2}}{\mathcal{B}}_i\\
    &+r_\mathrm{s}\sum_{i=1}^{N^2-1}\Omega_i({\mathcal{A}}_ie^{-i\frac{\hat{\Lambda}\hbar}{2}}{\mathcal{B}}_0+{\mathcal{A}}_0e^{-i\frac{\hat{\Lambda}\hbar}{2}}{\mathcal{B}}_i)\nonumber\\
    &+\frac{r_\mathrm{s}}{2}\sum_{i,j,k=1}^{N^2-1}\Omega_i{\mathcal{A}}_je^{-i\frac{\hat{\Lambda}\hbar}{2}}{\mathcal{B}}_k(d_{ijk}+if_{ijk}),\nonumber
\end{align}
where
\begin{equation}
\hat{\Lambda}=\frac{\overleftarrow{\partial}}{\partial P}\frac{\overrightarrow{\partial}}{\partial R}-\frac{\overleftarrow{\partial}}{\partial R}\frac{\overrightarrow{\partial}}{\partial P}
\end{equation}
is the negative Poisson operator associated with the nuclear DOF.\cite{groenewold1946,imre1967,hillery1984} Note that for convenience we write $\mathcal{A}_0$ instead of $\mathcal{A}_0(R,P)$, and the same applies to $\mathcal{A}_i(R,P)$, $\mathcal{B}_0(R,P)$ and $\mathcal{B}_i(R,P)$.

\section{The Kubo-transformed time-correlation function}\label{kubotcf}
The Kubo-transformed TCF is defined as follows
\begin{align}\label{KTCF}
&C_{AB}^\mathrm{K}(t)= \frac{1}{{\cal{Z}}\beta}\int_0^\beta d \lambda\mathrm{Tr}\big[e^{-(\beta-\lambda)\hat{H}}\hat{A}e^{-\lambda\hat{H}}e^{\frac{i}{\hbar}\hat{H}t}\hat{B}e^{-\frac{i}{\hbar}\hat{H}t}\big]\nonumber\\
&=\frac{1}{{\cal{Z}}{\mathcal{N}}}\sum_{\alpha=1}^{\mathcal{N}}\mathrm{Tr}\big[e^{-\beta_{\mathcal{N}}({\mathcal{N}}-\alpha)\hat{H}}\hat{A}e^{-\beta_{\mathcal{N}}\alpha\hat{H}}e^{\frac{i}{\hbar}\hat{H}t}\hat{B}e^{-\frac{i}{\hbar}\hat{H}t}\big],
\end{align}
where $\beta=1/k_\mathrm{B}T$,  $\beta_{\mathcal{N}}\equiv\beta/\mathcal{N}$, and $\mathrm{Tr}[\cdots]\equiv\mathrm{Tr_nTr_e}[\cdots]$ is a trace over nuclear and electronic DOFs and $\mathcal{Z}=\mathrm{Tr}[e^{-\beta\hat{H}}]$. Note that the index $\alpha$ is used to represent a discrete version of the imaginary time integral over $\lambda$, not to be confused with $\alpha_{nm}$, the label of the $\mathfrak{su}(N)$ generators used in Eq.~\ref{eq:Hk}. We will later identify that index $\alpha$ as the nuclear bead index. In this section, we use the mixed SW/W representation to re-express $C_{AB}^\mathrm{K}(t)$, and derive the exact quantum Liouvillian.

\subsection{The Kubo-transformed TCF in the mixed SW/W representation}
To evaluate the Kubo-transformed TCF, we follow the previous strategy\cite{chowdhury2021,althorpe2015,shi2003,hele2016} to use the discretized expression in the second line of $C_{AB}^\mathrm{K}(t)$ in Eq.~\ref{KTCF}. We write the Kubo-transformed TCF in a block form by inserting ${\mathcal{N}}-1$ identities, $\hat{{I}} = e^{\frac{i}{\hbar}\hat{H}t}e^{-\frac{i}{\hbar}\hat{H}t}$, into Eq.~\ref{KTCF}, leading to
\begin{align}\label{CABN1}
&C_{AB}^{[{\mathcal{N}}]}(t)= \frac{1}{{\cal{Z}}{\mathcal{N}}}\sum_{\alpha=1}^{\mathcal{N}}\mathrm{Tr_{e}Tr_{n}}\Big[\big(e^{-\beta_{\mathcal{N}}\hat{H}}e^{\frac{i}{\hbar}\hat{H}t}e^{-\frac{i}{\hbar}\hat{H}t}\big)^{{\mathcal{N}}-\alpha-1}\nonumber\\ 
&~~~~\times e^{-\beta_{\mathcal{N}}\hat{H}}\hat{A} e^{\frac{i}{\hbar}\hat{H}t}e^{-\frac{i}{\hbar}\hat{H}t}\big(e^{-\beta_{\mathcal{N}}\hat{H}}e^{\frac{i}{\hbar}\hat{H}t}e^{-\frac{i}{\hbar}\hat{H}t}\big)^{\alpha-1}\nonumber\\
&~~~~\times e^{-\beta_{\mathcal{N}}\hat{H}}e^{\frac{i}{\hbar}\hat{H}t}\hat{B}e^{-\frac{i}{\hbar}\hat{H}t}\Big].
\end{align}

To evaluate the above trace, we insert the nuclear identity $\hat{\mathds{1}}_{R_1'}=\int d R_1'|R_1'\rangle\langle R_1'|\otimes\hat{\mathcal{I}}$ (where $\hat{\mathcal{I}}$ is the electronic identity) and use the property of the SW transform in Eq.~\ref{SW_TrA} to compute electronic traces, leading to the following expression
\begin{align}
&C_{AB}^{[\mathcal{N}]}(t)= \frac{1}{{\cal{Z}}{\mathcal{N}}}\sum_{\alpha=1}^{\mathcal{N}}\int d R_1'\int d \mathbf{\Omega}^{(1)}\Big[\big\langle R_1'\big|e^{\frac{i}{\hbar}\hat{H}t}e^{-\frac{i}{\hbar}\hat{H}t}\nonumber\\
&~~\times\big(e^{-\beta_{\mathcal{N}}\hat{H}}e^{\frac{i}{\hbar}\hat{H}t}e^{-\frac{i}{\hbar}\hat{H}t}\big)^{{\mathcal{N}}-\alpha-2}  e^{-\beta_{\mathcal{N}}\hat{H}}\hat{A}e^{\frac{i}{\hbar}\hat{H}t}e^{-\frac{i}{\hbar}\hat{H}t} \nonumber \\
&~~\times\big(e^{-\beta_{\mathcal{N}}\hat{H}}e^{\frac{i}{\hbar}\hat{H}t}e^{-\frac{i}{\hbar}\hat{H}t}\big)^{\alpha-1} e^{-\beta_{\mathcal{N}}\hat{H}}e^{\frac{i}{\hbar}\hat{H}t}\hat{B}e^{-\frac{i}{\hbar}\hat{H}t}\nonumber\\
&~~\times e^{-\beta_{\mathcal{N}}\hat{H}}\big|R_1'\big\rangle\Big]_\mathrm{s}^{(1)},
\end{align}
where we denote $[\cdots]_\mathrm{s}^{(\alpha)}$ as the SW transform of the $\alpha$-th coherent state basis $|\boldsymbol{\Omega}^{(\alpha)}\rangle$ as follows
\begin{subequations}
\begin{align}
&\big[\hat{A}\big]^{(\alpha)}_\mathrm{s}=\mathrm{Tr_e}\big[\hat{A}\cdot\hat{w}_\mathrm{s}^{(\alpha)}\big],\\
&\hat{w}^{(\alpha)}_\mathrm{s}=\frac{1}{N}\hat{\mathcal{I}}+r_\mathrm{s}\cdot \frac{2}{\hbar}\sum_{k=1}^{N^2-1}\Omega^{(\alpha)}_{k}\cdot\hat{\mathcal{S}}_k\label{kernel-bead}\\
&\Omega^{(\alpha)}_{k}\equiv\big\langle\mathbf{\Omega}^{(\alpha)}\big|\hat{\mathcal{S}}_{k}\big|\mathbf{\Omega}^{(\alpha)}\big\rangle.\label{Omega-mapping}
\end{align}
\end{subequations}

We then use the following electronic identities (see Eq.~\ref{SW_Id}) and nuclear identities
\begin{subequations}
    \begin{align}
        \hat{\mathds{1}}_{R_\gamma',\mathbf{\Omega}^{(\gamma)}} = &\int d R_\gamma'|R_\gamma'\rangle\langle R_\gamma'|\otimes \int d \mathbf{\Omega}^{(\gamma)}\hat{w}_\mathrm{s}^{(\gamma)} \\
        \hat{\mathds{1}}_{R_\gamma''} = &\int d R_\gamma''|R_\gamma''\rangle\langle R_\gamma''|\otimes\hat{\mathcal I},
    \end{align}
\end{subequations}
where $\gamma = 1,\cdots, {\mathcal{N}}$. The identity $\hat{\mathds{1}}_{R_\gamma',\mathbf{\Omega}^{(\gamma)}}$ is inserted after each $e^{-\beta_{\mathcal{N}}\hat{H}}$ and $\hat{\mathds{1}}_{R_\gamma''}$ is inserted after each $e^{\frac{i}{\hbar}\hat{H}t}e^{-\frac{i}{\hbar}\hat{H}t}$, leading to
\begin{align}\label{eq:kubo}
    C_{AB}^{[\mathcal{N}]}(t) = &\frac{1}{{\cal{Z}}}\int d\{R_\alpha'\}\int d\{R_\alpha''\}\int d\{\mathbf{\Omega}^{(\alpha)}\}\\
    &\times\frac{1}{{\mathcal{N}}}\sum_{\alpha=1}^{\mathcal{N}}\Big[\big\langle R_1'\big|e^{i\hat{H}t/\hbar}e^{-i\hat{H}t/\hbar}\hat{\mathds{1}}_{R_1''}\nonumber\\
    &\times\prod_{\gamma=2}^{{\mathcal{N}}-\alpha-1} e^{-\beta_{\mathcal{N}}\hat{H}}\hat{\mathds{1}}_{R_\gamma',\mathbf{\Omega}^{(\gamma)}}e^{\frac{i}{\hbar}\hat{H}t}e^{-\frac{i}{\hbar}\hat{H}t}\hat{\mathds{1}}_{R_\gamma''}\nonumber\\
    &\times e^{-\beta_{\mathcal{N}}\hat{H}}\hat{A}\hat{\mathds{1}}_{R_{{\mathcal{N}}-\alpha}',\mathbf{\Omega}^{({\mathcal{N}}-\alpha)}}e^{\frac{i}{\hbar}\hat{H}t}e^{-\frac{i}{\hbar}\hat{H}t}\hat{\mathds{1}}_{R_{{\mathcal{N}}-\alpha}''} \nonumber \\
    &\times\prod_{\gamma={\mathcal{N}}-\alpha+1}^{{\mathcal{N}}-1}e^{-\beta_{\mathcal{N}}\hat{H}}\hat{\mathds{1}}_{R_\gamma',\mathbf{\Omega}^{(\gamma)}}e^{\frac{i}{\hbar}\hat{H}t}e^{-\frac{i}{\hbar}\hat{H}t}\hat{\mathds{1}}_{R_\gamma''}\nonumber\\
    &\times e^{-\beta_{\mathcal{N}}\hat{H}}\hat{\mathds{1}}_{R_{\mathcal{N}}',\mathbf{\Omega}^{{(\mathcal{N})}}}e^{\frac{i}{\hbar}\hat{H}t}\hat{B}e^{-\frac{i}{\hbar}\hat{H}t}\hat{\mathds{1}}_{R_{\mathcal{N}}''}\nonumber\\
    &\times e^{-\beta_{\mathcal{N}}\hat{H}}\big|R_1'\big\rangle\Big]_\mathrm{s}^{(1)},\nonumber
\end{align}
where we have introduced the shorthand notation $d \{R_\alpha'\}\equiv\prod_{\alpha=1}^{\mathcal{N}}dR_\alpha'$, $d \{R_\alpha''\}\equiv\prod_{\alpha=1}^{\mathcal{N}}dR_\alpha''$ and $d \{\mathbf{\Omega}^{(\alpha)}\}\equiv\prod_{\alpha=1}^{\mathcal{N}}d\mathbf{\Omega}^{(\alpha)}$, with each $d\mathbf{\Omega}^{(\alpha)}$ expressed in Eq.~\ref{dOmega}.

Using the properties of the SW transform (Eq.~\ref{A-sw-map} and Eq.~\ref{SW_TrA}), we can rewrite Eq.~\ref{eq:kubo} as follows
\begin{align}\label{CAB-bead}
    C_{AB}^{[{\mathcal{N}}]}(t)=& \frac{1}{\cal{Z}}\int d \{R_\alpha'\}\int d\{R_\alpha''\}\int d\{\mathbf{\Omega}^{(\alpha)}\} \\
    & \times\frac{1}{{\mathcal{N}}}\sum_{\alpha=1}^{\mathcal{N}}\mathrm{Tr_e}\Big[\prod_{\gamma\neq \alpha}^{\mathcal{N}}\big\langle R_{\gamma-1}''\big|e^{-\beta_{\mathcal{N}}\hat{H}}\big|R_\gamma'\big\rangle\hat{w}_\mathrm{\bar s}^{(\gamma)}\nonumber\\
    &\times\big\langle R_{\alpha-1}''\big|e^{-\beta_{\mathcal{N}}\hat{H}}\hat{A}\big|R_\alpha'\big\rangle\hat{w}_\mathrm{\bar s}^{(\alpha)}\Big] \nonumber \\
    & \times\frac{1}{{\mathcal{N}}}\sum_{\alpha=1}^{\mathcal{N}}\prod_{\gamma\neq \alpha}^{\mathcal{N}}\Big[\big\langle R_{\gamma}''\big|e^{\frac{i}{\hbar}\hat{H}t}e^{-\frac{i}{\hbar}\hat{H}t}\big|R_\gamma'\big\rangle\Big]_\mathrm{s}^{(\gamma)}\nonumber\\
    &\times\Big[\big\langle R_{\alpha}''\big|e^{\frac{i}{\hbar}\hat{H}t}\hat{B}e^{-\frac{i}{\hbar}\hat{H}t}\big|R_\alpha'\big\rangle\Big]_\mathrm{s}^{(\alpha)},\nonumber
\end{align}
where we have used the cyclic-symmetric property to write the operator $\hat{B}$ into a bead-averaged fashion (the index $\alpha$ can be viewed as a bead index of a ring-polymer of $\mathcal{N}$ beads). Note that the two sums with respect to $\alpha$ are two independent ones. The details of the derivation are provided in Appendix~\ref{CABN-detail}.

We can now change the nuclear variables\cite{shi2003,poulsen2003} as the following mean and difference variables
\begin{subequations}
    \begin{align}
        R_\alpha = & \frac{1}{2}(R_\alpha'+R_\alpha''), \\
        D_\alpha = & R_\alpha' - R_\alpha'',
    \end{align}
\end{subequations}
 and insert the following identity for each block
\begin{align}
    1 = \int d D_\alpha'\delta(D_\alpha+D_\alpha') = \frac{1}{2\pi\hbar}\int d D_\alpha'\int d P_\alpha e^{\frac{i}{\hbar}P_\alpha(D_\alpha+D_\alpha')}.
\end{align}
This leads to the Wigner representation of the nuclear DOFs, and the TCF in Eq.~\ref{CAB-bead} becomes
\begin{align}\label{CABN}
C_{AB}^{[{\mathcal{N}}]}(t)=&\frac{1}{{\cal{Z}}(2\pi\hbar)^{\mathcal{N}}}\int d \{R_\alpha\}\int d \{P_\alpha\}\int d \{\mathbf{\Omega}^{(\alpha)}\}\\
&~~~~~~~~~~~~~~~~~~~~~~\times\big[e^{-\beta\hat{H}}\hat{A}\big]_\mathrm{{w\bar s}}\big[\hat{B}(t)\big]_\mathrm{ws},\nonumber
\end{align}
where $\big[e^{-\beta\hat{H}}\hat{A}\big]_\mathrm{{w\bar s}}$ is the SW/W transformed Boltzmann operator expressed as
\begin{align}
    &\big[e^{-\beta\hat{H}}\hat{A}\big]_\mathrm{{w\bar s}}= \int d \{D_\alpha\}\frac{1}{{\mathcal{N}}}\sum_{\alpha=1}^{\mathcal{N}}\prod_{\gamma=1}^{\mathcal{N}}e^{iP_\gamma D_\gamma/\hbar}\\\
    &\times\mathrm{Tr_e}\Big[\prod_{\gamma\neq \alpha}^{\mathcal{N}}\big\langle R_{\gamma-1}-\frac{1}{2}D_{\gamma-1}\big|e^{-\beta_{\mathcal{N}}\hat{H}}\big|R_\gamma+\frac{1}{2}D_\gamma\big\rangle\hat{w}_\mathrm{\bar s}^{(\gamma)} \nonumber \\
    &\times\big\langle R_{\alpha-1}-\frac{1}{2}D_{\alpha-1}\big|e^{-\beta_{\mathcal{N}}\hat{H}}\hat{A}\big|R_\alpha+\frac{1}{2}D_\alpha\big\rangle\hat{w}_\mathrm{\bar s}^{(\alpha)}\Big],\nonumber
\end{align}
and $\big[\hat{B}(t)\big]_{\mathrm{ws}}$ is the bead-averaged SW/W transform of operator $\hat{B}$ expressed as follows
\begin{align}\label{estB}
\big[\hat{B}(t)\big]_{\mathrm{ws}}=&\int d \{D_\alpha'\}\frac{1}{{\mathcal{N}}}\sum_{\alpha=1}^{\mathcal{N}}\prod_{\gamma=1}^{\mathcal{N}}e^{iP_\gamma D_\gamma'/\hbar}\\
&\times\prod_{\gamma\neq \alpha}^{\mathcal{N}}\Big[\big\langle R_\gamma-\frac{1}{2}D_\gamma'\big|e^{\frac{i}{\hbar}\hat{H}t}e^{-\frac{i}{\hbar}\hat{H}t}\big|R_\gamma+\frac{1}{2}D_\gamma'\big\rangle\Big]^{(\gamma)} \nonumber \\
&\times\Big[\big\langle R_\alpha-\frac{1}{2}D_\alpha'\big|e^{\frac{i}{\hbar}\hat{H}t}\hat{B}e^{-\frac{i}{\hbar}\hat{H}t}\big|R_\alpha+\frac{1}{2}D_\alpha'\big\rangle\Big]^{(\alpha)}\nonumber\\
=&\frac{1}{\mathcal{N}}\sum_{\alpha=1}^{\mathcal{N}}\big[\hat{B}(t)\big]_{\mathrm{ws}}^{(\alpha)}\nonumber.
\end{align}
Note that we use the notation $\big[\hat{B}(t)\big]_{\mathrm{ws}}$ to represent the bead averaged SW/W transform. When introduced in Eq.~\ref{W-SWprod}, it was only for a single bead.

We can formally express the Kubo-transformed TCF in Eq.~\ref{CABN} as follows
\begin{align}\label{CAB_N}
    C_{AB}^{[{\mathcal{N}}]}(t)=&\frac{1}{{\cal{Z}}(2\pi\hbar)^{\mathcal{N}}}\int d \{R_\alpha\}\int d \{P_\alpha\}\int d \{\mathbf{\Omega}^{(\alpha)}\}\\
    &\times \big[e^{-\beta\hat{H}}\hat{A}\big]_\mathrm{w\bar s}e^{\mathcal{L}t}\big[\hat{B}\big]_\mathrm{ws},\nonumber
\end{align}
where the time-evolution is governed by the quantum Liouvillian
\begin{equation}\label{Bt-Liou}
    \big[\hat{B}(t)\big]_\mathrm{ws}=e^{{\mathcal{L}}t}\big[\hat{B}\big]_\mathrm{ws},
\end{equation}
which will be derived in Sec.~\ref{Qliou}.

If in addition $\hat{A}$ is linear in $\hat{R}$ (and equivalently for $\hat{P}$), then
\begin{equation}
\big[e^{-\beta\hat{H}}\hat{A}\big]_\mathrm{w\bar s}= \big[\hat{A}\big]_\mathrm{w}\big[e^{-\beta\hat{H}}\big]_\mathrm{w\bar s}=\mathcal{A}_0(R)\big[e^{-\beta\hat{H}}\big]_\mathrm{w\bar s},
\end{equation}
where we have used $\big[\hat{A}(\hat{R})\big]_\mathrm{w\bar s}=\big[\mathcal{A}_0(\hat{R})\big]_\mathrm{w}=\mathcal{A}_0(R)$. Due to the cyclic property of the beads, the estimator $\mathcal{A}_0(R)$ is expressed as a bead-average $\mathcal{A}_0(R)\equiv\frac{1}{\mathcal{N}}\sum_{\alpha=1}^{\mathcal{N}}\mathcal{A}_0(R_\alpha)$. In this case, the Kubo-transformed TCF is
\begin{align}\label{CAB_R}
    C_{AB}^{[{\mathcal{N}}]}(t)=&\frac{1}{{\cal{Z}}(2\pi\hbar)^{\mathcal{N}}}\int d \{R_\alpha\}\int d \{P_\alpha\}\int d \{\mathbf{\Omega}^{(\alpha)}\}\\
    &\times \mathcal{A}_0(R)\big[e^{-\beta\hat{H}}\big]_\mathrm{w\bar s}e^{\mathcal{L}t}\big[\hat{B}\big]_\mathrm{ws},\nonumber
\end{align}

The estimator $[\hat{B}]_\mathrm{ws}$ (in Eq.~\ref{estB}) can be expressed (using Eq.~\ref{A-w-sw}) as 
\begin{align}
\big[\hat{B}\big]_{\mathrm{ws}}&=\frac{1}{\mathcal{N}}\sum_{\alpha=1}^{\mathcal{N}}\big[\hat{B}\big]_{\mathrm{ws}}^{(\alpha)}\\
&=\frac{1}{\mathcal{N}}\sum_{\alpha=1}^{\mathcal{N}}\big({\mathcal{B}}_{0}^{(\alpha)}+r_{\mathrm{s}}\sum_{k=1}^{N^2-1}{\mathcal{B}}_{k}^{(\alpha)}\cdot \Omega^{(\alpha)}_{k}\big),\nonumber
\end{align}
with the short notation $\big[\hat{B}\big]_\mathrm{ws}^{(\alpha)}\equiv \big[\hat{B}(\hat{R}_\alpha,\hat{P_\alpha})\big]_\mathrm{ws}$, $\mathcal{B}_0^{(\alpha)}\equiv\mathcal{B}_0(R_\alpha,P_\alpha)$ and $\mathcal{B}_k^{(\alpha)}\equiv\mathcal{B}_k(R_\alpha,P_\alpha)$. This expression is a bead average of the mixed SW/W transform of the operator $\hat{B}$. If it is a position operator, $\hat{B}=\hat{R}$, 
\begin{equation}\label{Bpos}
    \big[\hat{B}\big]_{\mathrm{ws}}=\frac{1}{\mathcal{N}}\sum_{\alpha=1}^{\mathcal{N}}\big[\hat{R}_\alpha\big]_\mathrm{w}=\frac{1}{\mathcal{N}}\sum_{\alpha=1}^{\mathcal{N}}R_\alpha.
\end{equation}
If $\hat{B}$ is a projection operator $\hat{B}=|n\rangle\langle m|$, then the corresponding expression becomes
\begin{equation}\label{pop_est}
    \big[\hat{B}\big]_{\mathrm{ws}}=\frac{1}{\mathcal{N}}\sum_{\alpha=1}^{\mathcal{N}}\big[|n\rangle\langle m|\big]_{\mathrm{s}}^{(\alpha)}.
\end{equation}  
This projection operator can be expressed in terms of different mapping variables, which can be found in our previous work (Eq.~44-Eq.~45, Eq.~62-Eq.~64 and Eq.~D8-Eq.~D10 in Ref.~\citenum{bossion2022}). In terms of the spin mapping variables $\{\Omega_{k}^{(\alpha)}\}$ in Eq.~\ref{Omegak} (for $n>m$) we have
\begin{subequations}\label{proj-sw}
    \begin{align}
        &\big[|n\rangle\langle n|\big]_{\mathrm{s}}^{(\alpha)}\\
        &=\frac{1}{N}+r_\mathrm{s}\sum_{m=n+1}^N\sqrt{\frac{2}{m(m-1)}}\Omega_{\gamma_{m}}^{(\alpha)}-r_\mathrm{s}\sqrt{\frac{2(n-1)}{n}}\Omega_{\gamma_{n}}^{(\alpha)},\nonumber\\
        &\big[|n\rangle\langle m|\big]_\mathrm{s}^{(\alpha)}
        =r_\mathrm{s}\big(\Omega_{\alpha_{nm}}^{(\alpha)}-i\Omega_{\beta_{nm}}^{(\alpha)}\big),\\
        &\big[|m\rangle\langle n|\big]^{(\alpha)}_\mathrm{s}
        =r_\mathrm{s}\big(\Omega_{\alpha_{nm}}^{(\alpha)}+i\Omega_{\beta_{nm}}^{(\alpha)}\big),
    \end{align}
\end{subequations}
where the detailed expressions of $\Omega_{\alpha_{nm}}$, $\Omega_{\beta_{nm}}$, and $\Omega_{\gamma_{m}}$ are provided in Eq.~\ref{eq:Omega_a}-Eq.~\ref{eq:Omega_g}.

\subsection{Expression of the Quantum Liouvillian}\label{Qliou}
To obtain the quantum Liouvillian $\mathcal{L}$ (in Eq.~\ref{Bt-Liou}), we use the time-derivative of $\big[\hat{B}\big]_\mathrm{ws}$ through the Heisenberg EOMs as follows
\begin{align}\label{Liou_a}
    &\frac{\mathrm{d}}{\mathrm{d}t}\big[\hat{B}\big]_{\mathrm{ws}}={\mathcal L}\big[\hat{B}\big]_{\mathrm{ws}}\nonumber\\
    &=\frac{1}{\mathcal{N}}\sum_{\alpha=1}^{\mathcal{N}}\frac{\mathrm{d}}{\mathrm{d}t}\big[\hat{B}\big]_{\mathrm{ws}}^{(\alpha)}=\frac{1}{\mathcal{N}}\sum_{\alpha=1}^{\mathcal{N}}{\mathcal L}^{(\alpha)}\big[\hat{B}\big]_{\mathrm{ws}}^{(\alpha)},
\end{align}
leading to an expression of bead specific Liouvillian components, ${\mathcal L}^{(\alpha)}$, evolving each bead. Each of these Liouvillian component is identical to what we have previously derived for a regular TCF.\cite{bossion2022} The expression of ${\mathcal L}^{(\alpha)}$ is obtained through 
\begin{align}\label{Liou_smnrpmd}
\frac{\mathrm{d}}{\mathrm{d}t}\big[\hat{B}\big]_{\mathrm{ws}}^{(\alpha)}=&\frac{\mathrm{d}}{\mathrm{d}t}\big[{\mathcal{B}}_0\hat{\mathcal{I}}\big]_{\mathrm{ws}}^{(\alpha)}+\sum_{i=1}^{N^2-1}\frac{\mathrm{d}}{\mathrm{d}t}\big[{\mathcal{B}}_i\cdot\frac{1}{\hbar}\hat{\mathcal{S}}_i\big]_\mathrm{ws}^{(\alpha)}\\
=&\frac{\mathrm{d}}{\mathrm{d}t}\big[{\mathcal{B}}_0\big]_{\mathrm{w}}^{(\alpha)}+\sum_{i=1}^{N^2-1}\frac{\mathrm{d}}{\mathrm{d}t}\Big(\big[{\mathcal{B}}_i]_{\mathrm{w}}^{(\alpha)}\cdot r_\mathrm{s}\Omega_{i}^{(\alpha)}\Big)\nonumber\\
\equiv&{{\mathcal{L}}}_{0}^{(\alpha)}{\mathcal{B}}_{0}^{(\alpha)}+r_\mathrm{s}\sum_{i=1}^{N^2-1}{\mathcal{L}}_{i}^{(\alpha)}\big(\Omega_i^{(\alpha)} {\mathcal{B}}_i^{(\alpha)}\big).\nonumber
\end{align}

The state-independent Liouvillian ${\mathcal{L}}_0^{(\alpha)}$ is expressed as
\begin{equation}\label{liou_l0}
{\mathcal{L}}_0^{(\alpha)}\mathcal{B}_0^{(\alpha)}\equiv\frac{2}{\hbar}H_\mathrm{s}(R_\alpha,P_\alpha)\sin\left(\frac{\hbar}{2}\hat{\Lambda}_\alpha\right)\mathcal{B}_0^{(\alpha)},
\end{equation}
with the SW/W transform of the Hamiltonian 
\begin{subequations}
\begin{align}
H_\mathrm{s}(R_\alpha,P_\alpha)&\equiv\big[\hat{H}(\hat{R}_\alpha,\hat{P}_\alpha)\big]_\mathrm{ws}\\
&=\mathcal{H}_0(R_\alpha,P_\alpha)+r_\mathrm{s}\sum_{k=1}^{N^2-1}\mathcal{H}_k(R_\alpha)\Omega_k^{(\alpha)},\nonumber\\
\mathcal{H}_0(R_\alpha,P_\alpha)&=\frac{P_\alpha^2}{2m}+U_0(R_\alpha)+\frac{1}{N}\sum_{n=1}^NV_{nn}(R_\alpha),\label{eq:h0}\\
\mathcal{H}_k({R}_{\alpha})&=\frac{2}{\hbar}\mathrm{Tr_e}\big[\hat{V}_\mathrm{e}({R}_{\alpha})\cdot\hat{\mathcal S}_k\big],
\end{align}
\end{subequations}
and the negative Poisson operator for the $\alpha_\mathrm{th}$ bead is 
\begin{equation}
\hat{\Lambda}_\alpha=\frac{\overleftarrow{\partial}}{\partial P_\alpha}\frac{\overrightarrow{\partial}}{\partial R_\alpha}-\frac{\overleftarrow{\partial}}{\partial R_\alpha}\frac{\overrightarrow{\partial}}{\partial P_\alpha}.
\end{equation}

The state-dependent Liouvillian is split into two terms, ${\mathcal{L}}_i^{(\alpha)}={\mathcal{L}}_i^{\mathrm{e}(\alpha)}+{\mathcal{L}}_i^{\mathrm{n}(\alpha)}$. The first term  ${\mathcal{L}}_i^{\mathrm{e}(\alpha)}$ evolves the spin mapping variables (electronic DOFs) as
\begin{align}\label{liou_sm}
&{\mathcal{L}}_i^{\mathrm{e}(\alpha)}\big(r_\mathrm{s}\Omega_i^{(\alpha)}\mathcal{B}_i^{(\alpha)}\big)\nonumber\\
&\equiv\frac{r_\mathrm{s}}{\hbar}\sum_{j,k=1}^{N^2-1}f_{ijk}{\mathcal{H}}_j(R_\alpha)\Omega_k^{(\alpha)}\cos\left(\frac{\hbar}{2}\hat{\Lambda}_\alpha\right)\mathcal{B}_i^{(\alpha)},
\end{align}
and the second term ${\mathcal{L}}_i^{\mathrm{n}(\alpha)}$ evolves the nuclear DOFs and couples the spin-mapping variables and the nuclear DOFs as follows
\begin{align}\label{Liou_end}
&{\mathcal{L}}_i^{\mathrm{n}(\alpha)}\big(r_\mathrm{s}\Omega_i^{(\alpha)}\mathcal{B}_i^{(\alpha)}\big)\equiv\frac{1}{\hbar}\Bigg[\frac{1}{N}{\mathcal{H}}_i(R_\alpha)+2r_\mathrm{s}\Omega_i^{(\alpha)}{\mathcal{H}}_0(R_\alpha,P_\alpha)\nonumber\\
&~~~~~~~+r_\mathrm{s}\sum_{j,k=1}^{N^2-1}d_{ijk}{\mathcal{H}}_j(R_\alpha)\Omega_k^{(\alpha)}\Bigg]\sin\left(\frac{\hbar}{2}\hat{\Lambda}_\alpha\right)\mathcal{B}_i^{(\alpha)}.
\end{align}
and note that the $1/N$ in Eq.~\ref{Liou_end} comes from the SW transform (where $N$ is the number of electronic states, not to be confused with the total number of beads $\mathcal{N}$).

These Liouvillian expressions give rise to the time evolution of each bead, $\alpha$ as follows
\begin{align}\label{eq:exactL}
&{\mathcal L}\big[\hat{B}\big]_{\mathrm{ws}}=\frac{1}{\mathcal{N}}\sum_{\alpha=1}^{\mathcal{N}}{\mathcal L}^{(\alpha)}\big[\hat{B}\big]_{\mathrm{ws}}^{(\alpha)}\\
&=\frac{1}{\mathcal{N}}\sum_{\alpha=1}^{\mathcal{N}}\Bigg[{\mathcal{L}}_0^{(\alpha)}\mathcal{B}_0^{(\alpha)}+r_\mathrm{s}\sum_{i=1}^{N^2-1}\big({\mathcal{L}}_i^{\mathrm{e}(\alpha)}+{\mathcal{L}}_i^{\mathrm{n}(\alpha)}\big)\Omega_i^{(\alpha)}\mathcal{B}_{i}^{(\alpha)}\Bigg].\nonumber
\end{align}

We emphasize that the Kubo-transformed TCF in Eq.~\ref{CAB_N} as well as the Liouvillian in Eq.~\ref{eq:exactL} are in principle exact. Directly evaluating these expressions numerically will be as difficult as the exact quantum mechanics.

\section{Non-adiabatic Matsubara Dynamics and SM-NRPMD}\label{matnrpmd}
In this section, we follow the original work on the Matsubara dynamics in Ref.~\citenum{althorpe2015} as well as our recent work on non-adiabatic Matsubara dynamics based on the MMST mapping variables to derive the spin mapping based non-adiabatic Matsubara and the spin-mapping non-adiabatic RPMD (SM-NRPMD) approach.

\subsection{Non-adiabatic Matsubara Dynamics}
We follow the procedure of our previous work\cite{chowdhury2021} on deriving the non-adiabatic Matsubara dynamics. First, a transformation from the bead to the normal mode representation is introduced as follows
\begin{equation}
    {\mathcal{R}}_l=\sum_{\alpha=1}^{\mathcal{N}}\frac{T_{\alpha l}}{\sqrt{\mathcal{N}}}R_\alpha,~~~~~~~~~~{\mathcal{P}}_l=\sum_{\alpha=1}^{\mathcal{N}}\frac{T_{\alpha l}}{\sqrt{\mathcal{N}}}P_\alpha,\label{bead-to-mode}
\end{equation}
and the corresponding inverse transform is 
\begin{equation}
R_\alpha=\sum_{l=-\frac{{\mathcal{N}}-1}{2}}^{\frac{{\mathcal{N}}-1}{2}}T_{\alpha l}\sqrt{\mathcal{N}}{\mathcal{R}}_l,~~P_\alpha=\sum_{l=-\frac{{\mathcal{N}}-1}{2}}^{\frac{{\mathcal{N}}-1}{2}}T_{\alpha l}\sqrt{\mathcal{N}}{\mathcal{P}}_l,\label{norm-to-bead}
\end{equation}
with the transformation matrix element (with an odd number ${\mathcal{N}}$) defined as 
\begin{align}
    T_{\alpha l} = \left\{\begin{matrix}
\sqrt{\frac{1}{\mathcal{N}}}, & ~~~~~~~~~l=0, \\ 
\sqrt{\frac{2}{\mathcal{N}}}\sin\frac{2\pi\alpha l}{\mathcal{N}}, & ~~~~~~~~~1\leq l\leq \frac{{\mathcal{N}}-1}{2}, \\
\sqrt{\frac{2}{\mathcal{N}}}\cos\frac{2\pi\alpha l}{\mathcal{N}}, & -\frac{{\mathcal{N}}-1}{2}\leq l\leq -1.
\end{matrix}\right.
\end{align}
The new variables $\{\mathcal{R}_{l},\mathcal{P}_{l}\}$ are normal modes of the free ring polymer Hamiltonian
\begin{align}
\mathcal{H}_{\mathrm{rp}}^{0}=&\sum_{\alpha=1}^{\mathcal{N}}\Big[\frac{P^2_{\alpha}}{2m}+\frac{m}{2\beta_{\mathcal{N}}^2\hbar^2}(R_\alpha-R_{\alpha-1})^2\Big]\\
=&\mathcal{N}\sum_{l=-\frac{{\mathcal{N}}-1}{2}}^{\frac{{\mathcal{N}}-1}{2}}\frac{\mathcal{P}^2_{l}}{2m}+\frac{m}{2}\omega^2_{l}\mathcal{R}_l^2,\nonumber
\end{align}
where the normal mode frequency is \begin{equation}\label{eqn:normfreq}
{\omega}_{l}=\frac{2}{\beta_{N}\hbar}\sin\left(\frac{l\pi}{\mathcal{N}}\right),
\end{equation}
and $l = 0,\cdots,\pm ({\mathcal{N}}-1)/2$ is the index of normal modes. Using this transformation, the Liouvillian in Eqs.~\ref{liou_l0}-\ref{Liou_end} can be expressed in the normal mode representation using the Poisson operator expressed in terms of the normal modes. The estimator $[\hat{B}]_\mathrm{ws}^{(\alpha)}$ is also expressed in terms of the normal modes through Eq.~\ref{norm-to-bead} as $\mathcal{B}_0^{(\alpha)}\equiv\mathcal{B}_0\big(R_\alpha(\boldsymbol{\mathcal{R}}),P_\alpha(\boldsymbol{\mathcal{P}})\big)$, and equivalently for $\mathcal{B}_i^{(\alpha)}$.

To transform the bead-specific Liouvillian into normal mode representation, we first write the Poisson operator $\frac{\hbar}{2}\hat{\Lambda}_\alpha$ inside the sines and cosines of Eq.~\ref{liou_l0}-Eq.~\ref{Liou_end} as a sum over {\it all bead indices} $\sum_{\alpha}\frac{\hbar}{2}\hat{\Lambda}_\alpha$, leading to the following replacement
\begin{align}
    \sin\left(\frac{\hbar}{2}\hat{\Lambda}_\alpha\right)\rightarrow\sin\left(\frac{\hbar}{2}\sum_{\alpha=1}^{\mathcal{N}}\hat{\Lambda}_\alpha\right),\nonumber\\
    \cos\left(\frac{\hbar}{2}\hat{\Lambda}_\alpha\right)\rightarrow\cos\left(\frac{\hbar}{2}\sum_{\alpha=1}^{\mathcal{N}}\hat{\Lambda}_\alpha\right).\nonumber
\end{align}
This results in an identical Liouvillian compared to the original expression,\cite{willatt2017,althorpe2015} because there is no cross-bead term in $H_\mathrm{s}(R_\alpha,P_\alpha)$ nor in ${\mathcal{H}}_j(R_\alpha)$, $\mathcal{B}_0^{(\alpha)}$, and $\mathcal{B}_k^{(\alpha)}$. With these expressions, the Liouvillian can be transformed into the normal mode representation\cite{althorpe2015} as follows
\begin{subequations}\label{Liou_mode}
\begin{align}
&{\mathcal{L}}_0^{(\alpha)}\mathcal{B}_0^{(\alpha)}=\frac{2}{\hbar}H_\mathrm{s}\big(R_\alpha(\boldsymbol{\mathcal{R}}),P_\alpha(\boldsymbol{\mathcal{P}})\big)\sin\left(\frac{\hbar}{2{\mathcal{N}}}\hat{\Lambda}^{[\mathcal{N}]}\right)\mathcal{B}_0^{(\alpha)},\\
&{\mathcal{L}}_i^{\mathrm{e}(\alpha)}\big(r_\mathrm{s}\Omega_i^{(\alpha)}\mathcal{B}_i^{(\alpha)}\big)\\
&=\frac{r_\mathrm{s}}{\hbar}\sum_{j,k=1}^{N^2-1}f_{ijk}{\mathcal{H}}_j\big(R_\alpha(\boldsymbol{\mathcal{R}})\big)\Omega_k^{(\alpha)}\cos\left(\frac{\hbar}{2{\mathcal{N}}}\hat{\Lambda}^{[\mathcal{N}]}\right)\mathcal{B}_i^{(\alpha)},\nonumber\\
&{\mathcal{L}}_i^{\mathrm{n}(\alpha)}\big(r_\mathrm{s}\Omega_i^{(\alpha)}\mathcal{B}_i^{(\alpha)}\big)\label{liou_neg}\\
&=\Bigg[\frac{1}{N}{\mathcal{H}}_i\big(R_\alpha(\boldsymbol{\mathcal{R}})\big)+2r_\mathrm{s}\Omega_i^{(\alpha)}{\mathcal{H}}_0\big(R_\alpha(\boldsymbol{\mathcal{R}}),P_\alpha(\boldsymbol{\mathcal{P}})\big)\nonumber\\
&+r_\mathrm{s}\sum_{j,k=1}^{N^2-1}d_{ijk}{\mathcal{H}}_j(\mathcal{R}_\alpha(\boldsymbol{\mathcal{R}}))\Omega_k^{(\alpha)}\Bigg]\sin\left(\frac{\hbar}{2\mathcal{N}}\hat{\Lambda}^{[\mathcal{N}]}\right)\mathcal{B}_i^{(\alpha)},\nonumber
\end{align}
\end{subequations}
where $\hat{\Lambda}^{[\mathcal{N}]}$ is the Poisson operator  with the normal mode representation expressed as follows
\begin{equation}
\hat{\Lambda}^{[\mathcal{N}]}=\sum_{l=-\frac{\mathcal{N}-1}{2}}^{\frac{\mathcal{N}-1}{2}}\frac{\overleftarrow{\partial}}{\partial \mathcal{P}_l}\frac{\overrightarrow{\partial}}{\partial \mathcal{R}_l}-\frac{\overleftarrow{\partial}}{\partial \mathcal{R}_l}\frac{\overrightarrow{\partial}}{\partial \mathcal{P}_l}.
\end{equation}

The correlation function in Eq.~\ref{CAB_N} can be expressed in the normal mode representation as follows
\begin{align}\label{CAB_Mat}
    C_{AB}^{[{\mathcal{N}}]}(t)=&\frac{1}{{\cal{Z}}(2\pi\hbar)^{\mathcal{N}}}\int d \{\mathcal{R}_{l}\}\int d \{\mathcal{P}_{l}\}\int d \{\mathbf{\Omega}^{(\alpha)}\}\\
    &\times \big[e^{-\beta\hat{H}}\hat{A}\big]_\mathrm{w\bar s}e^{\mathcal{L}t}\big[\hat{B}\big]_\mathrm{ws},\nonumber,
\end{align}
where $d \{\mathcal{R}_{l}\}\equiv\prod_{l=-(\mathcal{N}-1)/2}^{(\mathcal{N}-1)/2}d \mathcal{R}_{l}$, $d \{\mathcal{P}_{l}\}\equiv\prod_{l=-(\mathcal{N}-1)/2}^{(\mathcal{N}-1)/2}d \mathcal{P}_{l}$, both $[e^{-\beta\hat{H}}\hat{A}]_\mathrm{w\bar s}$ and $[\hat{B}\big]_\mathrm{ws}$ are expressed in the normal mode representation, and the Liouvillian $\mathcal{L}$ is expressed using the normal modes, with the detailed expressions in Eq.~\ref{Liou_mode}.

In the limit of ${\mathcal{N}}\rightarrow\infty$, only the $\mathcal{M}$ lowest frequencies of the free ring-polymer of ${\mathcal{N}}$ beads (when $\mathcal{M}\ll{\mathcal{N}}$) contribute to the initial quantum Boltzmann distribution.\cite{freeman1984,chakravarty1997,chakravarty1998} The frequencies of these modes are referred to as the Matsubara frequencies,\cite{ceperley1995,althorpe2015} 
\begin{equation}\label{matfreq}
\tilde{\omega}_l=\frac{2l\pi}{\beta\hbar},~~|l|\leq(\mathcal{M}-1)/2,
\end{equation}
which corresponds to the $\mathcal{M}\ll\mathcal{N}$ limit (under the ${\mathcal{M}}\rightarrow\infty$ and ${\mathcal{N}}\rightarrow\infty$ limits) of the normal mode frequencies in Eq.~\ref{eqn:normfreq}.

The {\it main idea} of the Matsubara dynamics\cite{althorpe2015} is to {\it only use the Matsubara modes} to evolve the quantum dynamics in the Kubo-transformed TCF, because these modes completely determine the initial quantum statistics. Hence, the central approximation of the Matsubara dynamics is to {\it discard the non-Matsubara modes} in the quantum Liouvillian.\cite{althorpe2015} To briefly discuss this, we introduce the following mode-to-bead transformation that only considering the Matsubara modes as follows
\begin{subequations}
\begin{align}\label{Fourier_mat}
R_\alpha^{[\mathcal{M}]}\equiv R_\alpha({\boldsymbol{\mathcal{R}}}_\mathcal{M})&=\sum_{l=-\frac{\mathcal{M}-1}{2}}^{\frac{\mathcal{M}-1}{2}}\frac{T_{\alpha l}}{\sqrt{\mathcal{N}}}{\mathcal{R}}_l,\\
P_\alpha^{[\mathcal{M}]}\equiv P_\alpha({\boldsymbol{\mathcal{P}}}_\mathcal{M})&=\sum_{l=-\frac{\mathcal{M}-1}{2}}^{\frac{\mathcal{M}-1}{2}}T_{\alpha l}\sqrt{\mathcal{N}}{\mathcal{P}}_l,
\end{align}
\end{subequations}
which have the same expression as those in  Eq.~\ref{norm-to-bead}, except a truncation to the normal modes only including the Matsubara modes. In the above equation, we introduced the notation $\boldsymbol{\mathcal{R}}_{\mathcal{M}}=\{\mathcal{R}_{-\frac{\mathcal{M}-1}{2}},\cdots,\mathcal{R}_{\frac{\mathcal{M}-1}{2}}\}$ and $\boldsymbol{\mathcal{P}}_{\mathcal{M}}=\{\mathcal{P}_{-\frac{\mathcal{M}-1}{2}},\cdots,\mathcal{P}_{\frac{\mathcal{M}-1}{2}}\}$ which are the Matsubara modes.

Following the Matsubara approximation\cite{althorpe2015}, we approximate the Liouvillian in Eq.~\ref{Liou_smnrpmd} by only considering the Matsubara modes, leading to
\begin{equation}\label{Matsubara}
\mathcal{L}( \hat{\Lambda}^{[\mathcal{N}]})\approx\mathcal{L}^{[\mathcal{M}]} (\hat{\Lambda}^{[\mathcal{M}]})
\end{equation}
where we have discarded all non-Matsubara modes. The corresponding approximate Liouvillian has the same formal expression as the one given in Eq.~\ref{Liou_mode}, except by replacing $\hat{\Lambda}^{[\mathcal{N}]}$ with $\hat{\Lambda}^{[\mathcal{M}]}$ as follows
\begin{equation}\label{LambdaM}
    \hat{\Lambda}^{[\mathcal{N}]}\rightarrow\hat{\Lambda}^{[\mathcal{M}]}=\sum_{l=-\frac{\mathcal{M}-1}{2}}^{\frac{\mathcal{M}-1}{2}}\frac{\overleftarrow{\partial}}{\partial {\mathcal{P}}_l}\frac{\overrightarrow{\partial}}{\partial {\mathcal{R}}_l}-\frac{\overleftarrow{\partial}}{\partial {\mathcal{R}}_l}\frac{\overrightarrow{\partial}}{\partial {\mathcal{P}}_l}.
\end{equation}
One can further expand the sines and cosines in the Liouvillian expression $\mathcal{L}^{[\mathcal{M}]}$, leading to\cite{willatt2017,chowdhury2021}
\begin{subequations}\label{mat-exp}
    \begin{align}
        \sin\left(\frac{\hbar}{2{\mathcal{N}}}\hat{\Lambda}^{[\mathcal{M}]}\right)=&\frac{\hbar}{2{\mathcal{N}}}\hat{\Lambda}^{[\mathcal{M}]}+\mathcal{O}\Big(\frac{\mathcal{M}^3\hbar^3}{\mathcal{N}^3}\Big),\\
        \cos\left(\frac{\hbar}{2{\mathcal{N}}}\hat{\Lambda}^{[\mathcal{M}]}\right)=&1+\mathcal{O}\Big(\frac{\mathcal{M}^2\hbar^2}{\mathcal{N}^2}\Big).
    \end{align}
\end{subequations}
Under the limit $\mathcal{M}\ll\mathcal{N}$, the Planck constant is effectively scaled as $\hbar\rightarrow\hbar\mathcal{M}/\mathcal{N}$ (note the overall scaling of $\mathcal{O}(\hat{\Lambda}^{[\mathcal{M}]})\sim\mathcal{M}$ due to the sum in Eq.~\ref{LambdaM}). This means that inside the Matsubara subspace, one can effectively scale the Planck constant as small as one desires, such that the linearization of the sines and cosines becomes {\it exact} (but remains an approximation in the full normal mode space).\cite{althorpe2015} The Matsubara Liouvillian $\mathcal{L}^{[\mathcal{M}]} (\hat{\Lambda}^{[\mathcal{M}]})$ will only update the Matsubara modes in $\mathcal{B}_0^{(\alpha)}$ and $\mathcal{B}_i^{(\alpha)}$ in Eq.~\ref{Liou_a}, and will no longer evolve the non-Matsubara modes. Eventually, one can analytically integrate out all non-Matsubara modes\cite{althorpe2015} in the Kubo-transformed TCF expression (including those inside $[e^{-\beta\hat{H}}\hat{A}]_\mathrm{w\bar s}$ as well as  inside $\big[\hat{B}\big]_\mathrm{ws}$), as shown in Appendix B of Ref.~\citenum{chowdhury2021} as well as in Ref.~\citenum{althorpe2015}.

Using the argument in Eq.~\ref{mat-exp}, {\it i.e.}, only the leading order of the quantum Liouvillian in the Matsubara space is needed, the Liouvillian in Eq.~\ref{Matsubara} can be expressed as
\begin{align}\label{LM}
&{\mathcal{L}}^{[\mathcal{M}]}\big[\hat{B}\big]_\mathrm{ws}^{[\mathcal{M}]}\equiv\frac{1}{\mathcal{N}}\sum_{\alpha=1}^{\mathcal{N}}\sum_{l=-\frac{\mathcal{M}-1}{2}}^{\frac{\mathcal{M}-1}{2}}\Bigg[\frac{\mathcal{P}_l}{m}\frac{\overrightarrow{\partial}}{\partial \mathcal{R}_l}-\frac{1}{\mathcal{N}}\Big[\frac{\partial {\mathcal{H}}_0\big(R_\alpha^{[\mathcal{M}]}\big)}{\partial \mathcal{R}_l}\nonumber\\
&~~~+r_\mathrm{s}\sum_{j=1}^{N^2-1}\frac{\partial {\mathcal{H}}_j\big(R_\alpha^{[\mathcal{M}]}\big)}{\partial \mathcal{R}_l}\Omega_j^{(\alpha)}\Big]\frac{\overrightarrow{\partial}}{\partial \mathcal{P}_l}\Bigg]\mathcal{B}_{0}^{(\alpha)[\mathcal{M}]}\\
&~~~+\frac{1}{\mathcal{N}}\sum_{\alpha=1}^{\mathcal{N}}\sum_{i=1}^{N^2-1}\Bigg[\frac{r_\mathrm{s}}{\hbar}\sum_{j,k=1}^{N^2-1}f_{ijk}{\mathcal{H}}_j\big(R_\alpha^{[\mathcal{M}]}\big)\Omega_{k}^{(\alpha)}\Bigg]{\mathcal{B}}_i^{(\alpha)[\mathcal{M}]},\nonumber
\end{align}
where the SW/W transform of the operator $\hat{B}$ will only contain the Matsubara modes
\begin{equation}
\mathcal{B}_0^{(\alpha)[\mathcal{M}]}\equiv\mathcal{B}_0\big(R_\alpha^{[\mathcal{M}]},P_\alpha^{[\mathcal{M}]}\big) ;~\mathcal{B}_i^{(\alpha)[\mathcal{M}]}\equiv\mathcal{B}_i\big(R_\alpha^{[\mathcal{M}]},P_\alpha^{[\mathcal{M}]}\big).
\end{equation}
Note that in the Matsubara Liouvillian (Eq.~\ref{LM}) we have also ignored the ${\mathcal{L}}_i^{\mathrm{n}(\alpha)}\big(r_\mathrm{s}\Omega_i^{(\alpha)}\mathcal{B}_i^{(\alpha)}\big)$ term (Eq.~\ref{liou_neg}). In the linearized spin-mapping approach, this is an additional Liouvillian that is less straightforward to evaluate using independent trajectory method, and has been  ignored.\cite{bossion2022,runesonrichardson2020} This should be viewed as an independent approximation, and the consequences of making this approximation are subjects to future investigations.

Following the same procedure outlined in our early work in Ref.~\citenum{chowdhury2021}, we analytically integrate out the non-Matsubara modes (see Appendix B of Ref.~\citenum{chowdhury2021}) in the Kubo-transformed correlation function. A symmetric Trotter expansion (which is exact under the $\mathcal{N}\to\infty$ limit) is further used to split the Boltzmann operator into a state-dependent and a state-independent term to simplify the distribution expression (see Eq.~B4 in Appendix B of Ref.~\citenum{chowdhury2021}). Finally, We obtain the Matsubara dynamics expression of the Kubo-transformed TCF. When the operator $\hat{A}$ is a linear nuclear operator, the TCF in Eq.~\ref{CAB_Mat} has the final form of
\begin{align}\label{CABMR}
&C_{AB}^{[\mathcal{M}]}=\frac{\alpha_\mathcal{M}}{(2\pi\hbar)^{\mathcal{N}}{\mathcal{Z}}_\mathcal{M}}\int d \boldsymbol{\mathcal{R}}_{\mathcal{M}}\int d \boldsymbol{\mathcal{P}}_{\mathcal{M}}\int d \{\mathbf{\Omega}^{(\alpha)}\}\\
&\times e^{-\beta_{\mathcal{N}}(H_\mathcal{M}-i\mathcal{N}\phi_\mathcal{M})}\mathcal{A}_0\big(\boldsymbol{\mathcal{R}}_{\mathcal{M}}\big)\mathrm{Tr_e}\big[\hat{\mathbf{\Gamma}}_\mathrm{\bar s}^{[\mathcal{M}]}\big]e^{{\mathcal{L}}^{[\mathcal{M}]}t}\big[\hat{B}\big]_\mathrm{ws}^{[\mathcal{M}]},\nonumber
\end{align}
where $\alpha_\mathcal{M}=\frac{\hbar^{1-\mathcal{M}}}{(\mathcal{M}-1)/2!^2}$, with the shorthand notation $d \boldsymbol{\mathcal{R}}_{\mathcal{M}}\equiv\prod_{l=-(\mathcal{M}-1)/2}^{(\mathcal{M}-1)/2}d \mathcal{R}_l$ and  $d \boldsymbol{\mathcal{P}}_{\mathcal{M}}\equiv\prod_{l=-(\mathcal{M}-1)/2}^{(\mathcal{M}-1)/2}d \mathcal{P}_l$. Note that compared to Eq.~\ref{CAB_Mat} that contains integrals for all normal modes, in $C_{AB}^{[\mathcal{M}]}$ (Eq.~\ref{CABMR})
only the Matsubara modes are present. Furthermore, the state-independent Hamiltonian $H_\mathcal{M}$, the Matsubara phase $\phi_\mathcal{M}$, and the electronic phase $\hat{\mathbf{\Gamma}}_\mathrm{\bar s}^{[\mathcal{M}]}$ have the following expressions
\begin{subequations}\label{h-phi-gamma}
\begin{align}
&H_\mathcal{M}=\sum_{\alpha=1}^{\mathcal{N}}{\mathcal{H}}_0\big(R_\alpha^{[\mathcal{M}]},P_\alpha^{[\mathcal{M}]}\big),\\
&\phi_\mathcal{M}=\sum_{l=-\frac{\mathcal{M}-1}{2}}^{\frac{\mathcal{M}-1}{2}}{\mathcal{P}}_l\tilde{\omega}_l{\mathcal{R}}_{-l},\\
&\hat{\mathbf{\Gamma}}_\mathrm{\bar s}^{[\mathcal{M}]}=\prod_{\alpha=1}^{\mathcal{N}}e^{-\beta_\mathcal{N}\frac{1}{\hbar}\sum_k{\mathcal{H}}_k\big(R_\alpha^{[\mathcal{M}]}\big)\cdot\hat{\mathcal{S}}_k}\cdot\hat{w}_\mathrm{\bar s}^{(\alpha)},
\end{align}
\end{subequations}
where $\tilde{\omega}_l$ is the Matsubara frequency (Eq.~\ref{matfreq}), $\hat{w}_\mathrm{\bar s}^{(\alpha)}$ is the SW kernel defined in Eq.~\ref{kernel-bead} (with radius $r_{\bar{\mathrm{s}}}$). The Matsubara partition function $\mathcal{Z}_\mathcal{M}$ is expressed as
\begin{align}\label{eqn:partiton_mat}
\mathcal{Z}_\mathcal{M} =\frac{\alpha_\mathcal{M}}{(2\pi\hbar)^{\mathcal{N}}}&\int d \boldsymbol{\mathcal{R}}_{\mathcal{M}}\int d \boldsymbol{\mathcal{P}}_{\mathcal{M}}\int d \{\mathbf{\Omega}^{(\alpha)}\}\\
&\times e^{-\beta_\mathcal{N}(H_\mathcal{M}-i\mathcal{N}\phi_\mathcal{M})}\mathrm{Tr_e}\big[\hat{\mathbf{\Gamma}}_\mathrm{\bar s}^{[\mathcal{M}]}\big].\nonumber
\end{align}

If the operator $\hat{A}$ is a projection operator (that only depends on the electronic DOFs), then the TCF in Eq.~\ref{CAB_Mat} after the Matsubara approximation is expressed as
\begin{align}\label{CABMproj}
    C_{AB}^{[\mathcal{M}]}=&\frac{\alpha_\mathcal{M}}{(2\pi\hbar)^{\mathcal{N}}{\mathcal{Z}}_\mathcal{M}}\int d \boldsymbol{\mathcal{R}}_{\mathcal{M}}\int d \boldsymbol{\mathcal{P}}_{\mathcal{M}}\int d \{\mathbf{\Omega}^{(\alpha)}\}\\
    &\times e^{-\beta_\mathcal{N}(H_\mathcal{M}-i\mathcal{N}\phi_\mathcal{M})}\mathrm{Tr_e}\big[\hat{\mathbf{\Gamma}}_\mathrm{\bar s}^{[\mathcal{M}]}\hat{A}\big]e^{{\mathcal{L}}^{[\mathcal{M}]}t}\big[\hat{B}\big]_\mathrm{ws}^{[\mathcal{M}]},\nonumber
\end{align}
with the same expressions of $H_\mathcal{M}$, $\phi_\mathcal{M}$, $\hat{\mathbf{\Gamma}}_\mathrm{\bar s}^{[\mathcal{M}]}$ defined in Eq.~\ref{h-phi-gamma}, and $\mathcal{Z}_\mathcal{M}$ defined in Eq.~\ref{eqn:partiton_mat}. The trace $\mathrm{Tr_e}\big[\hat{\mathbf{\Gamma}}_\mathrm{\bar s}^{[\mathcal{M}]}\hat{A}\big]$ is expressed as 
\begin{align}\label{trgammaA}
&\mathrm{Tr_e}\big[\hat{\mathbf{\Gamma}}_\mathrm{\bar s}^{[\mathcal{M}]}\hat{A}\big]=\frac{1}{\mathcal{N}}\sum_{\alpha=1}^{\mathcal{N}}\mathrm{Tr_e}\Bigg[\prod_{\gamma\leq\alpha}^{\mathcal{N}}e^{-\beta_{\mathcal{N}}\frac{1}{\hbar}\sum_k{\mathcal{H}}_{k}(R_\gamma^{[\mathcal{M}]})\cdot\hat{\mathcal{S}}_k}\hat{w}_\mathrm{\bar s}^{(\gamma)}\nonumber\\
&~~~~~~~~~~~\times\hat{A}\prod_{\gamma>\alpha}^{\mathcal{N}}e^{-\beta_{\mathcal{N}}\frac{1}{\hbar}\sum_k{\mathcal{H}}_{k}(R_\gamma^{[\mathcal{M}]})\cdot\hat{\mathcal{S}}_k}\cdot\hat{w}_\mathrm{\bar s}^{(\gamma)}\Bigg].
\end{align}

\subsection{The RPMD Approximation}\label{RPMD}
As discussed in the previous works,\cite{althorpe2015,chowdhury2021} numerically evaluating the Matsubara phase $\phi_\mathcal{M}$ remains computationally challenging due to the severe sign problem. One possible way to avoid directly evaluating this phase is to make the transformation ${\mathcal{P}}_l\rightarrow\mathcal{P}_l-im\tilde{\omega}_l{\mathcal{R}}_{-l}$ together with the change in the contour of integration\cite{althorpe2015,hele2015_2} $\int_{-\infty}^{\infty} d {\mathcal{P}}_l\rightarrow\int_{-\infty-im\tilde{\omega}_l{\mathcal{R}}_{-l}}^{\infty-im\tilde{\omega}_l{\mathcal{R}}_{-l}} d {\mathcal{P}}_l$. The change of variables leads to a complex Liouvillian 
\begin{equation}\label{Lbar}
\mathcal{L}^{[\mathcal{M}]}\rightarrow\bar{\mathcal{L}}^{[\mathcal{M}]}=\bar{\mathcal{L}}_\mathrm{rp}^{[\mathcal{M}]}+ i\bar{\mathcal{L}}_\mathrm{I}^
{[\mathcal{M}]},
\end{equation}
where $\bar{\mathcal{L}}_\mathrm{rp}^{[\mathcal{M}]}$ is the real part of the Liouvillian in the normal mode representation as follows
\begin{align}\label{liou_rp_M}
&\bar{\mathcal{L}}_\mathrm{rp}^{[\mathcal{M}]}\big[\hat{B}\big]_\mathrm{ws}^{[\mathcal{M}]}\equiv\frac{1}{\mathcal{N}}\sum_{\alpha=1}^{\mathcal{N}}\sum_{l=-\frac{\mathcal{M}-1}{2}}^{\frac{\mathcal{M}-1}{2}}\Bigg[\Big(\frac{\mathcal{P}_l}{m}\frac{\overrightarrow{\partial}}{\partial \mathcal{R}_l}-m\tilde{\omega}_l^2\mathcal{R}_l\frac{\overrightarrow{\partial}}{\partial \mathcal{P}_l}\Big)\nonumber\\
&-\frac{1}{\mathcal{N}}\Big[\frac{\partial {\mathcal{H}}_0\big(R_\alpha^{[\mathcal{M}]}\big)}{\partial \mathcal{R}_l}+r_\mathrm{s}\sum_{j=1}^{N^2-1}\frac{\partial {\mathcal{H}}_j\big(R_\alpha^{[\mathcal{M}]}\big)}{\partial \mathcal{R}_l}\Omega_j^{(\alpha)}\Big]\frac{\overrightarrow{\partial}}{\partial \mathcal{P}_l}\Bigg]\mathcal{B}_0^{(\alpha)[\mathcal{M}]}\nonumber\\
&+\frac{1}{\mathcal{N}}\sum_{\alpha=1}^{\mathcal{N}}\sum_{i=1}^{N^2-1}\Bigg[\frac{r_\mathrm{s}}{\hbar}\sum_{j,k=1}^{N^2-1}f_{ijk}{\mathcal{H}}_j\big(R_\alpha^{[\mathcal{M}]}\big)\Omega_{k}^{(\alpha)}\Bigg]{\mathcal{B}}_i^{(\alpha)[\mathcal{M}]},
\end{align}
and $i\bar{\mathcal{L}}_\mathrm{I}^
{[\mathcal{M}]}$ is the imaginary part of the Liouvillian\cite{chowdhury2021,althorpe2015} 
\begin{equation}\label{eqn:imaginary_liouv}
\bar{\mathcal{L}}_{\mathrm{I}}^{[M]}\mathcal{B}_0^{[\mathcal{M}]}=\frac{1}{\mathcal{N}}\sum_{\alpha=1}^{\mathcal{N}}\sum_{l=-\frac{\mathcal{M}-1}{2}}^{\frac{\mathcal{M}-1}{2}}\tilde{\omega}_l\Big(\mathcal{P}_l\frac{\overrightarrow{\partial}}{\partial \mathcal{P}_{-l}} - \mathcal{R}_{l}\frac{\overrightarrow{\partial}}{\partial \mathcal{R}_{-l}}\Big)\mathcal{B}_0^{(\alpha)[\mathcal{M}]}.
\end{equation}
Note that there is no spin mapping variable-related derivative in the above imaginary Liouvillian $\bar{\mathcal{L}}_{\mathrm{I}}^{[M]}$, and its impact on the electronic dynamics should only come from its influence on the nuclear dynamics, which in turn couples to the electronic mapping DOFs via $\bar{\mathcal{L}}_{\mathrm{rp}}^{[M]}$.

As discussed in Ref.~\citenum{willatt2017} (Chapter 3) as well as Ref.~\citenum{hele2015_2}, one can gradually discard the imaginary Liouvillian $i\bar{\mathcal{L}}_{\mathrm{I}}^{[M]}$ in Eq.~\ref{Lbar} and make the following approximation, 
\begin{equation}\label{rpmd-approx}
\bar{\mathcal{L}}^{[\mathcal{M}]}\approx\bar{\mathcal{L}}_\mathrm{rp}^{[\mathcal{M}]},
\end{equation}
while gradually pushing each $\int d{\mathcal{P}}_l$ integral toward the real axis of $\mathcal{P}_{l}$.  This procedure\cite{hele2015_2} is referred to as the ``RPMD approximation''. After applying the RPMD approximation, the correlation function in Eq.~\ref{CABMR} (when $\hat{A}$ is a linear operator of $\hat{R}$) is expressed as
\begin{align}\label{CAB_mat}
    C_{AB}^{[\mathcal{M}]}=&\frac{\alpha_\mathcal{M}}{(2\pi\hbar)^{\mathcal{N}}{\mathcal{Z}}_\mathcal{M}}\int d {\boldsymbol{\mathcal{R}}}_\mathcal{M}\int d {\boldsymbol{\mathcal{P}}}_\mathcal{M}\int d \{\mathbf{\Omega}^{(\alpha)}\}\\
    &\times e^{-\beta_\mathcal{N} H_\mathrm{rp}^{[\mathcal{M}]}}\mathcal{A}_0(\boldsymbol{\mathcal{R}}_{\mathcal{M}})\mathrm{Tr_e}\big[\hat{\mathbf{\Gamma}}_\mathrm{\bar s}^{[\mathcal{M}]}\big]e^{\bar{\mathcal{L}}_\mathrm{rp}^{[\mathcal{M}]}t}\big[\hat{B}\big]_\mathrm{ws}^{[\mathcal{M}]},\nonumber
\end{align}
where the dynamics is only evolved with $\bar{\mathcal{L}}_\mathrm{rp}^{[\mathcal{M}]}$ defined in Eq.~\ref{liou_rp_M}, and the state-independent Hamiltonian in the initial distribution $e^{-\beta_\mathcal{N} H_\mathrm{rp}^{[\mathcal{M}]}}$ now includes the spring term of the ring-polymer as follows
\begin{align}\label{HrpM}
H_\mathrm{rp}^{[\mathcal{M}]}=&\sum_{\alpha=1}^{\mathcal{N}}{\mathcal{H}}_0\big(R_\alpha^{[\mathcal{M}]},P_\alpha^{[\mathcal{M}]}\big)+\mathcal{N}\sum_{l=-\frac{\mathcal{M}-1}{2}}^{\frac{\mathcal{M}-1}{2}}\frac{1}{2}m\tilde{\omega}_l^2{\mathcal{R}}_l^2,
\end{align}
where $\tilde{\omega}_l$ is the Matsubara frequency introduced in Eq.~\ref{matfreq}.

In the case of $\hat{A}$ being a projection operator, the same approximation can be performed, and Eq.~\ref{CABMproj} becomes
\begin{align}\label{CAB_mat2}
    C_{AB}^{[\mathcal{M}]}(t)=&\frac{\alpha_\mathcal{M}}{(2\pi\hbar)^{\mathcal{N}}{\mathcal{Z}}_\mathcal{M}}\int d \boldsymbol{\mathcal{R}}_{\mathcal{M}}\int d \boldsymbol{\mathcal{P}}_{\mathcal{M}}\int d \{\mathbf{\Omega}^{(\alpha)}\}\nonumber\\
    &\times e^{-\beta_{\mathcal{N}} H_\mathrm{rp}^{[\mathcal{M}]}}\mathrm{Tr_e}\big[\hat{\mathbf{\Gamma}}_\mathrm{\bar s}^{[\mathcal{M}]}\hat{A}\big]e^{\bar{\mathcal{L}}_\mathrm{rp}^{[\mathcal{M}]}t}\big[\hat{B}\big]_\mathrm{ws}^{[\mathcal{M}]},
\end{align}
with the same $H_\mathrm{rp}^{[\mathcal{M}]}$ expressed in Eq.~\ref{HrpM} and $\bar{\mathcal{L}}_\mathrm{rp}^{[\mathcal{M}]}$ expressed in Eq.~\ref{liou_rp_M}. The dynamics in these Kubo-transformed TCFs (Eq.~\ref{CAB_mat} and Eq.~\ref{CAB_mat2}) can be seen as governed by a $\mathcal{N}$-beads ring-polymer $\{R_\alpha({\boldsymbol{\mathcal{R}}}_\mathcal{M})\}$ that only contains the $\mathcal{M}$ Matsubara modes for the nuclear DOFs, and $\mathcal{N}$ mapping beads for the electronic DOFs. 

\subsection{The Spin-Mapping NRPMD Method}\label{SM-NRPMD_sec}
Directly evaluating $C_{AB}^{[\mathcal{M}]}$ in Eq.~\ref{CAB_mat} and Eq.~\ref{CAB_mat2} remains numerically challenging, because it often requires a large  number of beads $\mathcal{N}$, such that the total number of Matsubara modes $\mathcal{M}$ is large enough to converge the initial quantum statistics.\cite{althorpe2015} This is a well-known numerical fact for the Matsubara-based path-integral molecular dynamics or Monte-Carlo approaches.\cite{ceperley1995} For the non-adiabatic case investigated here, the further complication also comes from the initial non-adiabatic phase $\mathrm{Tr_e}\big[\hat{\mathbf{\Gamma}}_\mathrm{\bar s}^{[\mathcal{M}]}\big]$ (in Eq.~\ref{CAB_mat}) or $\mathrm{Tr_e}\big[\hat{\mathbf{\Gamma}}_\mathrm{\bar s}^{[\mathcal{M}]}\hat{A}\big]$ (in Eq.~\ref{CAB_mat2}), which induces a (separate) sign problem as $\mathcal{N}$ gets very large.

We thus consider an additional approximation by replacing the Matsubara frequencies $\tilde{\omega}_l$ in Eq.~\ref{CAB_mat} and Eq.~\ref{CAB_mat2} with the normal mode frequencies of the ring polymer ${\omega}_{l}$ (Eq.~\ref{eqn:normfreq}) as suggested by the original adiabatic Matsubara dynamics.\cite{althorpe2015} This should be viewed as a separate approximation in addition to the RPMD approximation in Sec.~\ref{RPMD}. Expressing the TCF in the bead representation, we obtain the approximate TCF as
\begin{align}\label{CAB_rpmd}
    C_{AB}^{[{\mathcal{N}}]}(t)=&\frac{1}{{\cal{Z}_\mathcal{N}}(2\pi\hbar)^{\mathcal{N}}}\int d \{R_\alpha\}\int d \{P_\alpha\}\int d \{\mathbf{\Omega}^{(\alpha)}\}\nonumber\\
    &\times e^{-\beta_{\mathcal{N}} H_\mathrm{rp}^{[\mathcal{N}]}}\mathcal{A}_0(R)\mathrm{Tr_e}\big[\hat{\mathbf{\Gamma}}_\mathrm{\bar s}\big]e^{\mathcal{L}_\mathrm{rp}^{[\mathcal{N}]}t}\big[\hat{B}\big]_\mathrm{ws},
\end{align}
where $\hat{A}$ is linear in $\hat{R}$ (or $\hat{P}$), and the SM-NRPMD partition function is 
\begin{equation}
\mathcal{Z}_\mathcal{N}=\frac{1}{(2\pi\hbar)^{\mathcal{N}}}\int d \{R_\alpha\}\int d \{P_\alpha\}\int d \{\mathbf{\Omega}^{(\alpha)}\} e^{-\beta_{\mathcal{N}} H_\mathrm{rp}^{[\mathcal{N}]}}\mathrm{Tr_e}\big[\hat{\mathbf{\Gamma}}_\mathrm{\bar s}\big].
\end{equation}
The ring-polymer Hamiltonian $H_\mathrm{rp}^{[\mathcal{N}]}$ in the bead representation is expressed as
\begin{equation}\label{eq:Hrpn}
    H_\mathrm{rp}^{[\mathcal{N}]}=\sum_{\alpha=1}^{\mathcal{N}}\Big[{\mathcal{H}}_0(R_\alpha,P_\alpha)+\frac{m}{2\beta_{\mathcal{N}}^2\hbar^2}(R_\alpha-R_{\alpha-1})^2\Big],
\end{equation}
and the electronic phase term is
\begin{equation}\label{eq:Gamma-sbar}
    \hat{\mathbf{\Gamma}}_\mathrm{\bar s}=\prod_{\alpha}^{\mathcal{N}}e^{-\beta_{\mathcal{N}}\frac{1}{\hbar}\sum_k{\mathcal{H}}_k(R_\alpha)\cdot\hat{\mathcal{S}}_k}\cdot\hat{w}_\mathrm{\bar s}^{(\alpha)}.
\end{equation}
When $\hat{A}$ is a projection operator, the TCF is expressed as
\begin{align}\label{CAB_rpmd2}
    C_{AB}^{[{\mathcal{N}}]}(t)=&\frac{1}{{\cal{Z}}_\mathcal{N}(2\pi\hbar)^{\mathcal{N}}}\int d \{R_\alpha\}\int d \{P_\alpha\}\int d \{\mathbf{\Omega}^{(\alpha)}\}\nonumber\\
    &\times e^{-\beta_{\mathcal{N}} H_\mathrm{rp}^{[\mathcal{N}]}}\mathrm{Tr_e}\big[\hat{\mathbf{\Gamma}}_\mathrm{\bar s}\hat{A}\big]e^{\mathcal{L}_\mathrm{rp}^{[\mathcal{N}]}t}\big[\hat{B}\big]_\mathrm{ws},
\end{align}
and the electronic phase $\mathrm{Tr_e}\big[\hat{\mathbf{\Gamma}}_\mathrm{\bar s}\hat{A}\big]$ is expressed as 
\begin{align}\label{ba_trace}
\mathrm{Tr_e}\big[\hat{\mathbf{\Gamma}}_\mathrm{\bar s}\hat{A}\big]=&\frac{1}{\mathcal{N}}\sum_{\alpha=1}^{\mathcal{N}}\mathrm{Tr_e}\Bigg[\prod_{\gamma\leq\alpha}^{\mathcal{N}}e^{-\beta_{\mathcal{N}}\frac{1}{\hbar}\sum_k{\mathcal{H}}_k(R_\gamma)\cdot\hat{\mathcal{S}}_k}\hat{w}_\mathrm{\bar s}^{(\gamma)}\nonumber\\
&\times\hat{A}\prod_{\gamma>\alpha}^{\mathcal{N}}e^{-\beta_{\mathcal{N}}\frac{1}{\hbar}\sum_k{\mathcal{H}}_{k}(R_\gamma)\cdot\hat{\mathcal{S}}_k}\hat{w}_\mathrm{\bar s}^{(\gamma)}\Bigg].
\end{align}

The Liouvillian of SM-NRPMD (see Eq.~\ref{liou_rp_M}) in the bead representation is expressed as
\begin{align}\label{liou_rp}
&{\mathcal{L}}_\mathrm{rp}^{[\mathcal{N}]}\big[\hat{B}\big]_\mathrm{ws}\nonumber\\
&=\frac{1}{\mathcal{N}}\sum_{\alpha=1}^{\mathcal{N}}\Bigg[\frac{P_\alpha}{m}\frac{\overrightarrow{\partial}}{\partial R_\alpha}-\frac{m}{\beta_{\mathcal{N}}^2\hbar^2}\big(2R_\alpha-R_{\alpha-1}-R_{\alpha+1}\big)\frac{\overrightarrow{\partial}}{\partial P_\alpha}\nonumber\\
&-\Big[\frac{\partial {\mathcal{H}}_0(R_\alpha)}{\partial R_\alpha}+r_\mathrm{s}\sum_{j=1}^{N^2-1}\frac{\partial {\mathcal{H}}_j(R_\alpha)}{\partial R_\alpha}\Omega_j^{(\alpha)}\Big]\frac{\overrightarrow{\partial}}{\partial P_\alpha}\Bigg]{\mathcal{B}}_0^{(\alpha)}\nonumber\\
&+\frac{1}{\mathcal{N}}\sum_{\alpha=1}^{\mathcal{N}}\sum_{i=1}^{N^2-1}\Bigg[\frac{r_\mathrm{s}}{\hbar}\sum_{j,k=1}^{N^2-1}f_{ijk}{\mathcal{H}}_j(R_\alpha)\Omega_{k}^{(\alpha)}\Bigg]{\mathcal{B}}_i^{(\alpha)},
\end{align}
where $\mathcal{B}_0^{(\alpha)}$ and $\mathcal{B}_i^{(\alpha)}$ are defined in Eq.~\ref{Liou_smnrpmd}.

The Liouvillian in Eq.~\ref{liou_rp} leads to the SM-NRPMD Hamiltonian as follows
\begin{align}\label{HN-Omega}
\mathcal{H}_{\mathcal{N}}=&\sum_{\alpha=1}^{\mathcal{N}}\Big[{\mathcal{H}}_0(R_\alpha,P_\alpha)+\frac{m}{2\beta_{\mathcal{N}}^2\hbar^2}(R_\alpha-R_{\alpha-1})^2\nonumber\\
&~~~~~~~~~~~~~~~~~~~+\sum_{i=1}^{N^2-1}r_\mathrm{s}{\mathcal{H}}_i(R_\alpha)\Omega_i^{(\alpha)}\Big]\nonumber\\
=&H_\mathrm{rp}^{[\mathcal{N}]}+\sum_{\alpha=1}^{\mathcal{N}}\sum_{i=1}^{N^2-1}r_\mathrm{s}{\mathcal{H}}_i(R_\alpha)\Omega_i^{(\alpha)}
\end{align}
where ${\mathcal{H}}_0$ is expressed in Eq.~\ref{eq:h0}, and $H_\mathrm{rp}^{[\mathcal{N}]}$ is expressed in Eq.~\ref{eq:Hrpn}.

The EOMs for the SM-NRPMD method based on the Liouvillian in Eq.~\ref{liou_rp} are
\begin{subequations}\label{eom-omega}
    \begin{align}
        &\dot{R}_{\alpha}=\frac{P_{\alpha}}{m},\\
        &\dot{P}_{\alpha}=-\frac{\partial{{H}}_\mathrm{rp}^{[\mathcal{N}]}}{\partial R_{\alpha}}-r_\mathrm{s}\sum_{i=1}^{N^2-1}\frac{\partial{\mathcal{H}}_i}{\partial R_{\alpha}}\cdot \Omega_i^{(\alpha)},\\
        &\dot{\Omega}_i^{(\alpha)}=\frac{1}{\hbar}\sum_{j,k=1}^{N^2-1}f_{ijk}{\mathcal{H}}_{j}(R_\alpha)\cdot \Omega_k^{(\alpha)},
    \end{align}
\end{subequations}
which are the Hamilton's EOMs of the Hamiltonian $\mathcal{H}_{\mathcal{N}}$ in Eq.~\ref{HN-Omega}. This can be viewed as the ring-polymer version of the linearized spin mapping EOMs derived in Ref.~\citenum{bossion2022}, as well as  generalization of the spin precession to $N$-dimensions with $SU(N)$ symmetry\cite{hioeeberly1981} introduced by Hioe and Eberly (without the presence of any nuclear DOF).

The correlation function in Eq.~\ref{CAB_rpmd} and Eq.~\ref{CAB_rpmd2} and the Liouvillian ${\mathcal{L}}_\mathrm{rp}^{[\mathcal{N}]}$ in Eq.~\ref{liou_rp} constitute the SM-NRPMD approach to compute Kubo-transformed TCFs. The approximations we have made to obtain this expression are: (1) Matsubara approximation (Eq.~\ref{Matsubara}), (2) ignoring the ${\mathcal{L}}_i^{\mathrm{n}(\alpha)}$ term (Eq.~\ref{Liou_mode}c) in EOMs, (3) the RPMD approximation (Eq.~\ref{rpmd-approx}), (4) replacing the Matsubara frequencies (Eq.~\ref{matfreq}) with the normal mode frequencies of the ring polymer (Eq.~\ref{eqn:normfreq}) in the correlation function expression. Approximations (1), (3), and (4) are related to the original derivation of the adiabatic version of RPMD, whereas approximation (2) is related specifically to the SM-NRPMD approach.

There are several interesting connections between the current SM-NRPMD formalism and previous works. First, for the two level special case ($N=2$), both the EOMs and the Kubo-transformed TCF of the current formalism reduce back to those proposed in our previous work.\cite{bossion2021} Note that in Ref.~\citenum{bossion2021} the initial distribution (partition function) was derived based on the imaginary-time path-integral approach and the dynamics were proposed. Here, the Kubo-transformed TCF and dynamics are derived based on the Matsubara approximation and the RPMD approximation of the exact TCF. The main equations of SM-NRPMD for the two-state special case are provided in Appendix~\ref{2state}. Second, in the one-bead limit, $\mathcal{N}=1$, the Hamiltonian $\mathcal{H}_{\mathcal{N}}$ (Eq.~\ref{HN-Omega}) and the corresponding EOMs in Eq.~\ref{eom-omega} reduce back to the same expressions for the linearized spin mapping dynamics\cite{runesonrichardson2020,bossion2022} (Eq.~86 in Ref.~\citenum{bossion2022}) for a regular TCF. Third, under the state-independent limit or the electronically adiabatic limit (such that the ground electronic state is well separated from other electronic states), this formalism reduces back to the original adiabatic RPMD.\cite{craig2004,manolopoulos2013} Finally, when choosing only one bead for the nuclear DOF and $\mathcal{N}$ beads for the electronic mapping DOFs, the current formalism is closely connected to the recently proposed spin mapping path-integral approach.\cite{runesonrichardson2021}

Note that under the $t\to0$ limit, Eq.~\ref{CAB_rpmd2} becomes
\begin{align}
    C_{AB}^{[{\mathcal{N}}]}(0)=&\frac{1}{{\cal{Z}}_\mathcal{N}(2\pi\hbar)^{\mathcal{N}}}\int d \{R_\alpha\}\int d \{P_\alpha\}\int d \{\mathbf{\Omega}^{(\alpha)}\}\nonumber\\
    &\times e^{-\beta_{\mathcal{N}} H_\mathrm{rp}^{[\mathcal{N}]}}\mathrm{Tr_e}\big[\hat{\mathbf{\Gamma}}_\mathrm{\bar s}\hat{A}\big]\big[\hat{B}\big]_\mathrm{ws}.
\end{align}
This expression can also be obtained from a standard imaginary-time path-integral technique,\cite{} which will be exact under the $\mathcal{N}\to\infty$ limit. The two-state system ($N=2$) example for such a derivation is provided in our previous work,\cite{bossion2021} which is referred to as the spin coherent state (SCS) partition function (Eq.~35 in Ref.~\citenum{bossion2021}). The difference between the path-integral derivation and the Matsubara derivation is that the Matsubara dynamics introduce the nuclear momenta from a multidimensional Wigner transform (Eq.~\ref{CABN}), which can be viewed as the physical momenta, whereas the imaginary-time path-integral approach introduces these nuclear momenta as fictitious variables.


The original adiabatic version of RPMD has a desirable property which preserves quantum Boltzmann distribution and the detailed balance,  $\big\langle\hat{A}(t)\big\rangle=\big\langle\hat{A}(0)\big\rangle$. The key to achieve the detailed balance condition is
\begin{equation}\label{QBD}
    {\mathcal{L}}_\mathrm{rp}^{[{\mathcal{N}}]}\Big(K_n^{(\alpha)} e^{-\beta_\mathcal{N} H_\mathrm{rp}^{[\mathcal{N}]}}\mathrm{Tr_e}\big[\hat{\mathbf{\Gamma}}_\mathrm{\bar s}\big]\Big)=0,
\end{equation}
where $K_n^{(\alpha)}$ is the Jacobian determinant (see Eq.~\ref{domega-alpha}) in the differential phase space volume element $\int d \{\mathbf{\Omega}^{(\alpha)}\}$. Unfortunately, due to the complexity of the distribution which includes an electronic trace (Eq.~\ref{eq:Gamma-sbar}), we do not have an analytical proof of Eq.~\ref{QBD}. Nevertheless, in our previous work on SM-NRPMD in the 2-state special case, we have shown that the expectation values of position and population are conserved for a sufficiently large number of beads (Fig.~4 of Ref.~\citenum{bossion2021}).

It is also possible to simulate non-equilibrium TCF with SM-NRPMD, similar to the case of the MMST version of NPRMD, as discussed in Ref.~\citenum{chowdhury2021}. In Appendix~\ref{non-eq}, we provide details on how to compute the time-dependent reduced density matrix dynamics upon photo-excitation.

\subsection{Equations of Motion in the Cartesian Mapping Variables}\label{smnrpmd-cart}
There are multiple ways to write the EOMs using various conjugated mapping variables.\cite{bossion2022,runesonrichardson2020} To simplify the expression of the EOMs and reduce the number of mapping variables, one can express the mapping variables $\boldsymbol{\Omega}^{(\alpha)}$ (in Eq.~\ref{Omega-mapping}) as Cartesian mapping variables. This can be accomplished by representing the expansion coefficients of the spin coherent states in Eq.~\ref{eq:omega-expand} by their real and imaginary parts\cite{heslot1985,runesonrichardson2020} for every copy ($\alpha$-bead) of the coherent state as follows
\begin{equation}\label{c-pq-transform}
\langle n|\mathbf{\Omega}^{(\alpha)}\rangle\cdot e^{i\Phi^{(\alpha)}}=\frac{1}{\sqrt{2 r_\mathrm{s}}}(q^{(\alpha)}_{n}+ip^{(\alpha)}_{n}),
\end{equation}
where $q^{(\alpha)}_{n}/\sqrt{2r_\mathrm{s}}$ is the real part of the expansion coefficients, $p^{(\alpha)}_{n}/\sqrt{2r_\mathrm{s}}$ is the imaginary part of the expansion coefficients, and $e^{i\Phi^{(\alpha)}}$
is a {\it constant global phase} to {\it all} of the coefficients $\langle n|\mathbf{\Omega}^{(\alpha)}\rangle$ associated with the $\alpha_\mathrm{th}$ copy of the coherent state $|\mathbf{\Omega}^{(\alpha)}\rangle$. The above relation provides a canonical transformation from the previously defined spin-mapping variables $\{\Omega_{k}^{(\alpha)}\}$ to a set of Cartesian mapping variables
\begin{align}\label{gen-transform}
2r_\mathrm{s}\Omega^{(\alpha)}_k&=2r_\mathrm{s}\langle\mathbf{\Omega^{(\alpha)}}|\hat{\mathcal{S}}_k|\mathbf{\Omega}^{(\alpha)}\rangle\\
&=\sum_{n,m}\langle n|\hat{\mathcal{S}}_{k}|m\rangle\cdot (q^{(\alpha)}_n-ip^{(\alpha)}_{n})\big(q^{(\alpha)}_m+ip^{(\alpha)}_{m}\big).\nonumber
\end{align}

Using this transform (Eq.~\ref{gen-transform}), the estimator of the projection operator (Eq.~\ref{proj-sw}) becomes
\begin{equation}\label{eq:proj-sw}
    \big[|n\rangle\langle m|\big]_\mathrm{s}^{(\alpha)}=\frac{1-r_\mathrm{s}}{{N}}\delta_{nm}+\frac{1}{2}\big(q_m^{(\alpha)}+ip_m^{(\alpha)}\big)\big(q_n^{(\alpha)}-ip_n^{(\alpha)}\big),
\end{equation}
with the details presented in Appendix D (Eq.~D7-Eq.~D10) of Ref.~\citenum{bossion2022}.

The Hamiltonian in Eq.~\ref{HN-Omega} can be transformed to Cartesian variables as follows
\begin{align}\label{Hnpq}
&H_{\mathcal{N}}=\sum_{\alpha=1}^{\mathcal{N}}\Big[{\mathcal{H}}_0(R_\alpha,P_\alpha)+\frac{m}{2\beta_{\mathcal{N}}^2\hbar^2}(R_\alpha-R_{\alpha-1})^2\nonumber\\
&~~+\sum_{n=1}^N [{V}_{nn}(R_\alpha)-\bar{V}(R_\alpha)]\frac{1}{2}\big([q_n^{(\alpha)}]^2+[p_n^{(\alpha)}]^2 -\gamma\big)\nonumber\\
&~~+\sum_{n<m}^N{V}_{nm}(R_\alpha)\big(q_n^{(\alpha)}q_m^{(\alpha)}+p_n^{(\alpha)}p_m^{(\alpha)}\big)\Big],
\end{align}
where $\bar{V}(R_\alpha)\equiv\frac{1}{N}\sum_{n=1}^NV_{nn}({R}_{\alpha})$, and the effective zero-point energy parameter\cite{bossion2022, runesonrichardson2020} 
\begin{equation}\label{eq:gamma}
\gamma=2(r_\mathrm{s}-1)/{N}.
\end{equation}
The expression of $H_{\mathcal{N}}$ in Eq.~\ref{Hnpq} is equivalent to the MMST version of the NRPMD Hamiltonian (when choosing $\gamma=1$), which was originally proposed in Ref.~\citenum{richardson2013} and later justified in Ref.~\citenum{chowdhury2021}. Note that even with the Hamiltonian being identical, the derivation of $H_{\mathcal{N}}$ is based on the $SU(N)$ mapping outlined in Sec.~\ref{SM-NRPMD_sec}, and is not based on the MMST mapping formalism.\cite{stock1997, thoss1999} The fundamental differences between the two mapping formalism are discussed in Appendix D of Ref.~\citenum{bossion2022}.

In terms of the Cartesian mapping variables, $\{q_n^{(\alpha)},p_n^{(\alpha)}\}$, the corresponding EOMs in Eq.~\ref{eom-omega} are transformed as\cite{runesonrichardson2020}
\begin{subequations}\label{eom-pq}
    \begin{align}
        &\dot{R}_\alpha=\frac{P_\alpha}{m},\label{eoma}\\
        &\dot{P}_\alpha=-\frac{\partial{{\mathcal{H}}}_\mathrm{rp}^{[\mathcal{N}]}}{\partial R_{\alpha}}-\sum_{n=1}^{N}\frac{\partial (V_{nn}-\bar{V})}{\partial R_\alpha}\frac{1}{2}\Big([{q_n^{(\alpha)}}]^2+[{p_n^{(\alpha)}}]^2-\gamma\Big)\nonumber\\
        &~~~~~~~-\sum_{n<m}^{N}\frac{\partial V_{nm}}{\partial R_\alpha}\Big(q_n^{(\alpha)}q_m^{(\alpha)}+p_n^{(\alpha)}p_m^{(\alpha)}\Big),\label{eomb}\\
        &\dot{q}_n^{(\alpha)}=\sum_{m=1}^{N}V_{nm}p_m^{(\alpha)},\label{eomc}\\
        &\dot{p}_n^{(\alpha)}=-\sum_{m=1}^{N}V_{nm}q_m^{(\alpha)}\label{eomd},
    \end{align}
\end{subequations}
where $V_{nm}(R_\alpha)$ is defined in Eq.~\ref{eq:Ham}, and $\mathcal{H}_\mathrm{rp}^{[\mathcal{N}]}$ is expressed in Eq.~\ref{eq:Hrpn}. The above EOMs are indeed the Hamilton's EOMs of $H_{\mathcal{N}}$ in Eq.~\ref{Hnpq}.

\section{Computational Details}\label{comp}
{\bf Summary of the SM-NRPMD approach.} The SM-NRPMD approach computes the Kubo-transformed TCF for $\hat{A}=\hat{R}$ (or any linear nuclear operator) based upon Eq.~\ref{CAB_rpmd}, and for $\hat{A}=|n\rangle\langle n|$ (or any projection operator) using Eq.~\ref{CAB_rpmd2}. The initial distribution of the nuclear variables $\{R_\alpha,P_\alpha\}$ and the mapping variables $\{\boldsymbol{\theta}^{(\alpha)},\boldsymbol{\varphi}^{(\alpha)}\}$ (see Appendix~\ref{ps-sm}) are governed by the initial distribution 
\begin{equation}
d \{R_\alpha\} d \{P_\alpha\} d \{\mathbf{\Omega}^{(\alpha)}\}e^{-\beta_\mathcal{N} H_\mathrm{rp}^{[\mathcal{N}]}}\mathrm{Tr_e}\big[\hat{\mathbf{\Gamma}}_\mathrm{\bar s}\big]\nonumber
\end{equation}
for Eq.~\ref{CAB_rpmd} and 
\begin{equation}
d \{R_\alpha\} d \{P_\alpha\} d \{\mathbf{\Omega}^{(\alpha)}\}e^{-\beta_\mathcal{N} H_\mathrm{rp}^{[\mathcal{N}]}}\mathrm{Tr_e}\big[\hat{\mathbf{\Gamma}}_\mathrm{\bar s}\hat{A}\big].\nonumber
\end{equation}
for Eq.~\ref{CAB_rpmd2}, with each $d\mathbf{\Omega}^{(\alpha)}$ expressed in Eq.~\ref{domega-alpha} (with $\{\theta^{(\alpha)}_{n},\varphi^{(\alpha)}_{n}\}$. Note that $K^{(\alpha)}_{n}$ in Eq.~\ref{domega-alpha} (which is a function of $\{\theta^{(\alpha)}_{n}\}$) needs to be included in the sampling function for an efficient numerical convergence. The electronic quantities $\mathrm{Tr_e}\big[\hat{\mathbf{\Gamma}}_\mathrm{\bar s}\big]$ and $\mathrm{Tr_e}\big[\hat{\mathbf{\Gamma}}_\mathrm{\bar s}\hat{A}\big]$ are in general complex.

The dynamics are propagated using the EOMs in Eq.~\ref{eom-pq}. These EOMs are identical to the original MMST-based NRPMD approach.\cite{richardson2013,richardson2017,chowdhury2021} Thus, one can take advantage of using the existing algorithms\cite{richardson2017,kelly2012} for integrating these EOMs. Because the initial conditions are sampled with $\{\varphi^{(\alpha)}_{n},\theta^{(\alpha)}_{n}\}$, these variables are transformed into the Cartesian mapping variables, $\{q^{(\alpha)}_n,p^{(\alpha)}_n\}$, for the dynamics propagation. This can be done based on Eq.~\ref{c-pq-transform} as follows
\begin{subequations}\label{pq-omega}
\begin{align}
q^{(\alpha)}_{n}&=\sqrt{2 r_\mathrm{s}} \cdot \mathrm{Re}\big[\langle n|\mathbf{\Omega^{(\alpha)}}\rangle\cdot e^{i\Phi^{(\alpha)}}\big],\\
p^{(\alpha)}_{n}&=\sqrt{2 r_\mathrm{s}} \cdot \mathrm{Im}\big[\langle n|\mathbf{\Omega^{(\alpha)}}\rangle\cdot e^{i\Phi^{(\alpha)}}\big],
\end{align}
\end{subequations}
where the explicit expression of $\langle n|\mathbf{\Omega^{(\alpha)}}\rangle$ as a function of $\{\theta^{(\alpha)}_{n},\varphi^{(\alpha)}_{n}\}$ can be found in Eq.~\ref{absscs} (for all $\alpha$). The correlation function can then be explicitly calculated through trajectory average. The global phase $e^{i\Phi^{(\alpha)}}$ does not influence the results. For the results presented in this work, we use a particular Bloch sphere radius (referred to as the W method\cite{runesonrichardson2020,bossion2022})
\begin{equation}
r_\mathrm{s}=r_\mathrm{\bar s}=\sqrt{N+1}. 
\end{equation}
This choice has shown to give the overall most accurate results for linearized method\cite{runesonrichardson2020} as well as for the two level case for SM-NRPMD\cite{bossion2021} compared to other radii.

{\bf Numerical algorithm to perform SM-NRPMD simulations.} The SM-NRPMD method evaluates the Kubo-transformed TCFs expressed in Eq.~\ref{CAB_rpmd} or Eq.~\ref{CAB_rpmd2}. The initial distribution is obtained using the Metropolis algorithm\cite{metropolis1953,hastings1970} with the nuclear variables $\{R_\alpha,P_\alpha\}$ and the mapping variables $\{\boldsymbol{\theta}^{(\alpha)},\boldsymbol{\varphi}^{(\alpha)}\}$ as those define the initial phase space. The Metropolis algorithm requires a positive-definite sampling function. As done previously,\cite{bossion2021,ananth2010,ananth2013,richardson2013,chowdhury2021} the absolute value of the initial distribution is sampled and the remaining phase is included in the estimator. The distribution function used for sampling is chosen to be 
\begin{align}\label{dist}
&\rho_\mathrm{\bar s}\big(\{R_\alpha,P_\alpha,\boldsymbol{\theta}^{(\alpha)},\boldsymbol{\varphi}^{(\alpha)}\}\big)\\
&~~~~=\Big[\prod_{\alpha=1}^{\mathcal{N}}\prod_{n=1}^{N-1}K^{(\alpha)}_{n} \Big]\cdot e^{-\beta_{\mathcal{N}} H_\mathrm{rp}^{[\mathcal{N}]}}\big|\mathrm{Tr_e}\big[\hat{\mathbf{\Gamma}}_\mathrm{\bar s}\big]\big|,\nonumber
\end{align}
where the elements $K_n^{(\alpha)}$ (which are functions of $(\theta_{n}^{(\alpha)})$) are expressed in Eq.~\ref{domega-alpha}, and $\hat{\mathbf{\Gamma}}_\mathrm{\bar s}$ is expressed in Eq.~\ref{eq:Gamma-sbar}. In $\rho_\mathrm{\bar s}$, every quantity is expressed in terms of $\{R_\alpha,P_\alpha\}$ and $\{\boldsymbol{\theta}^{(\alpha)},\boldsymbol{\varphi}^{(\alpha)}\}$. The expression of $\hat{\mathbf{\Gamma}}_\mathrm{\bar s}$ (Eq.~\ref{eq:Gamma-sbar}) contains the SW kernel $\hat{w}_\mathrm{\bar s}^{(\alpha)}$ (Eq.~\ref{A-sw-map2}b) as a function of $\{\Omega^{(\alpha)}_{k}\}$. These $\{\Omega^{(\alpha)}_{k}\}$ are also functions of $\{\boldsymbol{\theta}^{(\alpha)},\boldsymbol{\varphi}^{(\alpha)}\}$, with the detailed expressions in Eq.~\ref{eq:Omega_a}-Eq.~\ref{eq:Omega_g} in Appendix~\ref{ps-sm}. 

Using the absolute value of $\mathrm{Tr_e}\big[\hat{\mathbf{\Gamma}}_\mathrm{\bar s}\big]$ as part of the distribution $\rho_\mathrm{\bar s}$, the remaining phase associated with  $\mathrm{Tr_e}\big[\hat{\mathbf{\Gamma}}_\mathrm{\bar s}\big]$ is expressed as 
\begin{equation}\label{Xi-Gamma}
    \Xi_\mathrm{\bar s}\big(\{R_\alpha,P_\alpha,\boldsymbol{\theta}^{(\alpha)},\boldsymbol{\varphi}^{(\alpha)}\}\big)=\frac{\mathrm{Tr_e}\big[\hat{\mathbf{\Gamma}}_\mathrm{\bar s}\big]}{\big|\mathrm{Tr_e}\big[\hat{\mathbf{\Gamma}}_\mathrm{\bar s}\big]\big|},
\end{equation}
which has to be included as part of the estimator when computing ensemble averages (see Eq.~\ref{CRR})

To propagate the dynamics, the spin-mapping variables $\{\boldsymbol{\theta}^{(\alpha)},\boldsymbol{\varphi}^{(\alpha)}\}$ (generalized Euler angles) are converted into the Cartesian variables $\{\bf{p}^{(\alpha)},\bf{q}^{(\alpha)}\}$ after the initial sampling governed by $\rho_\mathrm{\bar s}$ (Eq.~\ref{dist}). The global phase is chosen to be $\Phi^{\alpha}=0$ for all initially sampled configurations. Specifically, we use the normal mode integrator\cite{ceriotti2010} to solve the ring-polymer part of the EOMs in Eqs.~\ref{eoma}-\ref{eomb}, and the symplectic integrator for the mapping variables\cite{nandini2018,kelly2012} in Eqs.~\ref{eomc}-\ref{eomd}. Note that a simple Verlet algorithm\cite{richardson2017} with a sufficiently small time-step to propagate the mapping variables gives an identical result.

The Kubo-transformed nuclear position auto-correlation function is obtained by choosing $\hat{A}=\hat{B}=\hat{R}$. The SM-NRPMD approximation is evaluated as
\begin{equation}\label{CRR}
    C^{[\mathcal{N}]}_{RR}(t)=\frac{\Big\langle\mathrm{Re}\{\Xi_\mathrm{\bar s}\}\bar{R}(0)\bar{R}(t)\Big\rangle}{\Big\langle\mathrm{Re}\{\Xi_\mathrm{\bar s}\}\Big\rangle},
\end{equation}
where $\bar{R}=\frac{1}{\mathcal{N}}\sum_{\alpha=1}^{\mathcal{N}}R_\alpha$. The brackets $\langle\cdots\rangle$ indicate an ensemble average according to the initial distribution governed by $\rho_\mathrm{\bar s}$. Only the real part of the phase estimator in Eq.~\ref{Xi-Gamma} is included as the estimator, because the correlation function is purely real. 

We also consider $\hat{A}=|m\rangle\langle m|$ and $\hat{B}=|n\rangle\langle n|$ for the Kubo-transformed TCF of population. The SM-NRPMD approximation of this TCF is
\begin{equation}
    C^{[\mathcal{N}]}_{mn}(t)=\frac{\Big\langle\mathrm{Re}\{\xi_\mathrm{\bar s}\}\big[|n\rangle\langle n|\big]_\mathrm{s}(t)\Big\rangle}{\Big\langle\mathrm{Re}\{\Xi_\mathrm{\bar s}\}\Big\rangle},
\end{equation}
where the initial distribution function is always $\rho_\mathrm{\bar s}$ in Eq.~\ref{dist}, and 
\begin{equation}
\xi_\mathrm{\bar s}\big(\{R_\alpha,P_\alpha,\boldsymbol{\theta}^{(\alpha)},\boldsymbol{\varphi}^{(\alpha)}\}\big)=\frac{\mathrm{Tr_e}\big[\hat{\mathbf{\Gamma}}_\mathrm{\bar s}|m\rangle\langle m|\big]}{\big|\mathrm{Tr_e}\big[\hat{\mathbf{\Gamma}}_\mathrm{\bar s}\big]\big|},
\end{equation}
where $\mathrm{Tr_e}\big[\hat{\mathbf{\Gamma}}_\mathrm{\bar s}|m\rangle\langle m|\big]$ is evaluated using the expression in Eq.~\ref{ba_trace} (with $\hat{A}=|m\rangle\langle m|$), and the time-evolved projection operator (Eq.~\ref{pop_est}) is evaluated (based on Eq.~\ref{eq:proj-sw}) as follows
\begin{equation}\label{n-bead}
\big[|n\rangle\langle n|\big]_\mathrm{s}(t)=\frac{1}{\mathcal N}\sum_{\alpha=1}^{\mathcal N}\frac{1}{2}\Big([q_n^{(\alpha)}(t)]^2+[p_n^{(\alpha)}(t)]^2-\gamma\Big),
\end{equation}
where $\gamma$ is defined in Eq.~\ref{eq:gamma}.

{\bf Model Systems.} The accuracy of SM-NRPMD dynamics for $N=2$ special case has already been extensively tested in our previous work.\cite{bossion2021} In this work, we focus on $N=3$ model systems with electronic coupling parameters $\Delta$, ranging from electronically adiabatic regime ($\beta \Delta\gg 1$) to non-adiabatic regime ($\beta \Delta\ll 1$). The model contains three diabatic electronic states $|1\rangle$, $|2\rangle$, and $|3\rangle$, as well as one nuclear DOF $\hat{R}$, with the Hamiltonian expressed as
\begin{align}\label{model}
\hat{H} &= \frac{\hat{P}^2}{2m} + \frac{1}{2}m\omega\hat{R}^2+\begin{pmatrix}
    k_1\hat{R}+\varepsilon_1 & \Delta_{12} & 0 \\ 
    \Delta_{12} & k_2\hat{R}+\varepsilon_2 & \Delta_{23} \\ 
    0 & \Delta_{23} & k_3\hat{R}+\varepsilon_3
    \end{pmatrix},\nonumber\\
&\equiv\frac{\hat{P}^2}{2m} +\hat{V}.
\end{align}
In this model, state $|2\rangle$ is coupled to state $|1\rangle$ and state $|3\rangle$, but there is no direct electronic coupling between state $|1\rangle$ and $|3\rangle$. This model is similar to a charge transfer model from donor state to acceptor state, with a bridge state mediating the charge transfer process. We set $\hbar=\omega=\beta=1$. Three sets of parameters are considered in this work, which are summarized in Tab.~\ref{tab:tw_param} and referred to as Model I, II, and III.
\begin{table}[htbp]
    \caption{Parameters for the three-state model (in a.u.).}
    \begin{tabular*}{\columnwidth}{c @{\extracolsep{\fill}} c @{\extracolsep{\fill}} c @{\extracolsep{\fill}} c}
        \hline\hline
        Parameter & Model I & Model II & Model III \\
        \hline
        $m$ & 1.0 & 1.0 & 2.0 \\
        $k_1$ & 1.0 & 2.0 & 2.0 \\
        $k_2$ & -1.0 & -1.0 & 0.0 \\
        $k_3$ & 2.0 & 0.0 & -2.0 \\
        $\epsilon_1$ & 0.0 & 0.0 & 0.0 \\
        $\epsilon_2$ & 0.0 & 0.0 & -2.5 \\
        $\epsilon_3$ & 0.0 & 0.0 & 0.0 \\
        $\Delta_{12}$ & 10.0 & 5.0 & 0.5 \\
        $\Delta_{23}$ & 10.0 & 1.0 & 0.5 \\ [0.5ex]
        \hline\hline
    \end{tabular*}
    \label{tab:tw_param}
\end{table}

{\bf Computational Details.} All the simulations presented here require $\mathcal{N}=6$ beads for converged results. The nuclear time-step used in all the simulations is $\Delta t =10^{-2}$ a.u., and the electronic time-step for the mapping variables is $\mathrm{d}t=\Delta t/10$. To obtain converged results, a total of $5\times10^5$ trajectories is used for the $C^{[\mathcal{N}]}_{RR}(t)$ calculations for model I, and up to $5\times10^6$ trajectories are required for models II, and $10^7$ for model III. The population correlation function $C^{[\mathcal{N}]}_{mn}(t)$ requires 5-10 times more trajectories to converge compared to the $C^{[\mathcal{N}]}_{RR}(t)$ calculations for each model systems.

We present numerical comparisons between the SM-NRPMD approach and the previously proposed MMST-based NRPMD method,\cite{richardson2013} and benchmark against the exact calculation of the Kubo-transformed TCF, with the details of these other two approaches provided in Appendix \ref{exact-nrpmd}. The numerical convergence of the MMST-based NRPMD method\cite{richardson2013} is similar to that of SM-NRPMD. For both of the NRPMD approaches, it requires more trajectories to converge the three-level model systems studied here compare to the two-level systems investigated previously,
\cite{richardson2013} due to the more severe sign problem in the $\mathrm{Tr_e}\big[\hat{\mathbf{\Gamma}}_\mathrm{\bar s}\big]$ term.

\section{Results and discussion}\label{result}
\begin{figure}
    \centering
    \includegraphics[width=\linewidth]{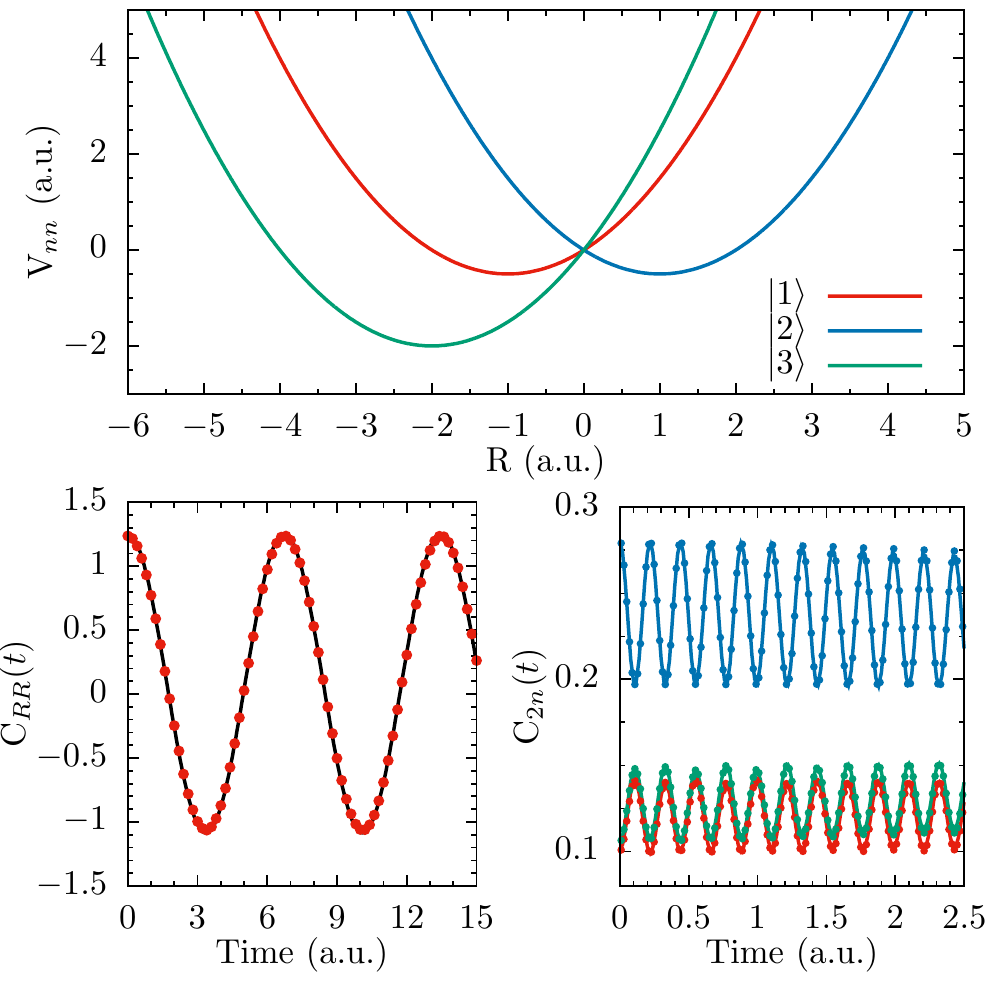}
    \caption{Top panel: The diagonal potential $V_{nn}=\langle n|\hat{V}|n\rangle$ of Model I, with $\hat{V}$ defined in Eq.~\ref{model}. Bottom left panel: Kubo-transformed position auto-correlation function obtained from SM-NRPMD (black solid line) compared to the numerically exact result (red dots). Bottom right panel: Kubo-transformed electronic population correlation functions, $C^{[\mathcal{N}]}_{21}(t)$ (red), $C^{[\mathcal{N}]}_{22}(t)$ (blue) and $C^{[\mathcal{N}]}_{23}(t)$ (green), obtained with SM-NRPMD (solid lines) and compared to exact results (dots). The results from MMST-NRPMD are indistinguishable with the SM-NRPMD, and thus not presented here.}
    \label{fig1}
\end{figure}

Fig.~\ref{fig1} presents the results of the Kubo-transformed TCF for model I. This model is in the electronically adiabatic regime ($\beta \Delta_{12}=\beta\Delta_{23}\gg 1$). The top panel depicts the diagonal potential $V_{nn}=\langle n|\hat{V}|n\rangle$ with $\hat{V}$ defined in Eq.~\ref{model}. The bottom left panel presents the auto-correlation function of the nuclear position operator $C^{[\mathcal{N}]}_{RR}(t)$, computed using the SM-NRPMD approach (black solid line) and the numerically exact result (red dots). The result of the MMST-based NRPMD approach is visually identical to the result of SM-NRPMD, and thus is not shown here. The bottom right panel presents the population correlation functions $C^{[\mathcal{N}]}_{21}(t)$ (red), $C^{[\mathcal{N}]}_{22}(t)$ (blue), and $C^{[\mathcal{N}]}_{23}(t)$ (green), obtained with SM-NRPMD (solid lines) and numerically exact simulations (filled circles). The MMST-based NRPMD approach is also visually identical to the results of SM-NRPMD. Similar to the previous studies for two level systems,\cite{richardson2013} under the electronically adiabatic limit, SM-NRPMD agrees with the exact answer. Furthermore, the SM-NRPMD approach provides accurate initial quantum statistics (exact value of $C^{[\mathcal{N}]}_{AB}(0)$) as well as the correct electronic Rabi oscillations in $C^{[\mathcal{N}]}_{mn}(t)$.

\begin{figure}
    \centering
    \includegraphics[width=\linewidth]{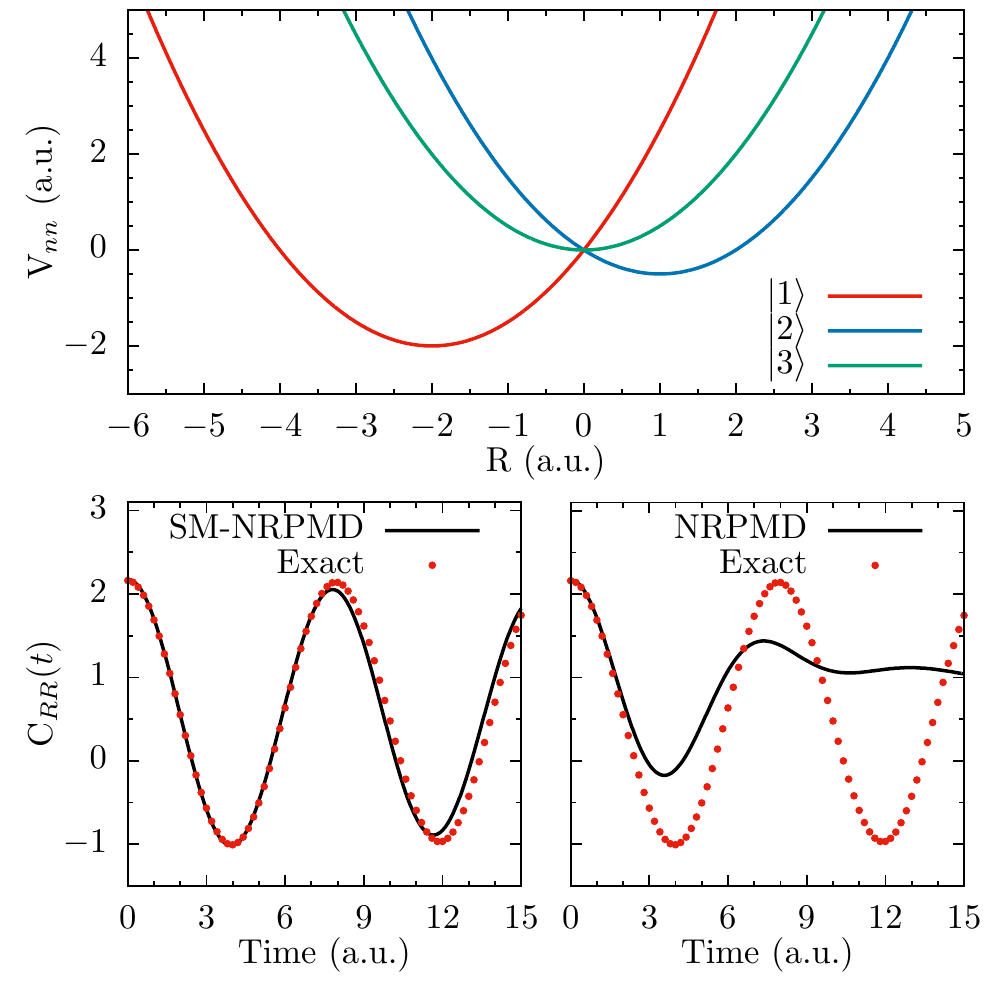}
    \caption{Top: The diagonal potential $V_{nn}$ of Model II. Bottom left panel: auto-correlation function of position obtained with SM-NRPMD (black solid line) compared to exact result (red dots). Bottom right panel: auto-correlation function of position obtained with NRPMD (solid lines).}
    \label{fig2}
\end{figure}

Fig.~\ref{fig2} presents the results for model II, where $\beta\Delta_{12}>1$ (adiabatic limit) and $\beta\Delta_{23}=1$ (intermediate regime). The top panel depicts the diagonal potential $V_{nn}$. The bottom panels present the $C^{[\mathcal{N}]}_{RR}(t)$ obtained from SM-NRPMD (bottom left) and the MMST-based NRPMD (bottom right). The dynamics are reproduced at early times by SM-NRPMD (bottom left) and matches the exact result. For a longer time (for $t>6$ a.u.) the SM-NRPMD dynamics starts to deviate from the exact answer but remain accurate. On the other hand, NRPMD (bottom right) is only able to accurately capture the dynamics for a very short time. The magnitude of the oscillations of the NRPMD approach dampens and the auto-correlation function remains positive, whereas the exact result oscillates around zero. 

\begin{figure}
    \centering
    \includegraphics[width=\linewidth]{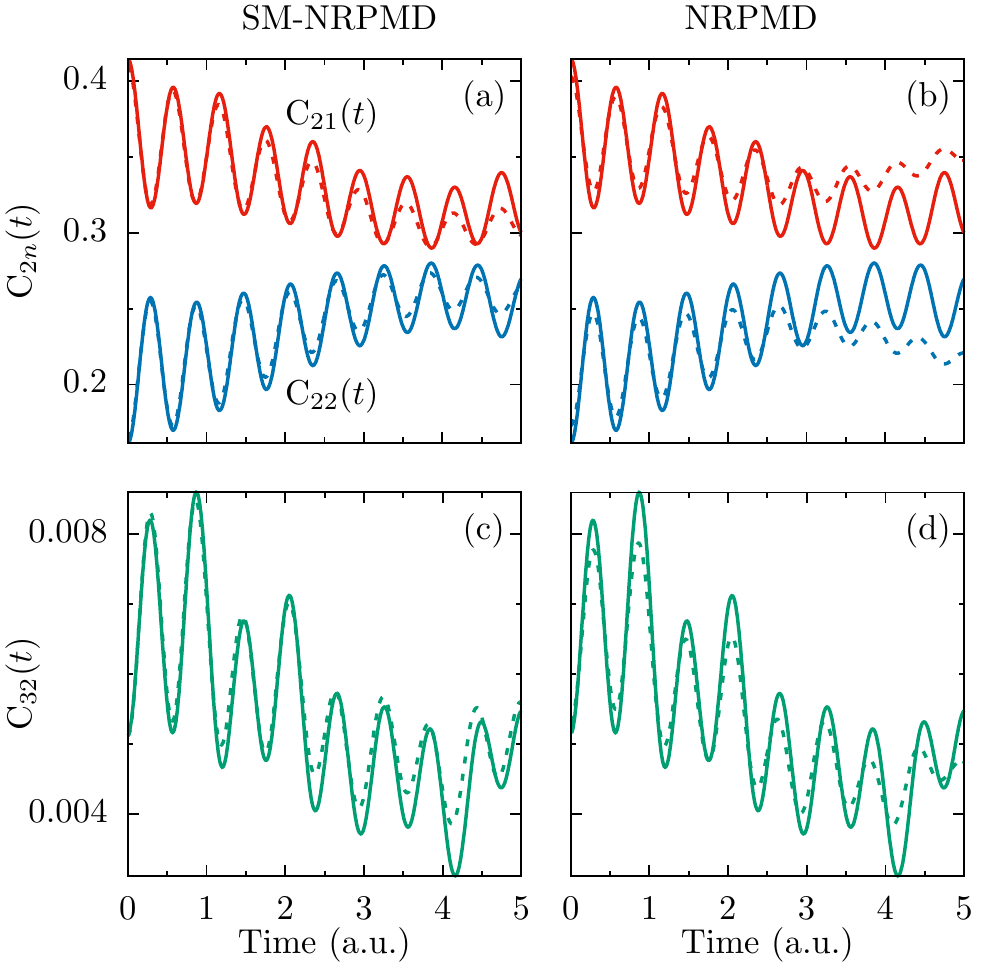}
    \caption{Kubo-transformed population TCF (a)-(b) $C^{[\mathcal{N}]}_{21}(t)$ (red) and $C^{[\mathcal{N}]}_{22}(t)$ (blue), as well as $C^{[\mathcal{N}]}_{32}(t)$ (green) for model II. Results are obtained with SM-NRPMD (dashed lines in panels (a) and (c)) and NRPMD (dashed lines in panels (b) and (d)) compared with exact calculations (solid lines).}
    \label{fig3}
\end{figure}

Fig.~\ref{fig3} presents the population correlation functions of model II. Although both SM-NRPMD and NRPMD are able to capture the basic features of the exact dynamics, there are deviations between them with the exact results. Nevertheless, SM-NRPMD is more accurate than NRPMD, where the latter deviates from the correct value in panel (b) for $C^{[\mathcal{N}]}_{21}(t)$ (red) and $C^{[\mathcal{N}]}_{22}(t)$ (blue) although the coupling $\Delta_{12}$ is under the adiabatic regime  $\beta\Delta_{12}>1$. This deviation is likely due to the presence of the other intermediate coupling $\Delta_{23}$, causing numerical challenges for NRPMD. This deviation was also seen on the position auto-correlation function in the bottom right panel of Fig~\ref{fig2}. Panels (c) and (d) show $C^{[\mathcal{N}]}_{32}(t)$, which present a correlation between the population of states $|2\rangle$ and $|3\rangle$ where the electronic coupling between them is in the intermediate regime ($\beta\Delta_{23}=1$). In this case, the early dynamics is exactly reproduced by SM-NRPMD, but NRPMD is underestimating the oscillation magnitude for $t<0.5$ a.u. At longer times, both methods overly dampen the oscillation magnitudes compared to the exact results.

\begin{figure}
    \centering
    \includegraphics[width=\linewidth]{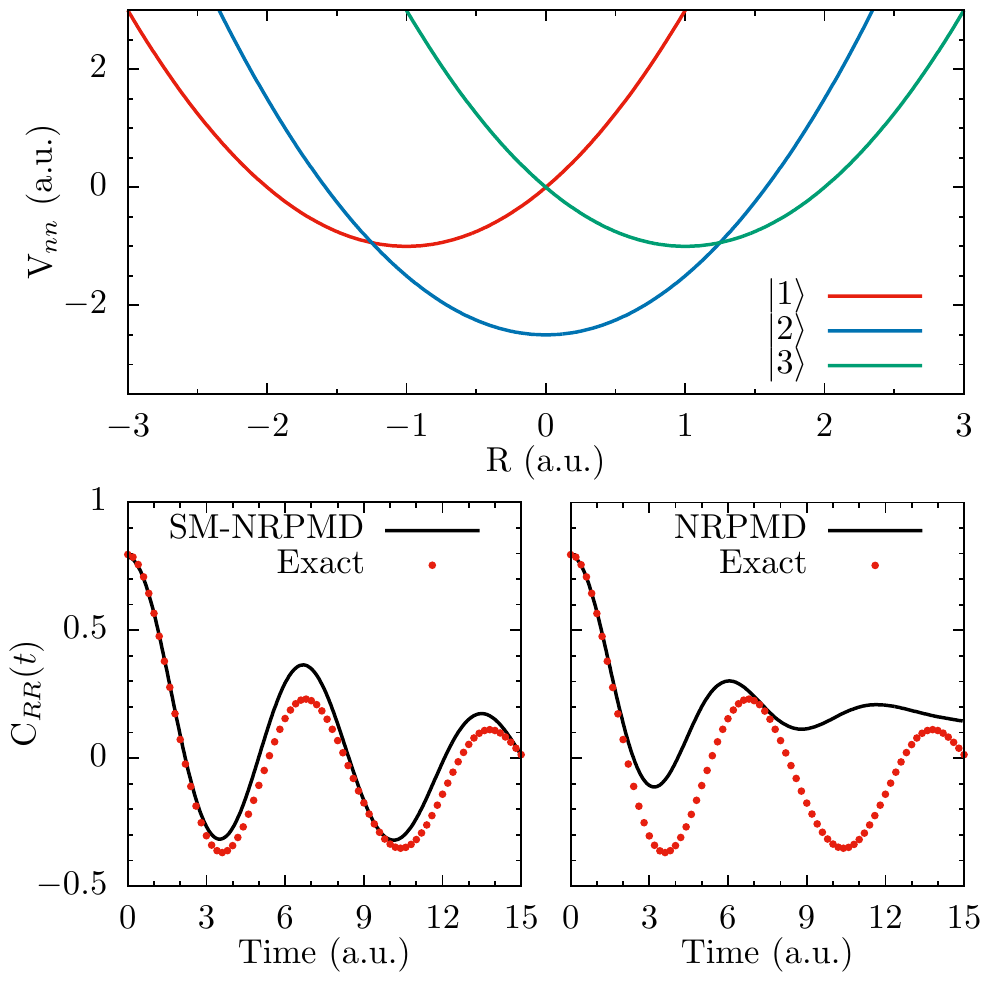}
    \caption{Top: The diagonal potential $V_{nn}$ of Model III. Bottom left panel: Kubo-transformed auto-correlation function of position obtained with SM-NRPMD (black solid line) compared to exact result (red dots). Bottom right panel: auto-correlation function of position obtained with NRPMD (solid lines).}
    \label{fig4}
\end{figure}

Fig.~\ref{fig4} presents the results for model III, with the diagonal potential presented in the top panel. Model III is under the electronically non-adiabatic regime ($\beta \Delta_{12}=\beta\Delta_{23}<1$). This model is numerically challenging for both  trajectory-based methods, similarly to the previously investigated two-level systems under the same regime.\cite{richardson2013,ananth2013} Both SM-NRPMD (bottom left) and NRPMD (bottom right) are unable to reproduce the dynamics exactly except at short times. While the NRPMD results quickly deviates from the exact answer (bottom right), SM-NRPMD (bottom left) provides a much better agreement of the period of oscillation for the correlation function, and only deviates slightly from the exact answer in terms of oscillation amplitude. All the numerical results  demonstrate that SM-NRPMD provides more accurate dynamics compared to the original NRPMD approach.\cite{richardson2013} 

\section{Conclusion}\label{conclusion}
In this work, we derive the SM-NRPMD method for a general $N$-level system. Our formalism is based on the Kubo-transformed time correlation function (TCF). We use the spin-mapping representation and Stratonovich-Weyl transform that map the electronic DOFs onto continuous variables in the phase space of $SU(N)$ Lie group, which explicitly addresses the known challenges  \cite{runesonrichardson2020,bossion2022} in the original MMST mapping formalism.\cite{stock1997,thoss1999} We then use the multidimensional Wigner transform to describe the nuclear DOFs in the Kubo-transformed TCF. We further derive the spin mapping non-adiabatic Matsubara dynamics using the Matsubara approximation that removes the high frequency nuclear normal modes in the TCF. Further, discarding the imaginary part of the Liouvillian (i.e., the RPMD approximation) from the non-adiabatic Matsubara dynamics, we derive the expression of the SM-NRPMD method. The two-state special case of SM-NRPMD was first proposed in Ref.~\citenum{bossion2021}. For the one bead limit, the EOMs of the SM-NRPMD method reduce back to the EOMs used in the spin-LSC method for a regular TCF. Even though the EOMs for SM-NRPMD (Eq.~\ref{eom-pq}) are formally identical to those of the MMST-based NRPMD,\cite{richardson2013,chowdhury2021} the initial distribution of the mapping variables (see Eq.~\ref{dist} for SM-NRPMD and Eq.~\ref{eqn:NRPMD_original} for NRPMD) and the choice of zero-point energy parameters, or rather of Bloch sphere radius (Eq.~\ref{eq:gamma} for SM-NRPMD and $\gamma=1$ for NRPMD\cite{richardson2013}) are indeed different.

We use numerical simulations to demonstrate the accuracy of the SM-NRPMD approach, with a three-state system coupled to one nuclear DOF, with the electronic couplings ranging from adiabatic to the non-adiabatic limit. Using the SM-NRPMD approach, we computed Kubo-transformed nuclear position auto-correlation function, and population time-correlation functions. The results generated from SM-NRPMD are very accurate compared to the numerically exact simulations in all parameter regimes, and provide a significant improvement compared to the NRPMD method. Future applications of SM-NRPMD could be computing rate constants for non-adiabatic systems based on the flux-side correlation function formalism,\cite{huo2013, duke2016,menzeleev2014,Lawrence2019,Lawrence2020} where nuclear quantum effects and electronic non-adiabatic effects are both present.

We want to point out that the spin mapping based non-adiabatic Matsubara dynamics method (Eq.~\ref{CABMR} and Eq.~\ref{CABMproj}) is a general theoretical framework, which can be used to derive other state-dependent path-integral approaches, such as the non-adiabatic CMD\cite{liao2002} (through a mean field approximation\cite{hele2015_2,trenins2018} of the non-adiabatic Matsubara dynamics), or taking advantage of the recent progress in the developments of nuclear quantum dynamics based on various approximations of the Matsubara dynamics.\cite{althorpe2015,willatt2017,jung2019,jung2020,cao1994,hele2015_2} We hope that our current work provides a framework for developing accurate quantum dynamics approaches.

\section*{ACKNOWLEDGEMENTS}
This work was supported by the National Science Foundation
CAREER Award under Grant No. CHE-1845747. Computing
resources were provided by the Center for Integrated Research
Computing (CIRC) at the University of Rochester.

\section*{Conflict of Interest}
The authors have no conflicts to disclose.

\section*{Availability of Data}	
The data that support the findings of this study are available from the corresponding author upon a reasonable request.

\appendix
\section{Analytic Expressions of the Generators and structure constants in $\mathfrak{su}(N)$ Lie Algebra}\label{fijk}

We present the expression of the spin operators $\hat{\mathcal{S}}_{i}$ (equivalent to the generators up to a constant, $\hbar/2$) with $i\in \{1, \cdots, N^2-1\}$. There are $N(N-1)/2$ symmetric matrices
\begin{equation}\label{ssym}
    \hat{\mathcal{S}}_{\alpha_{nm}}=\frac{\hbar}{2}\big(|m\rangle\langle n|+|n\rangle\langle m|\big),
\end{equation}
$N(N-1)/2$ antisymmetric matrices,
\begin{equation}\label{sasym}
    \hat{\mathcal{S}}_{\beta_{nm}}=-i\frac{\hbar}{2}\big(|m\rangle\langle n|-|n\rangle\langle m|\big),
\end{equation}
and $N-1$ diagonal matrices,
\begin{equation}\label{sdiag}
    \hat{\mathcal{S}}_{\gamma_{n}}=\frac{\hbar}{\sqrt{2n(n-1)}}\Big(\sum_{l=1}^{n-1}|l\rangle\langle l|+(1-n)|n\rangle\langle n|\Big),
\end{equation}
where we introduced the indices $\alpha_{nm}$ related to the symmetric matrices, $\beta_{nm}$ related to the antisymmetric matrices, and $\gamma_{n}$ related to the diagonal matrices as follows
\begin{subequations}
    \begin{align}\label{index1}
        \alpha_{nm}=&n^2+2(m-n)-1,\\
        \beta_{nm}=&n^2+2(m-n),\\
        \gamma_{n}=&n^2-1.\label{index2}
    \end{align}
\end{subequations}
where $1\leq m<n\leq N$ and $2\leq n\leq N$, and the generators are ordered according to the conventions.~\cite{pfeifer2003,Krammer2008} 

All the non-zero totally antisymmetric structure constants are expressed as follows
\begin{align}\label{eq:fijk}
    &f_{\alpha_{nm}\alpha_{kn}\beta_{km}}=f_{\alpha_{nm}\alpha_{nk}\beta_{km}}=f_{\alpha_{nm}\alpha_{km}\beta_{kn}}=\frac{1}{2},\\
    &f_{\beta_{nm}\beta_{km}\beta_{kn}}=\frac{1}{2},\nonumber\\
    &f_{\alpha_{nm}\beta_{nm}\gamma_{m}}=-\sqrt{\frac{m-1}{2m}},~~~f_{\alpha_{nm}\beta_{nm}\gamma_{n}}=\sqrt{\frac{n}{2(n-1)}},\nonumber\\
    &f_{\alpha_{nm}\beta_{nm}\gamma_{k}}=\sqrt{\frac{1}{2k(k-1)}},~m<k<n.\nonumber
\end{align}

All the non-zero totally symmetric structure constants are expressed as follows
\begin{align}\label{eq:dijk}
    &d_{\alpha_{nm}\alpha_{kn}\alpha_{km}}=d_{\alpha_{nm}\beta_{kn}\beta_{km}}=d_{\alpha_{nm}\beta_{mk}\beta_{nk}}=\frac{1}{2},\\
    &d_{\alpha_{nm}\beta_{nk}\beta_{km}}=-\frac{1}{2},\nonumber\\
    &d_{\alpha_{nm}\alpha_{nm}\gamma_{m}}=d_{\beta_{nm}\beta_{nm}\gamma_{m}}=-\sqrt{\frac{m-1}{2m}},\nonumber\\
    &d_{\alpha_{nm}\alpha_{nm}\gamma_{k}}=d_{\beta_{nm}\beta_{nm}\gamma_{k}}=\sqrt{\frac{1}{2k(k-1)}},~m<k<n,\nonumber\\
    &d_{\alpha_{nm}\alpha_{nm}\gamma_{n}}=d_{\beta_{nm}\beta_{nm}\gamma_{n}}=\frac{2-n}{ \sqrt{2n(n-1)}},\nonumber\\
    &d_{\alpha_{nm}\alpha_{nm}\gamma_{k}}=d_{\beta_{nm}\beta_{nm}\gamma_{k}}=\sqrt{\frac{2}{k(k-1)}},~n<k,\nonumber\\
    &d_{\gamma_{n}\gamma_{k}\gamma_{k}}=\sqrt{\frac{2}{n(n-1)}},~k<n,\nonumber\\
    &d_{\gamma_{n}\gamma_{n}\gamma_{n}}=(2-n)\sqrt{\frac{2}{n(n-1)}}.\nonumber
\end{align}
Those expressions are valid for any dimension $N$ of the $\mathfrak{su}(N)$ Lie algebra without needing to explicitly compute the commutation and anti-commutation relations. The derivations are provided in the Supplementary Materials of Ref.~\citenum{bossion2022}.

\section{Phase space of the spin-mapping variables}\label{ps-sm}
The expansion coefficients of the generalized spin-coherent states in the diabatic basis are\cite{runesonrichardson2020,nemoto2000,tilma2002,tilma2004} 
\begin{equation}\label{absscs}
\langle n|\mathbf{\Omega}\rangle=
    \begin{cases}
        \cos\frac{\theta_1}{2}, & n=1, \\ 
        \cos\frac{\theta_n}{2}\prod_{l=1}^{n-1}\sin\frac{\theta_l}{2}e^{i\varphi_{l}}, & 1<n<N, \\ 
        \prod_{l=1}^{N-1}\sin\frac{\theta_l}{2}e^{i\varphi_l}, & n=N,
    \end{cases}
\end{equation}
with $\{\theta_n\}\in[0,\pi]$ and $\{\varphi_n\}\in[0,2\pi]$. The $N=2$ special case of the spin coherent state is expressed in Eq.~\ref{spin2}.

The expression of the differential phase-space volume element $ d \mathbf{\Omega}$ (which is also referred to as the invariant integration measure on the group), {\it i.e.}, the Haar measure \cite{GTM225} is
\begin{equation}\label{dOmega}
     d \mathbf{\Omega}=\frac{N!}{(2\pi)^{N-1}}\prod_{n=1}^{N-1}K_n(\theta_n) d \theta_n d \varphi_n,
\end{equation}
where
\begin{equation}\label{KOmega}
    K_n(\theta_n)=\cos\frac{\theta_n}{2} \Big(\sin\frac{\theta_n}{2}\Big)^{2(N-n)-1}.
\end{equation}

When generalizing the above expression for each bead $\alpha$, they have the same expressions as in Eq.~\ref{dOmega} and Eq.~\ref{KOmega}. More specifically,
\begin{subequations}\label{domega-alpha}
\begin{align}  
d \mathbf{\Omega}^{(\alpha)}&=\frac{N!}{(2\pi)^{N-1}}\prod_{n=1}^{N-1}K_n^{(\alpha)} d \theta^{(\alpha)}_n d \varphi^{(\alpha)}_n,\\
K_n^{(\alpha)}&=\cos\frac{\theta^{(\alpha)}_n}{2} \Bigg(\sin\frac{\theta^{(\alpha)}_n}{2}\Bigg)^{2(N-n)-1},
\end{align}
\end{subequations}
and $\langle n|\mathbf{\Omega}^{(\alpha)}\rangle$ as a function of $\{\theta^{(\alpha)}_{n},\varphi^{(\alpha)}_{n}\}$ has the same expression in Eq.~\ref{absscs} for every bead index $\alpha$.

The expectation values of the spin operators in terms of the angles $\{\theta_i,\varphi_i\},~i\in[1,N-1]$ are for the symmetric ones,
\begin{align}\label{eq:Omega_a}
    &\hbar\Omega_{\alpha_{nm}}\equiv\langle\mathbf{\Omega}|\hat{\mathcal{S}}_{\alpha_{nm}}|\mathbf{\Omega}\rangle\nonumber\\
    &=\hbar\prod_{j=1}^{m-1}\sin^2\frac{\theta_j}{2}\cos\frac{\theta_m}{2}\prod_{k=m}^{n-1}\sin\frac{\theta_k}{2}\cos\frac{(1-\delta_{nN})\theta_n}{2}\nonumber\\
    &~~~\times\cos\Big(\sum_{l=m}^{n-1}\varphi_l\Big),
\end{align}
with $1\leq m<n\leq N$. When $m=1$, $\prod_{j=1}^{m-1}\sin^2\frac{\theta_j}{2}$ is replaced by 1. Similarly, for the antisymmetric spin operator, we have
\begin{align}\label{eq:Omega_b}
    &\hbar\Omega_{\beta_{nm}}\equiv\langle\mathbf{\Omega}|\hat{\mathcal{S}}_{\beta_{nm}}|\mathbf{\Omega}\rangle\nonumber\\
    &=\hbar\prod_{j=1}^{m-1}\sin^2\frac{\theta_j}{2}\cos\frac{\theta_m}{2}\prod_{k=m}^{n-1}\sin\frac{\theta_k}{2}\cos\frac{(1-\delta_{nN})\theta_n}{2}\nonumber\\
    &~~~\times\sin\Big(\sum_{l=m}^{n-1}\varphi_{l}\Big),
\end{align}
and when $m=1$, the term $\prod_{j=1}^{m-1}\sin^2\frac{\theta_j}{2}$ is replaced by 1. For the diagonal spin operators there is only one index $1<n\leq N$ and the expression is
\begin{align}\label{eq:Omega_g}
    &\hbar\Omega_{\gamma_{n}}\equiv\langle\mathbf{\Omega}|\hat{\mathcal{S}}_{\gamma_{n}}|\mathbf{\Omega}\rangle\\
    &=\frac{\hbar}{\sqrt{2n(n-1)}}\Big(\sum_{j=1}^{n-1}\cos^2\frac{\theta_j}{2}\prod_{k=1}^{j-1}\sin^2\frac{\theta_k}{2}\nonumber\\
    &~~~+(1-n)\cos^2\frac{(1-\delta_{nN})\theta_n}{2}\prod_{j=1}^{n-1}\sin^2\frac{\theta_j}{2}\Big),\nonumber
\end{align}
where $\prod_{k=1}^{j-1}\sin^2\frac{\theta_k}{2}$ is replaced by 1 when $n=2$ (or $j=1$).

\section{Derivation of Eq.~\ref{CABN}}\label{CABN-detail}
From Eq.~\ref{eq:kubo}, we use the properties of the SW transform, as well as perform the Wigner transform on the nuclear DOF, to derive Eq.~\ref{CABN}. Here, we provide the details of this derivation. In Eq.~\ref{eq:kubo}, we identify 
\begin{equation}
\hat{\mathcal{A}}_\alpha\equiv~e^{-\beta_{\mathcal{N}}\hat{H}}\hat{\mathds{1}}_{R_\alpha'};~\hat{\mathcal{B}}_\alpha\equiv~e^{\frac{i}{\hbar}\hat{H}t}e^{-\frac{i}{\hbar}\hat{H}t}\hat{\mathds{1}}_{R_\alpha''},
\end{equation}
except for
\begin{equation}
\hat{\mathcal{A}}_{{\mathcal{N}}-\alpha}\equiv~e^{-\beta_{\mathcal{N}}\hat{H}}\hat{A}\hat{\mathds{1}}_{R_{{\mathcal{N}}-\alpha}'};~\hat{\mathcal{B}}_{\mathcal{N}}\equiv~e^{\frac{i}{\hbar}\hat{H}t}e^{-\frac{i}{\hbar}\hat{H}t}\hat{B}\hat{\mathds{1}}_{R_{\mathcal{N}}''}.
\end{equation}
We further use the property of the SW transform in Eq.~\ref{SWssb} to rearrange the integrand in $\int d \{\mathbf{\Omega}^{(\alpha)}\}$ as follows
\begin{equation}\label{re-arrange1}
\int d\{\mathbf{\Omega}^{(\alpha)}\}\big[\hat{\mathcal{B}}_1\hat{\mathcal{A}}_2\hat{w}_\mathrm{s}^{(2)}\hat{\mathcal{B}}_2\hat{\mathcal{A}}_3\hat{w}_\mathrm{s}^{(3)}\hat{\mathcal{B}}_3\cdots\hat{\mathcal{A}}_{\mathcal{N}}\hat{w}_\mathrm{s}^{({\mathcal{N}})}\hat{\mathcal{B}}_{\mathcal{N}}\hat{\mathcal{A}}_1\big]_\mathrm{s}^{(1)},
\end{equation}
where the competing structure is $\hat{\mathcal{A}}_{\alpha}\hat{w}_\mathrm{s}^{(\alpha)}\hat{\mathcal{B}}_{\alpha}$. Using the property in Eq.~\ref{SWssb}, we can re-express Eq.~\ref{re-arrange1} as follows,
\begin{align}\label{re-arrange2}
&\int d \{\mathbf{\Omega}^{(\alpha)}\}\big[\hat{\mathcal{B}}_1\big]_\mathrm{s}^{(1)}\big[\hat{\mathcal{A}}_2\hat{w}_\mathrm{s}^{(2)}\hat{\mathcal{B}}_2\hat{\mathcal{A}}_3\cdots\hat{\mathcal{A}}_{\mathcal{N}}\hat{w}_\mathrm{s}^{({\mathcal{N}})}\hat{\mathcal{B}}_{\mathcal{N}}\hat{\mathcal{A}}_1\big]_\mathrm{\bar s}^{(1)}. \nonumber\\
=& \int d \{\mathbf{\Omega}^{(\alpha)}\}\big[\hat{\mathcal{B}}_1\big]_\mathrm{s}^{(1)}\mathrm{Tr_{e}}\big[\hat{\mathcal{A}}_2\hat{w}_\mathrm{s}^{(2)}\hat{\mathcal{B}}_2\hat{\mathcal{A}}_3\cdots\hat{\mathcal{A}}_{\mathcal{N}}\hat{w}_\mathrm{s}^{({\mathcal{N}})}\hat{\mathcal{B}}_{\mathcal{N}}\hat{\mathcal{A}}_1\hat{w}_{\bar{\mathrm s}}^{(1)}\big]\nonumber\\
=&\int d \{\mathbf{\Omega}^{(\alpha)}\}\big[\hat{\mathcal{B}}_1\big]_\mathrm{s}^{(1)}\mathrm{Tr_{e}}\big[\hat{w}_\mathrm{s}^{(2)}\hat{\mathcal{B}}_2\hat{\mathcal{A}}_3\cdots\hat{\mathcal{A}}_{\mathcal{N}}\hat{w}_\mathrm{s}^{({\mathcal{N}})}\hat{\mathcal{B}}_{\mathcal{N}}\hat{\mathcal{A}}_1\hat{w}_{\bar{\mathrm s}}^{(1)}\hat{\mathcal{A}}_2\big],
\end{align}
where from the first  to the second line we used the property in Eq.~\ref{SW_TrA}, and for the last equality we used the property of the trace.

Using $\mathrm{Tr_{e}}\big[\hat{w}_\mathrm{s}^{(2)}\hat{O}\big]= [\hat{O}]_{\mathrm{s}}^{(2)}$ (Eq.~\ref{A-sw-map}), we can further express the last line of Eq.~\ref{re-arrange2} as follows
\begin{align}\label{split}
    & \int d \{\mathbf{\Omega}^{(\alpha)}\}\big[\hat{\mathcal{B}}_1\big]_\mathrm{s}^{(1)}\big[\hat{\mathcal{B}}_2\hat{\mathcal{A}}_3\cdots\hat{\mathcal{A}}_{\mathcal{N}}\hat{w}_\mathrm{s}^{({\mathcal{N}})}\hat{\mathcal{B}}_{\mathcal{N}}\hat{\mathcal{A}}_1\hat{w}_\mathrm{\bar s}^{(1)}\hat{\mathcal{A}}_2\big]_\mathrm{s}^{(2)} \nonumber \\
    = & \int d \{\mathbf{\Omega}^{(\alpha)}\}\big[\hat{\mathcal{B}}_1\big]_\mathrm{s}^{(1)}\big[\hat{\mathcal{B}}_2\big]_\mathrm{s}^{(2)}\big[\hat{\mathcal{A}}_3\cdots\hat{w}_\mathrm{s}^{({\mathcal{N}})}\hat{\mathcal{B}}_{\mathcal{N}}\hat{\mathcal{A}}_1\hat{w}_\mathrm{\bar s}^{(1)}\hat{\mathcal{A}}_2\big]_\mathrm{\bar s}^{(2)} \nonumber \\
    & \hspace{3.7cm}\vdots \nonumber \\
    = & \int d \{\mathbf{\Omega}^{(\alpha)}\}\prod_{\gamma=1}^{\mathcal{N}}\big[\hat{\mathcal{B}}_\gamma\big]_\mathrm{s}^{(\gamma)}\cdot\mathrm{Tr_e}\Big[\prod_{\gamma=1}^{\mathcal{N}}\hat{\mathcal{A}}_\gamma\hat{w}_\mathrm{\bar s}^{(\gamma)}\Big],
\end{align}
where from the second line to the last line of the above equation, we have repeated the procedure elaborated in Eq.~\ref{re-arrange2} for all indices, and the trace in the last line of Eq.~\ref{split} is introduced from $[\cdots]_\mathrm{\bar{s}}^{(\mathcal{N})}$.

Using Eq.~\ref{split} we can rewrite Eq.~\ref{eq:kubo} as follows
\begin{align}
    C_{AB}^{[{\mathcal{N}}]}(t)=& \frac{1}{\cal{Z}}\int d \{R_\alpha'\}\int d\{R_\alpha''\}\int d\{\mathbf{\Omega}^{(\alpha)}\} \\
    & \times\frac{1}{{\mathcal{N}}}\sum_{\alpha=1}^{\mathcal{N}}\mathrm{Tr_e}\Bigg[\prod_{\gamma\neq \alpha}^{\mathcal{N}}\big\langle R_{\gamma-1}''\big|e^{-\beta_{\mathcal{N}}\hat{H}}\big|R_\gamma'\big\rangle\hat{w}_\mathrm{\bar s}^{(\gamma)}\nonumber\\
    &\times\big\langle R_{\alpha-1}''\big|e^{-\beta_{\mathcal{N}}\hat{H}}\hat{A}\big|R_\alpha'\big\rangle\hat{w}_\mathrm{\bar s}^{(\alpha)}\Bigg] \nonumber \\
    & \times\prod_{\gamma\neq\mathcal{N}}\Big[\big\langle R_{\gamma}''\big|e^{\frac{i}{\hbar}\hat{H}t}e^{-\frac{i}{\hbar}\hat{H}t}\big|R_\gamma'\big\rangle\Big]_\mathrm{s}^{(\gamma)}\nonumber\\
    &\times\Big[\big\langle R_{\mathcal{N}}''\big|e^{\frac{i}{\hbar}\hat{H}t}\hat{B}e^{-\frac{i}{\hbar}\hat{H}t}\big|R_{\mathcal{N}}'\big\rangle\Big]_\mathrm{s}^{(\mathcal{N})}.\nonumber
\end{align}
Further using the cyclic-symmetric property to write the operator $\hat{B}$ into a bead-averaged fashion ({\it i.e.}, placing $\hat{B}$ in different blocks), we obtain Eq.~\ref{CABN} of the main text.

\section{SM-NRPMD for two-state systems}\label{2state}
We consider a two-level system $\hat{H} = \frac{\hat{P}^2}{2M}\hat{\mathcal{I}} + U_0(\hat{R})\hat{\mathcal{I}} + \hat{V}_e(\hat{R})$ where $\hat{\mathcal{I}}$ is the $2\times2$ identity matrix, and
\begin{equation}
    \hat{V}_\mathrm{e}(\hat{R}) =
    \begin{pmatrix}
        V_{11}(\hat{R}) & V_{12}(\hat{R})\\ 
        V_{21}(\hat{R}) & V_{22}(\hat{R})
    \end{pmatrix}.
\end{equation}
For this special case, $f_{ijk}=\varepsilon_{ijk}$ and $d_{ijk}=0$, all the equations in the main text remain general. Nevertheless, it will be beneficial to explicitly give several key equations under this special limit, whereas more detailed discussion on $SU(2)$ can be found in the previous work on spin-LSC\cite{runesonrichardson2019} as well as in spin-mapping non-adiabatic RPMD (SM-NRPMD).\cite{bossion2021}

Using the $SU(2)$ representation, one can express the original two-states Hamiltonian as follows\cite{runesonrichardson2019}
\begin{equation}\label{eq:spin_ham}
\hat{H} = \mathcal{H}_0\hat{\mathcal{I}} + \frac{1}{\hbar}\mathbf{H}\cdot\hat{\boldsymbol{\mathcal{S}}}=\mathcal{H}_0\hat{\mathcal{I}} + \frac{1}{\hbar}({\mathcal H}_{x}\cdot\hat{\mathcal S}_{x}+ {\mathcal H}_{y}\cdot\hat{\mathcal S}_{y}+ {\mathcal H}_{z}\cdot\hat{\mathcal S}_{z}),
\end{equation}
where $\hat{\mathcal{S}}_i = \frac{\hbar}{2}\hat{\sigma}_{i}~(\mathrm{for} \ i\in\{x,y,z\})$ with $\hat{\sigma}_i$ as the Pauli matrices, and $\mathcal{H}_0= \frac{\hat{P}^2}{2m} + U_0(\hat{R}) + \frac{1}{2}(V_{11}(\hat{R}) + V_{22}(\hat{R}))$, 
${\mathcal H}_x = 2 \mathrm{Re}(V_{12}(\hat{R}))$, 
${\mathcal H}_y = 2 \mathrm{Im}(V_{12}(\hat{R}))$,
${\mathcal H}_z= V_{11}(\hat{R}) - V_{22}(\hat{R})$, which is the $N=2$ limit of Eq.~\ref{eq:Hk}.

For $N=2$, the spin coherent state in Eq.~\ref{eq:omega-expand} is expressed as 
\begin{equation}\label{spin2}
|\boldsymbol{\Omega}\rangle=\cos\frac{\theta_1}{2}|1\rangle+\sin\frac{\theta_1}{2}e^{i\varphi_1}|2\rangle,
\end{equation}
and the expectation value of the spin operator is
\begin{equation}\label{eq:spin-exp}
\hbar\Omega^{(\alpha)}_{i}(\mathbf{\Omega})=\langle \mathbf{\Omega}^{(\alpha)}|\hat{S}_{i}|\mathbf{\Omega}^{(\alpha)}\rangle, \hspace{0.5cm} i\in\{x,y,z\},
\end{equation}
where $\Omega_x = \frac{1}{2}\sin\theta\cos\varphi$, $\Omega_y =\frac{1}{2}\sin\theta\sin\varphi$, $\Omega_z=\frac{1}{2}\cos\theta$ as the special case of Eqs.~\ref{eq:Omega_a}-\ref{eq:Omega_g}, and Eq.~\ref{dOmega} becomes $d\mathbf{\Omega}=\frac{1}{2\pi}\sin\theta d\theta d\varphi$.

The derivation procedure of the Kubo-transformed TCF and exact Liouvillian are same as outlined in the main text. All the approximations made to obtain the Matsubara and SM-NRPMD expressions of the TCFs are identical. For the $N=2$ special case (when $\hat{V}$ is purely real), the SM-NRPMD Hamiltonian is
\begin{align}
\mathcal{H}_\mathcal{N}&=H_\mathrm{rp}^{[\mathcal{N}]}+r_\mathrm{s}\sum_{\alpha=1}^{\mathcal{N}}\boldsymbol{\mathcal{H}}^{(\alpha)}\cdot\boldsymbol{\Omega}^{(\alpha)},\\
&=\sum_{\alpha=1}^{\mathcal{N}}\Big[{\mathcal{H}}_0(R_\alpha,P_\alpha)+\frac{m}{2\beta_{\mathcal{N}}^2\hbar^2}(R_\alpha-R_{\alpha-1})^2\nonumber\\
&~+\Big(\frac{1}{2}+r_\mathrm{s}\cos\theta\Big)\cdot V_{11}(R_{\alpha})+\Big(\frac{1}{2}-r_\mathrm{s}\cos\theta\Big)\cdot V_{22}(R_{\alpha})\nonumber\\
&~+2r_\mathrm{s}\sin\theta\cos\varphi\cdot V_{12}(R_{\alpha})]\Big].\nonumber
\end{align}
with $H_\mathrm{rp}^{[\mathcal{N}]}$ defined in the main text, and $\boldsymbol{\mathcal{H}}^{(\alpha)}=\{\mathcal{H}_{x}(R_{\alpha}),\mathcal{H}_{y}(R_{\alpha}),\mathcal{H}_{z}(R_{\alpha})\}$.

The electronic EOMs under the linearization approximation can be expressed as
\begin{equation}\label{domegadt}
\dot{{\bf \Omega}}^{(\alpha)}=\frac{1}{\hbar}\sum_{j,k=1}^3\varepsilon_{ijk}\mathcal{H}_j(R_{\alpha})\Omega^{(\alpha)}_k=\frac{1}{\hbar}{\bf H}(R_{\alpha})\times{\bf \Omega}^{(\alpha)},
\end{equation}
where $\times$ denotes the cross product of two vectors. This equation is the special case of Eq.~\ref{eom-omega}c for $N=2$. Here, $\varepsilon_{ijk}$ is the Levi-Civita symbol, which is the totally antisymmetric structure constant of the  $\mathfrak{su}(2)$ Lie algebra. One can also express Eq.~\ref{domegadt} in terms of the MMST mapping variables, which is Eq.~\ref{eom-pq} with  $N=2$. In addition, Eq.~\ref{domegadt} also has a rather simple expression using the Euler angles $\{\theta^{(\alpha)},\varphi^{(\alpha)}\}$ 
\begin{align}
    \dot{\theta}^{(\alpha)}&=\frac{1}{\hbar}\big(-\mathcal{H}_x\sin\varphi^{(\alpha)}+\mathcal{H}_y\cos\varphi^{(\alpha)}\big),\nonumber\\
    \dot{\varphi}^{(\alpha)}&=\frac{1}{\hbar}\Bigg(\mathcal{H}_z-\mathcal{H}_x\frac{\cos\varphi^{(\alpha)}}{\tan\theta^{(\alpha)}}-\mathcal{H}_y\frac{\sin\varphi^{(\alpha)}}{\tan\theta^{(\alpha)}}\Bigg),
\end{align}
whereas the equivalent equations for $N>2$ are rather complicated, with details provided in Appendix E of Ref.~\citenum{bossion2022}.

In the 2-state case, the initial electronic phase $\hat{\mathbf{\Gamma}}_\mathrm{\bar s}$ (Eq.~\ref{h-phi-gamma}) can be analytically expressed (thanks to the special property of the $SU(2)$ Lie group) as\cite{bossion2021}
\begin{align}
&\hat{\mathbf{\Gamma}}_\mathrm{\bar s}=\prod_{\alpha}^{\mathcal{N}}e^{-\beta_{\mathcal{N}}\frac{1}{\hbar}\sum_k{\mathcal{H}}_k(R_\alpha)\cdot\hat{\mathcal{S}}_k}\cdot\hat{w}_\mathrm{\bar s}^{(\alpha)}\nonumber\\
=&\prod_{\alpha}^{\mathcal{N}}\Bigg[\Bigg(\frac{1}{2}\cosh\eta-r_\mathrm{\bar s}\frac{\boldsymbol{\mathcal{H}}^{(\alpha)}}{\big|\boldsymbol{\mathcal{H}}^{(\alpha)}\big|}\cdot\mathbf{\Omega}^{(\alpha)}\sinh\eta\Bigg)\cdot\hat{\mathcal{I}}\\
    &+\Bigg(r_\mathrm{\bar s}\Omega^{(\alpha)}\cosh\eta\nonumber\\
    &-\Big(\frac{1}{2}\frac{\boldsymbol{\mathcal{H}}^{(\alpha)}}{\big|\boldsymbol{\mathcal{H}}^{(\alpha)}\big|}+ir_{\bar{\mathrm{s}}}\frac{\boldsymbol{\mathcal{H}}^{(\alpha)}}{\big|\boldsymbol{\mathcal{H}}^{(\alpha)}\big|}\times\mathbf{\Omega}^{(\alpha)}\Big)\sinh\eta\Bigg)\cdot\frac{2}{\hbar}\hat{\boldsymbol{\mathcal{S}}}\Bigg],\nonumber
\end{align}
where for convenience we defined
\begin{align}
&\boldsymbol{\mathcal{H}}^{(\alpha)}\cdot\mathbf{\Omega}^{(\alpha)}\equiv\sum_{k=1}^3{\mathcal{H}}_k(R_\alpha)\cdot\Omega_k^{(\alpha)},\nonumber\\
&\Big[\boldsymbol{\mathcal{H}}^{(\alpha)}\times\mathbf{\Omega}^{(\alpha)}\Big]_i\equiv\sum_{i,j,k=1}^3f_{ijk}{\mathcal{H}}_j(R_\alpha)\cdot\mathbf{\Omega}_k^{(\alpha)},\nonumber\\
&\big|\boldsymbol{\mathcal{H}}^{(\alpha)}\big|\equiv\sqrt{\sum_{k=1}^3{\mathcal{H}}^2_k(R_\alpha)}~~;~~\eta\equiv\frac{\beta_{\mathcal{N}}\big|\boldsymbol{\mathcal{H}}^{(\alpha)}\big|}{2}.
\end{align}
The details of the derivation can be found in Appendix D of Ref.~\citenum{bossion2021}. The advantage of having this analytic expression is that it avoids the numerical cost of evaluating the exponential by diagonalizing it. Unfortunately, we did not obtain the analytic expression for the general $N$-level case, due to the totally symmetric structure constants $d_{ijk}$ which do not cancel beyond $N=2$, making the exponential not exactly factorizable in terms of hyperbolic cosines and sines. That said, it might still exist alternative ways to evaluate it to get a closed analytic expression for $\hat{\mathbf{\Gamma}}_\mathrm{\bar s}$ (Eq.~\ref{h-phi-gamma}) for a general $N$-state system.

\section{Non-equilibrium Dynamics}\label{non-eq}
In our previous work, we have justified that  NRPMD is also capable of accurately describing the non-equilibrium TCF. Here, we carry the same procedure and show that the SM-NRPMD is also capable to describe the non-equilibrium TCF. For a given photo-induced process, we are often interested in the reduced density matrix dynamics upon an initial excitation of the molecular system. The reduced density matrix element can be expressed as 
\begin{equation}
\rho_{nm}(t)=\mathrm{Tr}[\hat{\rho}(0)e^{\frac{i}{\hbar}\hat{H}t}|n\rangle\langle m|e^{-\frac{i}{\hbar}\hat{H}t}],
\end{equation}
where the initial density operator $\hat{\rho}(0)=|a\rangle\langle a|\otimes\frac{1}{\mathcal{Z}}e^{-\beta \hat{H}_\mathrm{g}}$ is a tensor product of the electronic and nuclear DOFs, with $\mathcal Z=\mathrm{Tr}[e^{-\beta\hat{H}_\mathrm{g}}]$, and $\hat{H}_\mathrm{g}$ the ground state Hamiltonian 
\begin{equation}
\hat{H}_\mathrm{g}=\hat{T}_R+U_{g}(\hat{R}),
\end{equation}
with the ground state potential ${U}_{g}(\hat{R})$ associated with the ground electronic state $|g\rangle$.

The initial density $\hat{\rho}(0)$ evolves under the influence of the total Hamiltonian $\hat{H}$ of the system. The reduced density matrix elements can be equivalently expressed as a TCF
\begin{equation}
\rho_{nm}(t)=C_{AB}(t) = \frac{1}{\mathcal{Z}}\mathrm{Tr}[e^{-\beta \hat{H}_\mathrm{g}}\hat{A}e^{\frac{i}{\hbar}\hat{H}t}\hat{B}e^{-\frac{i}{\hbar}\hat{H}t}],
\end{equation}
where $\hat{A}=|a\rangle\langle a|$ is the initially occupied electronic state, and $\hat{B}=|n\rangle\langle m|$. Because $\hat{A}$ and $\hat{H}_\mathrm{g}$ commute, we have
\begin{align}
C^{[\mathcal{N}]}_{AB}(t)&=\frac{1}{\mathcal{Z}\beta}\int_{0}^{\beta} d\lambda \mathrm{Tr}[e^{-(\beta - \lambda)\hat{H}_\mathrm{g}}\hat{A}e^{-\lambda \hat{H}_\mathrm{g}}\hat{B}(t)]\\
&=\frac{\int_{0}^{\beta} d\lambda}{\beta}\cdot\frac{1}{\mathcal{Z}} \mathrm{Tr}[e^{-\beta\hat{H}_\mathrm{g}}\hat{A}\hat{B}(t)]=C_{AB}(t).\nonumber
\end{align}
Hence, one can rewrite the reduced density matrix elements $\rho_{nm}(t)$ into the Kubo-transformed time-correlation function $C^{[\mathcal{N}]}_{AB}(t)$. This Kubo-transformed TCF is {\it not} an equilibrium correlation function. Nevertheless, the Kubo-transformed structure allows us to express it as the discrete version of the time-correlation function. Following the same derivation outlined in the main text, we can obtain the out-of-equilibrium SM-NRPMD TCF
\begin{align}
\rho_{nm}(t)=&\frac{1}{{\cal{Z}}(2\pi\hbar)^{\mathcal{N}}}\int\{ d R\}\int\{ d P\}\int\{ d \mathbf{\Omega}\}e^{-\beta_{\mathcal{N}} H_\mathrm{rp}^{[\mathcal{N}]}}\nonumber\\
    &\times\mathrm{Tr_e}\Big[\prod_{\alpha}^{\mathcal{N}}\hat{w}_\mathrm{\bar s}^{(\alpha)}|a\rangle\langle a|\Big]e^{\mathcal{L}_\mathrm{rp}^{[\mathcal{N}]}t}\big[|n\rangle\langle m|\big]_\mathrm{s},
\end{align}
where $H_\mathrm{rp}^{[\mathcal{N}]}$ is expressed as
\begin{equation}\label{eq:Hrpn}
    H_\mathrm{rp}^{[\mathcal{N}]}=\sum_{\alpha=1}^{\mathcal{N}}\Big[{\mathcal{H}}_g(R_\alpha,P_\alpha)+\frac{m}{2\beta_{\mathcal{N}}^2\hbar^2}(R_\alpha-R_{\alpha-1})^2\Big],
\end{equation}
and the bead-averaged estimator $[|n\rangle\langle m|]_\mathrm{s}$ is expressed in Eq.~\ref{n-bead}. The initial electronic phase $\mathrm{Tr_e}\big[\prod_{\alpha}^{\mathcal{N}}\hat{w}_\mathrm{\bar s}^{(\alpha)}|a\rangle\langle a|\big]$ is expressed as
\begin{align}
&\mathrm{Tr_e}\Big[\prod_{\alpha}^{\mathcal{N}}\hat{w}_\mathrm{\bar s}^{(\alpha)}|a\rangle\langle a|\Big]=\frac{1}{\mathcal{N}}\sum_{\alpha=1}^{\mathcal{N}}\mathrm{Tr_e}\Big[\prod_{\gamma\leq\alpha}^{\mathcal{N}}\hat{w}_\mathrm{\bar s}^{(\gamma)}|a\rangle\langle a|\prod_{\gamma>\alpha}^{\mathcal{N}}\hat{w}_\mathrm{\bar s}^{(\gamma)}\Big].
\end{align}
Further, the Liouvillian $\mathcal{L}_\mathrm{rp}^{[\mathcal{N}]}$ corresponds to the EOM in Eq.~\ref{eom-pq}.

\section{MMST-based NRPMD approach and the Exact Simulation}\label{exact-nrpmd}
The original MMST-based NRPMD approach to compute position auto-correlation function was proposed as\cite{richardson2013} 
\begin{align}\label{eqn:NRPMD_original}
C^{[\mathcal{N}]}_{RR}(t)=& \frac{1}{\mathcal{Z}_\mathcal{N}}\int d \{R_\alpha\}\int d \{P_\alpha\}\int d \{{\bf q}^{(\alpha)}\}\int d \{{\bf p}^{(\alpha)}\} \nonumber\\
&\times e^{-\beta_\mathcal{N} H^{[\mathcal{N}]}_\mathrm{rp}}\phi e^{-\frac{\mathcal{G}_\mathcal{N}}{\hbar}}\mathrm{Tr_e}\big[\hat{\boldsymbol{\Gamma}}\big]  \bar{R}e^{\mathcal{L}_{\mathrm{rp}}^{[N]}t}\bar{R}(t),
\end{align}
where $d \{{\bf q}^{(\alpha)}\}=\prod_{\alpha=1}^{\mathcal{N}}d{\bf q}^{(\alpha)}$, $d \{{\bf p}^{(\alpha)}\}=\prod_{\alpha=1}^{\mathcal{N}}d{\bf p}^{(\alpha)}$, and ${\bf q}^{(\alpha)}=\{q_{1}^{(\alpha)},\cdots,q_{N}^{(\alpha)}\}$, ${\bf p}^{(\alpha)}=\{p_{1}^{(\alpha)},\cdots,p_{N}^{(\alpha)}\}$. Further, $\phi=\big(\frac{4}{\pi^N}\big)^{\mathcal{N}}$, $\mathcal{G}_\mathcal{N}=\sum_{\alpha=1}^{\mathcal{N}}([\mathbf{q}^{(\alpha)}]^{\mathrm{T}}\mathbf{q}^{(\alpha)}+[\mathbf{p}^{(\alpha)}]^{\mathrm{T}}\mathbf{p}^{(\alpha)})$, and $\bar{R}=\frac{1}{\mathcal{N}}\sum_{\alpha=1}^{\mathcal{N}}R_\alpha$ is the bead average position. In addition, $\hat{\boldsymbol{\Gamma}}$ is expressed as\cite{richardson2013}
\begin{align}
    \hat{\boldsymbol{\Gamma}}=\prod_{\alpha=1}^{\mathcal{N}}{\pmb{\mathcal M}}(R_\alpha){\bf q}^{(\alpha)}[{\bf q}^{(\alpha)}]^\mathrm{T}{\pmb{\mathcal M}}(R_\alpha){\bf p}^{(\alpha)}[{\bf p}^{(\alpha)}]^\mathrm{T}
\end{align}
where $\mathcal{M}_{nm}(R_\alpha) = \langle n|e^{-\frac{1}{2}\beta_\mathcal{N} \hat{V}_\mathrm{e}(R_\alpha)}|m\rangle$, and $[{\bf q}^{(\alpha)}]^\mathrm{T}$ represents the transpose of the ${\bf q}^{(\alpha)}$ row matrix. The Liouvillian $\mathcal{L}_{\mathrm{rp}}^{[\mathcal{N}]}$ (in terms of the MMST mapping variables) is identical to the corresponding one in SM-NRPMD, both corresponding to the EOMs in Eq.~\ref{eom-pq}.

To compute a population TCF, the NRPMD approach uses\cite{richardson2013}
\begin{align}
C^{[\mathcal{N}]}_{mn}(t)=& \frac{1}{\mathcal{Z}_\mathcal{N}}\int d \{R_\alpha\}\int d \{P_\alpha\}\int d \{{\bf q}_\alpha\}\int d \{{\bf p}_\alpha\} \nonumber\\
&\times e^{-\beta_\mathcal{N}H^\mathcal{N}_\mathrm{rp}}\phi e^{-\frac{\mathcal{G}_\mathcal{N}}{\hbar}}\mathrm{Tr_e}\big[\hat{\boldsymbol{\Gamma}}\hat{A}\big]e^{\mathcal{L}_{\mathrm{rp}}^{[\mathcal{N}]}t}\bar{B}_n,
\end{align}
where $\bar{B}_n$ is the bead average projection operator expressed in Eq.~\ref{n-bead}, with $\gamma=1$ corresponding to the ZPE parameter in the MMST mapping theory,\cite{stock1997,thoss1999} and $\mathrm{Tr_e}\big[\hat{\boldsymbol{\Gamma}}\hat{A}\big]$ is computed as
\begin{align}
&\mathrm{Tr_e}\big[\hat{\boldsymbol{\Gamma}}\hat{A}\big]=\frac{1}{\mathcal{N}}\sum_{\alpha}\mathrm{Tr_e}\Big[\prod_{\gamma\leq\alpha}^{\mathcal{N}}{\boldsymbol{\mathcal M}}(R_\gamma){\bf q}^{(\gamma)}[{\bf q}^{(\gamma)}]^\mathrm{T}\nonumber\\
&\times{\boldsymbol{\mathcal M}}(R_\gamma){\bf p}^{(\gamma)}[{\bf p}^{(\gamma)}]^\mathrm{T}\hat{A}\prod_{\gamma>\alpha}^{\mathcal{N}}{\boldsymbol{\mathcal M}}(R_\gamma){\bf q}^{(\gamma)}[{\bf q}^{(\gamma)}]^\mathrm{T}\nonumber\\
&\times{\boldsymbol{\mathcal M}}(R_\gamma){\bf p}^{(\gamma)}[{\bf p}^{(\gamma)}]^\mathrm{T}\Big].
\end{align}
Thus, the dynamics between the SM version and the MMST version of NRPMD are identical, and the difference between them comes from the initial distributions of the mapping variables and the choice and justification of the $\gamma$ value (so-called ZPE parameter).

Note that this is the original version of the MMST-based NRPMD approach,\cite{richardson2013,richardson2017} which has a different sampling than the NRPMD method derived from Kubo-transformed TCF in Ref.~\citenum{chowdhury2021}. In fact, when derived from the Matsubara approximation, the distribution is identical to the MV-RPMD method.\cite{ananth2013} Due to the severe sign problem encountered for the three states system, we could not converge our results with the MV-RPMD sampling approach. Nevertheless, the original MMST-based NRPMD method gives results very similar to the NRPMD method derived in Ref.~\citenum{chowdhury2021}, hence, we just use the original version of the NRPMD algorithm. The numerical details are referred back to the original NRPMD papers in Ref.~\citenum{richardson2013} and Ref.~\citenum{richardson2017}.

The exact results are obtained by explicitly computing the Kubo-transformed TCF as follows
\begin{align}\label{kubo-exact}
&C_{AB}^\mathrm{K}(t)=\frac{1}{{\cal{Z}}{\mathcal{N}}}\sum_{\alpha=1}^{\mathcal{N}}\mathrm{Tr}\big[e^{-\beta_{\mathcal{N}}({\mathcal{N}}-\alpha)\hat{H}}\hat{A}e^{-\beta_{\mathcal{N}}\alpha\hat{H}}e^{\frac{i}{\hbar}\hat{H}t}\hat{B}e^{-\frac{i}{\hbar}\hat{H}t}\big],
\end{align}
where the trace is for both electronic and nuclear DOFs. To evaluate the trace, we explicitly calculate the eigenstate of $\hat{H}$.

The full Hilbert space of the entire system is represented with a basis composed by a tensor product of the electronic subspace $\{|n\rangle\}$ and the Fock states $\{|\chi_{a}\rangle\}$ of the nuclear DOFs as follows
\begin{equation}
|n\rangle\otimes|\chi_{a}\rangle\equiv|n,\chi_a\rangle,
\end{equation}
where $|n\rangle$ is the electronic diabatic states and $|\chi_{a}\rangle$ is the eigenstate of $\frac{\hat{P}^2}{2m} + \frac{1}{2}m\omega\hat{R}^2$. We solve the eigenvalue problem of the total Hamiltonian as follows
\begin{equation}\label{TISE}
\hat{H}|\Psi_{\nu}\rangle=E_{\nu}|\Psi_{\nu}\rangle, 
\end{equation}
where $|\Psi_{\nu}\rangle=\sum_{n,a}c^{\nu}_{n,a}|n,\chi_{a}\rangle$ is the eigenstate of the {\it entire} system (including both electronic and nuclear DOFs), and $c^{\nu}_{n,a}=\langle n, \chi_{a}|\Psi_{\nu}\rangle$ is the expansion coefficient. Both $E_{\nu}$ and $c^{\nu}_{n,a}$ can be directly obtained from diagonalizing the matrix of $\hat{H}$ with the matrix elements $\langle \chi_{b}, m|\hat{H}|n,\chi_{a}\rangle$. Evaluating the $\mathrm{Tr}[\hat{O}]$ in Eq.~\ref{kubo-exact} as $\sum_{\nu}\langle \Psi_{\nu}|\hat{O}|\Psi_{\nu}\rangle$, the Kubo-transformed TCF is expressed as
\begin{align}\label{KTCFQ}
    C_{AB}^\mathrm{K}(t)=\frac{1}{{\cal{Z}}{\mathcal{N}}}\sum_\alpha^{\mathcal{N}}\sum_{\nu,\mu}&e^{-\beta_{\mathcal{N}}({\mathcal{N}}-\alpha)E_\nu}[\hat{A}]_{\nu\mu}e^{-\beta_{\mathcal{N}}\alpha E_\mu}\nonumber\\
    &\times e^{\frac{i}{\hbar}E_\mu t}[\hat{B}]_{\mu\nu}e^{-\frac{i}{\hbar}E_\nu t},
\end{align}
where the matrix $[\hat{A}]_{\nu\mu}$ is 
\begin{equation}
[\hat{A}]_{\nu\mu}=\langle \Psi_{\nu}|\hat{A}|\Psi_{\mu}\rangle=\sum_{n,a}\sum_{m,b}(c^{\nu}_{n,a})^{*}\langle \chi_{a},n|\hat{A}|m,\chi_{b}\rangle c^{\mu}_{m,b},
\end{equation}
and similarly for $[\hat{B}]_{\mu\nu}=\langle \Psi_{\mu}|\hat{B}|\Psi_{\nu}\rangle$. When $\hat{A}=\hat{R}$, $[\hat{A}]_{\nu\mu}=\sqrt{\frac{\hbar}{m\omega}}\sqrt{\frac{a+1}{{2}}}\delta_{m,n+1}$, and when $\hat{A}=|m\rangle\langle m|$, $[\hat{A}]_{\nu\mu}=\delta_{mn}\delta_{ab}$. For the partition function in Eq.~\ref{KTCFQ}, one can evaluate it exactly as
\begin{equation}
\mathcal{Z}=\mathrm{Tr}[e^{-\beta \hat{H}}]=\sum_{\nu}\langle \Psi_{\nu}|e^{-\beta \hat{H}}|\Psi_{\nu}\rangle=\sum_{\nu}e^{-\beta E_{\nu}}.\nonumber
\end{equation}

For the model systems studied here, solving the eigenvalue problem in Eq.~\ref{TISE} requires about 50 Fock states $\{|\chi_{a}\rangle\}$ to represent the nuclear operators, and the TCF in Eq.~\ref{KTCFQ} converges with $\mathcal{N}=6$.

%
%
%

%
%
\end{document}